\documentclass[aps,reprint,longbibliography]{revtex4-1}

\usepackage{floatrow}
\usepackage{graphicx}
\usepackage{here}
\usepackage{subcaption}

\DeclareCaptionLabelFormat{myformat}{#2}
\usepackage{amsmath,amssymb,amsfonts}
\usepackage{mathtools}
\usepackage{physics}

\usepackage{color}
\usepackage{xcolor}
\usepackage[T1]{fontenc}
\usepackage{textcomp}
\usepackage{hyperref}
\hypersetup{
     colorlinks   = true,
     linkcolor    = blue,
     citecolor    = blue,
     urlcolor     = blue
}

\DeclareUnicodeCharacter{2061}{}

\DeclareMathOperator{\Pf}{Pf}

\begin{document}

\title{Wiener-Hopf factorization and non-Hermitian topology for Amoeba formulation in one-dimensional multiband systems}

\author{Shin Kaneshiro}
\email{kaneshiro.shin.88a@st.kyoto-u.ac.jp}
\author{Robert Peters}
\affiliation{Department of Physics, Kyoto University, Kyoto 606-8502, Japan}
\date{\today}

\begin{abstract}
The non-Hermitian skin effect (NHSE), characterized by the extensive localization of bulk modes at the boundaries, has attracted significant attention as a hallmark feature of non-Hermitian topology.
This localization invalidates the conventional Bloch band theory, necessitating an analysis under open boundary conditions even in the thermodynamic limit.
The Amoeba formulation addresses this challenge by computing the spectral potential rather than the spectrum itself. 
Based on the (strong) Szeg\"o limit theorem and its topological generalization, this approach reduces the evaluation of the potential to an optimization problem involving the Ronkin function.
However, while the generalized Szeg\"o  limit theorem is formally applicable in arbitrary dimensions, its implementation is limited to single-band systems, and its applicability to multiband systems remains unclear even in one-dimensional systems.
In this paper, we establish the Wiener-Hopf factorization (WHF) of the non-Bloch Hamiltonian as a powerful framework, providing a unified and rigorous foundation for Amoeba analysis in one-dimensional multiband systems.
By combining the WHF with Hermitian doubling, we first elucidate the applicability criteria for the generalized Szeg\"o limit theorem in multiband systems. 
We then show that the WHF provides the natural mathematical origin for the symmetry-decomposed Ronkin function in symmetry class AII$^\dagger$, leading to a rigorous proof of the generalized Szeg\"o limit theorem for these systems and opening a path toward systematic generalizations to other symmetry classes. 
\end{abstract}

\maketitle

\section{Introduction}
Non-Hermitian matrices naturally arise in various kinds of systems, from classical waves to dissipative quantum systems \cite{YAshida_AdvPhys_2020_NonHermitianPhysics, EBergholtz_RMP_2021_ExceptionalTopologyOf, KDing_NatRevPhys_2022_NonHermitianTopology}
The interplay of non-Hermiticity and translation symmetry fundamentally alters bulk-band topology, introducing point-gap phases without Hermitian counterparts \cite{ZGong_PRX_2018_TopologicalPhasesOf, KKawabata_PRX_2019_SymmetryAndTopology}.
A prominent consequence is the non-Hermitian skin effect (NHSE) \cite{
VMartinezAlvarez_PRB_2018_NonHermitianRobust,
SYao_PRL_2018_EdgeStatesAnd, FKunst_PRL_2018_BiorthogonalBulkBoundary,
KYokomizo_PRL_2019_NonBlochBand, CLee_PRB_2019_AnatomyOfSkin, FKunst_PRB_2019_NonHermitianSystems, FSong_PRL_2019_NonHermitianTopological,
MBrandenbourger_NatCommun_2019_NonReciprocalRobotic,
LFoaTorres_JPhysMater_2019_PerspectiveOnTopological,
KZhang_PRL_2020_CorrespondenceBetweenWinding, ZYang_PRL_2020_NonHermitianBulk, NOkuma_PRL_2020_TopologicalOriginOf, LLi_NatCommun_2020_CriticalNonHermitian,
KKawabata_PRB_2020_HigherOrderNon,
SLonghi_PRL_2020_NonBlochBand,TYoshida_PRR_2020_MirrorSkinEffect,
NOkuma_PRB_2020_HermitianZeroModes,
AGhatak_PNAS_2020_ObservationOfNon, SWeidemann_Science_2020_TopologicalFunnelingOf, THelbig_NatPhys_2020_GeneralizedBulkBoundary, THofmann_PRR_2020_ReciprocalSkinEffect, LXiao_NatPhys_2020_NonHermitianBulkBoundary, WGou_PRL_2020_TunableNonreciprocalQuantum, DBorgnia_PRL_2020_NonHermitianBoundaryModes,
NOkuma_PRB_2021_QuantumAnomalyNon, KKawabata_PRL_2021_TopologicalFieldTheory, ROkugawa_PRB_2021_NonHermitianBand,
SLiu_Research_2021_NonHermitianSkin, XZhang_NatCommun_2021_ObservationOfHigher, LPalacios_NatCommun_2021_GuidedAccumulationOf,  
DWu_PRB_2022_ConnectionsBetweenThe, QLiang_PRL_2022_DynamicSignaturesOf, FSong_Proc_2022_NonBlochPt, SLonghi_PRB_2022_NonHermitianSkin, ZGu_NatCommun_2022_TransientNonHermitian,
KKawabata_PRX_2023_EntanglementPhaseTransition, CLi_PRL_2023_EnhancementOfSecond, SManna_CommunPhys_2023_InnerSkinEffects,
YZhang_arXiv_2024_HybridSkinTopological, DNakamura_PRL_2024_BulkBoundaryCorrespondence, YNakai_PRB_2024_TopologicalEnhancementOf,
SIshikawa_PRB_2024_NonHermitianZ4Skin,
XZhang_SciPostPhys_2024_ObservationOfNon, 
ZWei_arXiv_2025_GeneralizedNonHermitianSkin, RShen_NatCommun_2025_ObservationOfThe}
where bulk eigenmodes accumulate extensively near the system boundary.
This exponential localization invalidates the plane-wave expansions that underpin conventional Bloch band theory. 
As a result, the Fourier-based approach under periodic boundary conditions (PBC) no longer approximates systems under open boundary conditions (OBC), often necessitating a direct analysis of the OBC Hamiltonian.
However, a real-space treatment can obscure the underlying bulk topology and is susceptible to numerical instabilities.

For one-dimensional systems, the non-Bloch band theory \cite{SYao_PRL_2018_EdgeStatesAnd, KYokomizo_PRL_2019_NonBlochBand}, provides a comprehensive framework for describing OBC spectral, topological, and dynamical properties \cite{KKawabata_PRB_2020_NonBlochBand, 
HHu_PRL_2021_KnotsAndNon, KYokomizo_PRB_2021_NonBlochBand, TLi_PRR_2021_NonBlochQuench, GGuo_NJP_2021_NonHermitianBulkboundary, WXue_PRB_2021_SimpleFormulasOf, KYokomizo_PRB_2021_ScalingRuleFor,
YLi_PRB_2022_TopologicalEnergyBraiding, HLi_PRB_2022_ExactFormulasOf,
HLiu_PRB_2023_ModifiedGeneralizedBrillouin, TTai_PRB_2023_ZoologyOfNon, YHu_PRR_2023_GreensFunctionsOf,
YHu_PRB_2024_NonBlochBandTheory, KMatsushima_arXiv_2024_NonBlochBand, HWang_arXiv_2024_NonBlochSelf, KRoy_arXiv_2024_TopologicalCharacterizationOf, ZYang_PRB_2024_EntangelmentEntropyOn, YFu_PRB_2024_BraidingTopologyOf, SWang_PRB_2024_GeneralTheoryFor,  SVerma_arXiv_2024_NonBlochBand, YHu_PRL_2024_GeometricOriginOf,
QLi_arXiv_2025_PhaseSpaceGeneralized, JZhong_arXiv_2025_UnveilingNonHermitianBand, HMeng_arXiv_2025_GeneralizedBrillouinZone, KRoy_arXiv_2025_FloquetNonBlochFormalism, DNakamura_arxiv_2025_NonsymmorphicTopologicalPhases}
In this theory, the eigenmodes are expanded using exponentially modulated plane waves, known as non-Bloch states, whose complex wavenumbers trace out the generalized Brillouin zone (GBZ) \cite{SYao_PRL_2018_EdgeStatesAnd, KYokomizo_PRL_2019_NonBlochBand}.
Extending this framework to higher dimensions, however, remains a significant challenge, since the localization behavior can no longer be encoded by a single complex momentum. 
Indeed, higher-dimensional systems host exotic boundary physics, including skin-topological edge states~\cite{CLee_PRL_2019_HybridHigherOrder, WZhu_JPhysCondMat_2024_ABriefReview}, which defy a straightforward one-dimensional description.
As a result, several distinct formulations have been proposed 
\cite{SYao_PRL_2018_NonHermitianChern, TLiu_PRL_2019_SecondOrderTopological, KYokomizo_PRB_2023_NonBlochBands, HJiang_PRL_2023_DimensionalTransmutationFrom, ZXu_arXiv_2023_TwoDimensionalAsymptotic,
KZhang_PRB_2024_EdgeTheoryOf,
KZhang_PRX_2025_AlgebraicNonHermitian, CWang_arXiv_2025_UniversalTheoryFor, LLi_PRB_2025_ExactSolutionsDisentangle}.

The Amoeba formulation \cite{HWang_PRX_2024_AmoebaFormulationOf, SWang_PRB_2024_ConstraintsOfInternal, YXiong_PRB_2024_GraphMorphologyOf, YXiong_arXiv_2024_NonHermitianSkin, HHu_SciBull_2025_TopologicalOriginOf, AYang_CP_2025_TailoringBoundStateGeometry} addresses this challenge by focusing on the spectral potential, a thermodynamic-limit quantity derived from the characteristic polynomial under OBC. 
Through the (strong) Szeg\"o limit theorem \cite{GSzego_MathAnn_1915_EinGrenzwertsatzUber, HWidom_AdvMath_1974_AsymptoticBehaviorOf, HWidom_AdvMath_1976_AsymptoticBehaviorOf} and its topological generalization \cite{HWang_PRX_2024_AmoebaFormulationOf}, this OBC spectral potential can be related to a modulated PBC spectral potential, allowing the computation to be performed in momentum space. 
These results imply that the OBC spectral potential can be obtained by optimizing the Ronkin function, whose argument directly controls the inverse localization length of non-Bloch waves.

While the generalized Szeg\"o limit theorem is, in principle, applicable in arbitrary spatial dimensions, the current Amoeba formalism based on this theorem is limited to single-band systems.
In multiband settings, the Ronkin function mixes contributions across bands; when degenerate states with different localization lengths occur, these contributions compete during the optimization of the Ronkin function, causing the standard Amoeba formulation to fail even in one dimension.
This breakdown is unavoidable in the presence of symmetry-protected degeneracies, such as for a transpose-type time-reversal symmetry (TRS$^\dagger$), where Kramers pairs localize on opposite edges.

In multiband settings, one might hope to resolve the failure of the Amoeba formulation by decomposing the Ronkin function into bandwise contributions.
However, a straightforward mathematical decomposition is generally not feasible:
the mathematical properties of the Ronkin function are typically not preserved under bandwise separation.
In one-dimensional two-band system with TRS$^\dagger$, we previously demonstrated that a symmetry-guided, phenomenological decomposition consistent with non-Bloch band theory \cite{KKawabata_PRB_2020_NonBlochBand} can nevertheless be constructed, leading to the symmetry-decomposed Ronkin functions that separate the contributions of Kramers pairs while retaining the required structure \cite{SKaneshiro_PRB_2025_SymplecticAmoebaFormulation}.
Nevertheless, its relation to the generalized Szeg\"o limit theorem remained unclear, and extensions beyond TRS$^\dagger$, particularly to class~A multiband systems without protecting symmetries, were still lacking.

Originally, the Szeg\"o limit theorem was formulated using the Wiener-Hopf factorization (WHF) for topologically trivial Hamiltonians \cite{HWidom_AdvMath_1974_AsymptoticBehaviorOf, HWidom_AdvMath_1976_AsymptoticBehaviorOf}.
Recently, Ref.~\cite{AAlase_AnnPhys_2023_WienerHopfFactorization} demonstrated that the WHF itself offers a direct topological interpretation for one-dimensional Hermitian Hamiltonians: the WHF partial indices count the number of edge modes and thereby establish the bulk-boundary correspondence.
This intimate connection between WHF and topology is central to the generalization of the original theorem to topologically nontrivial Hermitian Hamiltonians, a result known as the Modified Szeg\"o limit theorem \cite{EBasor_JStatPhys_2019_ModifiedSzegoWidomAsymptotics}.

Inspired by these developments, we apply the WHF in this paper to non-Hermitian systems combined with Hermitian doubling \cite{ZGong_PRX_2018_TopologicalPhasesOf, KKawabata_PRX_2019_SymmetryAndTopology}.
We demonstrate that the WHF of a non-Hermitian non-Bloch Hamiltonian faithfully captures its topological properties.
Moreover, this framework allows us to identify the WHF partial indices as the exact criteria determining when the generalized Szeg\"o limit theorem remains valid in one-dimensional multiband systems.
By reinterpreting the generalized Szeg\"o limit theorem as a corollary of the modified Szeg\"o theorem, we derive the necessary correction terms to the optimization formalism.
Finally, this approach reveals the mathematical origin of the symmetry-decomposed Ronkin functions and provides a direct proof of the generalized Szeg\"o limit theorem for class~AII$^\dagger$ systems.

The remainder of this paper is organized as follows.
In Sec.~\ref{sec: review of Amoeba}, we introduce our notation and summarize the Amoeba formulation in one-dimensional class A and AII$^\dagger$ systems.
Sec.~\ref{sec: WHF} reviews the Wiener-Hopf factorization (WHF) for one-dimensional Hermitian Hamiltonians and relates it to band topology via partial indices. 
Next, we generalize this framework to non-Hermitian Hamiltonians via Hermitian doubling.
Section~\ref{sec: Szego in multiband} develops the connection between WHF and the Szeg\"o limit theorem in one-dimensional multiband systems, including the modified Szeg\"o theorem, and clarifies the precise criteria under which the OBC and PBC spectral potentials coincide.
Building on this groundwork, Sec.~\ref{sec: Amoeba multiband} presents the central theoretical results of this paper. We reformulate the generalized Szeg\"o limit theorem for one-dimension as a corollary of the modified Szeg\"o theorem via Hermitian doubling, derive the necessary correction terms for the multiband Amoeba formulation, and provide a rigorous mathematical foundation for the symmetry-decomposed Ronkin functions in class~AII$^\dagger$ systems. In Sec.~\ref{sec: Numerical verification}, we validate our theoretical framework through numerical studies on representative models.
Section~\ref{sec: conclusion} concludes with a summary and outlook.
Appendix~\ref{appendix: instability} analyzes the fragility of the $\kappa=2$ phase discussed in Sec.~\ref{sec: Numerical verification}, showing that it is unprotected by TRS$^\dagger$ and disappears under small TRS$^\dagger$-preserving perturbations.
Finally, Appendix~\ref{appendix: RPI and BBC} shows that our formulation is valid even in the presence of isolated edge modes, and the partial indices count the number of edge modes.

\section{Amoeba formulation in one-dimensional systems} \label{sec: review of Amoeba}
This section introduces the notation and provides a brief review of the Amoeba formulation in classes A \cite{HWang_PRX_2024_AmoebaFormulationOf} and AII$^\dagger$ \cite{SKaneshiro_PRB_2025_SymplecticAmoebaFormulation} systems.

\subsection{Notation}
We consider an $M$-band Bloch Hamiltonian of finite hopping range,
\begin{align}
    h(e^{ik}) = \sum_{j=-p}^{q} h_j\, e^{ikj},
\end{align}
where $p$ and $q$ denote the maximal hopping range to the left and right, respectively.
Each $h_j$ is an $M \times M$ matrix specifying the hopping amplitudes between orbitals separated by $j$ sites.

Substituting $\beta=e^{ik}$, $h(\beta)$ is a matrix-valued Laurent polynomial
\begin{align}
    h(\beta) = \sum_{j=-p}^q h_j \beta^j,
\end{align}
which we refer to as the non-Bloch Hamiltonian.

From the Fourier coefficients $\{h_j\}$, we construct the OBC Hamiltonian on a one-dimensional lattice with $N$ sites,
\begin{align}
    [H_N]_{x, y}=h_{x-y}\quad (1\le x,y\le N),
\end{align}
with the convention $h_l=0$ for $l \notin [-p,q]$. 
Equivalently, the inverse Fourier transform of $h$ generates an $N \times N$ Toeplitz matrix $\mathcal T_N[h]$
\begin{align} \label{eq: def FT}
    [H_N]_{x, y} = \mathcal T_N[h]_{x, y} 
    = \int_0^{2\pi} \frac{dk}{2\pi} h(e^{ik}) e^{-ik(x-y)}.
\end{align}
We define Hermitian conjugate, transpose, and complex conjugation on matrix-valued polynomials of $h(\beta)$ so that they commute with $\mathcal{T}_N$:
\begin{align}
    \mathcal{T}_N[h]^\dagger&=\mathcal{T}_N[h^\dagger] \qc
    \mathcal{T}_N[h]^\top=\mathcal{T}_N[h^\top], \nonumber \\
    \mathcal{T}_N[h]^*&=\mathcal{T}_N[h^*],
\end{align}
where
\begin{align} \label{eq: HC_Top_CC}
    h^\dagger(\beta)&=\sum_j h_j^\dagger \beta^{-j} \qc
    h^\top(\beta)=\sum_j h_j^\top \beta^{-j}, \nonumber \\
    h^*(\beta)&=\sum_j h_j^* \beta^{j}.
\end{align}
We stress that Hermitian conjugation and transpose reverse the Laurent powers $(\beta^{j}\!\mapsto\!\beta^{-j})$, whereas complex conjugation acts only on the coefficients and leaves the powers \(\beta^{j}\) unchanged.

\subsection{Amoeba formulation for class A}
We now focus on one-dimensional single-band systems ($M=1$) in class A, i.e., systems without symmetry constraints.
The Amoeba formulation determines the OBC spectrum through its DOS defined over the complex energy plane.
The DOS is defined using Dirac's delta functions as
\begin{align}
    \rho(E) = \lim_{N \to \infty} \frac{1}{N} \Tr \delta \qty(E-H_N).
\end{align}
Viewing $\rho(E)$ as a two-dimensional charge density, the spectral potential $\phi(E)$ is defined by the Poisson equation
\begin{align} \label{eq: DOS and Potential}
    \rho(E) &= \frac{1}{2\pi} \Delta \phi(E),
\end{align}
where $\Delta = {\partial^2}/{\partial (\Re E)^2} + {\partial^2}/{\partial (\Im E)^2}$.
We can easily find that $\phi(E)$ takes the form
\begin{align}
    \phi(E) = \lim_{N \to \infty} \frac{1}{N} \ln \abs{\det \mathcal T_N[E-h]}.
    \label{eq: potential 1}
\end{align}

Let $\sigma(\beta) = E-h(\beta)$ denote the non-Hermitian symbol of the system.
We assume that $\sigma(\beta)$ is invertible for all $\beta$ with absolute value $1$ and topologically trivial, i.e., the winding number
\begin{align}
    W[\sigma] = \int_0^{2\pi} \frac{dk}{2\pi i} \partial_k \ln \det \sigma(e^{ik}),
    \label{eq: Z topo. inv.}
\end{align}
is well-defined and vanishes.
Then $\mathcal{T}_N[\sigma]$ is invertible, and Szeg\"o limit theorem yields
\cite{GSzego_MathAnn_1915_EinGrenzwertsatzUber},
\begin{align}
    \lim_{N \to \infty} \frac{1}{N} \ln \det \mathcal T_N[\sigma]
    = \int_0^{2\pi} \frac{d k}{2\pi} \ln \det \sigma(e^{ik}).
    \label{eq: Szego limit theorem}
\end{align}
This theorem states that the OBC and PBC spectral potentials coincide for such
$E$.

For an arbitrary (possibly topologically nontrivial) energy \(E\), the diagonal similarity transformation
\begin{align} \label{eq: similarity transformation}
    D_N^{(\mu)} \mathcal{T}_N[\sigma] [D_N^{(\mu)}]^{-1}
    = \mathcal{T}_N[\sigma_\mu],
\end{align}
where \([D_N^{(\mu)}]_{x, y} = \delta_{x, y}\, e^{\mu x}\) and \(\sigma_\mu(e^{ik}) = \sigma(e^{\mu+ik})\), maps the Toeplitz matrix \(\mathcal{T}_N[\sigma]\) to one defined by the analytically continued symbol \(\sigma_\mu\).
Since the determinant is invariant under similarity transformations, \(\det \mathcal T_N[\sigma]=\det \mathcal T_N[\sigma_\mu]\).
Hence, if there exists a $\mu$ such that $W[\sigma_\mu]=0$, Szeg\"o limit theorem Eq.~(\ref{eq: Szego limit theorem}) can be applied for $\sigma_\mu$:
\begin{align}
    \lim_{N \to \infty} \frac{1}{N} \ln \det \mathcal T_N[\sigma] = \int_0^{2\pi} \frac{dk}{2\pi} \ln \det \sigma_\mu(e^{ik}).
\end{align}
This discussion is equivalent to the optimization problem
\begin{align}
    \phi(E) &= \min_\mu R_\sigma (\mu),
    \label{eq: OBC potential for class A}
\end{align}
where $R_\sigma$ is the Ronkin function defined as 
\begin{align} \label{eq: Ronkin function}
     R_\sigma (\mu) = \int_0^{2\pi} \frac{d k}{2\pi} \ln \abs{ \det \sigma_\mu(e^{ik}) },
\end{align}
and its derivative with respect to $\mu$ directly yields the winding number, $W[\sigma_\mu]$.

The Ronkin function can be rewritten in a compact form using the factorization
\begin{align}
    \det \sigma(\beta) = C_E \beta^{-p} \prod_{j=1}^{p+q} [\beta-\beta_j(E)],
\end{align}
where $C_E$ is a $\beta$-independent coefficient and the $\beta$-roots are ordered by increasing modulus.
With this factorization, the Ronkin function is written as
\begin{align}
    R_\sigma(\mu) = \ln \abs{C_E} - p\mu + \sum_{j=1}^{p+q} \max (\mu, \mu_j),
    \label{eq: Ronkin function in beta}
\end{align}
where $\mu_j = \ln |\beta_j|$.
The second and third terms are the contributions from the pole and roots of $\det \sigma$, respectively.

Using this compact expression, one finds that $R_\sigma(\mu)$ attains its minimum within the interval $[\mu_p, \mu_{p+1}]$, which precisely corresponds to the GBZ condition~\cite{SYao_PRL_2018_EdgeStatesAnd,KYokomizo_PRL_2019_NonBlochBand}. 
In this region, the transformed symbol $\sigma_\mu$ is topologically trivial, corresponding to a vanishing winding number $W[\sigma_\mu]=0$. 
For energies on the OBC spectrum, this interval collapses to a single value of $\mu$, 
whose magnitude gives the inverse localization length of the corresponding eigenstate.
Minimization over $\mu$ gives the compact expression of the potential
\begin{align}
    \phi(E) &= \min_\mu R_\sigma(\mu) = \ln \abs{C_E} + \sum_{j=p+1}^{p+q} \mu_j.
\end{align}

\subsection{Amoeba formulation for class AII$^\dagger$}
The Amoeba formulation requires modification in the presence of symmetry-protected degeneracies.
We recently addressed this issue in Ref.~\cite{SKaneshiro_PRB_2025_SymplecticAmoebaFormulation};
Below, we summarize the relevant results for later use.

In this section, we consider one-dimensional two-band systems ($M=2$) in the transpose-type time-reversal symmetry (class AII$^\dagger$).
The symmetry imposes the constraint
\begin{align}
    U_T \mathcal T_N[h]^\top U_T^{-1} = \mathcal T_N[h],
\end{align}
where \(U_T\) is a unitary matrix.
Using the non-Bloch representation and the convention in 
Eqs.~(\ref{eq: HC_Top_CC}), where transpose reverses the Laurent powers, the symmetry condition 
becomes
\begin{align}
    U_T h^\top(\beta^{-1}) U_T^{-1} = h(\beta^{-1}).
\end{align}

This symmetry constrains the polynomial $\det \sigma(\beta)$ to satisfy
\begin{align}
    \det \sigma(\beta) = \det \sigma(\beta^{-1}),
    \label{eq: TRS+ chp symmetry}
\end{align}
and hence it factorizes as
\begin{align}
    \det \sigma(\beta) = C_E \beta^{-2p} \prod_{j=1}^{2p} [\beta-\beta_j(E)] [\beta-\beta_j^{-1}(E)].
\end{align}
We order the $\beta_j$ by modulus as
\begin{align}
    \abs{\beta_{2p}^{-1}} \le \dots \le \abs{\beta_1^{-1}}
    \le 1 \le 
    \abs{\beta_{1}} \le \dots \le \abs{\beta_{2p}},
\end{align}
so that each pair $(\beta_j, \beta_j^{-1})$ encodes a Kramers pair.

Class AII$^\dagger$ is characterized by the $\mathbb{Z}_2$ topological invariant \cite{KKawabata_PRX_2019_SymmetryAndTopology} defined by
\begin{align}\label{eq: Z2 topological invariant}
    &(-1)^{\nu[\tau]} = \text{sgn} \left[
    \frac{\Pf \tau(-1)}{\Pf \tau(1)}
    \right . \nonumber \\
    &\times \left .
     \exp \Bqty{
        -\frac{1}{2} \int_0^\pi dk \partial_k \ln \det \tau(e^{ik})
     }
    \right],
\end{align}
with $\tau = (E-h)U_T$.
The sign function, $\mathrm{sgn}(x)$, takes the value $+1$ for $x>0$ and $-1$ for $x<0$, and $\Pf[A]$ corresponds to the Pfaffian of the skew-symmetric matrix $A$.

The Ronkin function for class AII$^\dagger$ is compactly expressed as
\begin{align}
    R_\sigma(\mu) = \ln \abs{C_E} -2p \mu + \sum_{j=1}^{2p} [\max(\mu, \mu_j) + \max(-\mu, \mu_j)],
    \label{eq: compacted Ronkin function in class AIId}
\end{align}
where $\mu_j = \ln \abs{\beta_j} \ge 0$.

In the \(\mathbb{Z}_2\)-nontrivial phase, two Kramers-related eigenstates localize on opposite edges with inverse localization lengths $\pm \mu_1$.
Consequently, their contributions compete in the minimization in Eq.~(\ref{eq: OBC potential for class A}) and the conventional optimization formalism cannot be used.

To resolve this competition, we introduced the symmetry-decomposed Ronkin functions $R_\sigma^{(\pm)}$ in Ref.~\cite{SKaneshiro_PRB_2025_SymplecticAmoebaFormulation}.
We impose that the decomposed Ronkin functions $R_\sigma^{(\pm)}$ inherit the key mathematical properties of $R_\sigma$, 
namely convexity, piecewise linearity, and integer-quantized derivatives. 
Moreover, $R_\sigma^{(\pm)}$ should satisfy the symmetry relations
\begin{align} \label{eq: partial Ronkin symmetry}
    R_\sigma(\mu) &= R_\sigma^{(+)}(\mu) + R_\sigma^{(-)}(\mu), \qquad
    R_\sigma^{(+)}(\mu) = R_\sigma^{(-)}(-\mu),
\end{align}
and remain consistent with the GBZ condition for class~AII$^\dagger$ systems~\cite{KKawabata_PRB_2020_NonBlochBand}, 
where $R_\sigma^{(+)}$ is optimized within the interval $[\mu_1, \mu_2]$.

Under these assumptions, the symmetry-resolved Ronkin functions take the forms:
\begin{align}\label{eq: sr Ronkin plus}
    R_\sigma^{(+)}(\mu) &= \frac{1}{2} \ln \abs{C_E} - \frac{1}{2} \sum_{j=1}^{2p} \mu_j -\mu \nonumber \\
    & + \sum_{j=1}^{2p} \max (\mu, \mu_j) \\
    \label{eq: sr Ronkin minus}
    R_\sigma^{(-)}(\mu) &= \frac{1}{2} \ln \abs{C_E} + \frac{1}{2} \sum_{j=1}^{2p} \mu_j -(2p-1) \mu \nonumber \\
    & + \sum_{j=1}^{2p} \max (-\mu, \mu_j).
\end{align}
The contributions from the roots, the third and fourth term in Eq.~(\ref{eq: compacted Ronkin function in class AIId}), are distributed symmetrically between $R_\sigma^{(+)}$ and $R_\sigma^{(-)}$, whereas the pole contribution $-2p \mu$ is split asymmetrically for consistency with the GBZ condition.

Optimizing yields the OBC spectral potential
\begin{align} \label{eq: OBC spectral potential}
    \phi(E) =
    \begin{cases}
    \min_\mu R_\sigma(\mu) & \nu[\tau] = 0 \\
    2 \min_\mu R_\sigma^{(+)}(\mu) & \nu[\tau] = 1
    \end{cases}
\end{align}
which can be written compactly as
\begin{align}
    \phi(E) &= \ln \abs{C_E} + (-1)^{\nu[\tau]} \mu_1 + \sum_{j=2}^{2p} \mu_j.
\end{align}
The $\mathbb Z_2$ invariant determines the sign of the leading root’s contribution, flipping the $\mu_1$ term between the topologically trivial and nontrivial phases.
This formulation reproduces the correct OBC spectral potential.

However, the above discussion relies on phenomenological intuition about assigning Kramers pairs to different decomposed functions, rather than on a rigorous mathematical basis.
In particular, the applicability criteria of the generalized Szeg\"o limit theorem in multiband systems have not been established, and the connection to this theorem therefore remains unclear.
This lack of a mathematical foundation prevents the extension of the Amoeba formulation to general multiband 
class~A systems.

\section{Wiener-Hopf factorization and its application to band topology} \label{sec: WHF}
In this section, we introduce the Wiener-Hopf factorization (WHF) and clarify its connection to the band topology of both Hermitian and non-Hermitian systems.
We first review the WHF of matrix-valued Laurent polynomials arising from Hermitian Bloch Hamiltonians and summarize its established correspondence with band topology~\cite{AAlase_AnnPhys_2023_WienerHopfFactorization}.
We then extend this framework to non-Hermitian Hamiltonians by employing the Hermitian doubling approach~\cite{ZGong_PRX_2018_TopologicalPhasesOf, KKawabata_PRX_2019_SymmetryAndTopology}, thereby demonstrating how the WHF naturally captures the topological structure of non-Hermitian bands.

\subsection{Wiener-Hopf factorization for matrix Laurent polynomials}
In this section, we consider a matrix-valued Laurent polynomial $A(\beta) = \sum_{j=-p}^q a_j \beta^j$, defined on the Riemann sphere $\mathbb C \cup \{\infty\}$.

A Wiener-Hopf factorization of $A$ is a decomposition of the form
\begin{align}
    A(\beta) = A_+(\beta) D(\beta) A_-(\beta).
\end{align}
where the factors are given by
\begin{align}
    A_+(\beta) = \sum_{j=0}^q a_{+, j} \beta^j \qc
    A_-(\beta) = \sum_{j=-p}^0 a_{-, j} \beta^j
\end{align}
and
\begin{align}
    D(\beta) = \mathrm{diag} \mqty(\beta^{\kappa_1}, \cdots, \beta^{\kappa_M}).
\end{align}
Here, $A_+(\beta)$ must be invertible for $|\beta|\le 1$ (including the origin),
while $A_-(\beta)$ must be invertible for $|\beta|\ge 1$ (including $\infty$).
The integers $\kappa_j$ are the partial indices of $A$, and we order them in descending order.
We denote the set of partial indices of a symbol $\sigma$ as
\begin{align}
    \mathcal K[\sigma] = (\kappa_1, \dots, \kappa_M).
\end{align}

We note that a WHF of $A$ exists if and only if $A(\beta)$ is invertible on the unit circle, 
that is, $\det A(\beta) \neq 0$ for $|\beta|=1$. 
This condition ensures that $\det A(\beta)$ has no zeros on the unit circle, 
so that analytic continuations of $A_+$ and $A_-$ can be defined uniquely up to an invertible constant matrix.
The factorization is not unique. 
Given an invertible matrix $U(\beta)$ with $\det U(\beta)=\text{const}$, another valid factorization can be found as
\begin{align}
    A_+ D A_- 
    &= (A_+ U) D (D^{-1} U^{-1} D A_-) \nonumber \\
    &= A_+' D A_-', \label{eq:gauge_freedom_WHF}
\end{align}
where $A_+' = A_+ U$ and $A_-' = D^{-1} U^{-1} D A_-$. 
However, after fixing an ordering of the partial indices $\{\kappa_j\}$, the diagonal factor $D$ is unique, so that the partial indices are uniquely defined.

We will often partition the index set $\{1, \cdots, M\}$ into $M_+$, $M_0$, and $M_-$ corresponding to positive, zero, and negative partial indices, respectively.
The corresponding partial indices are referred as $\mathcal K _+$, $\mathcal K _0$, and $\mathcal K _-$.

\subsection{Symmetric Wiener-Hopf factorization of Bloch Hamiltonians and band topology} \label{sec: SWHF of Bloch}

We now consider the WHF of Hermitian Bloch Hamiltonians.
As recently shown in Ref.~\cite{AAlase_AnnPhys_2023_WienerHopfFactorization}, a modified WHF, referred to as the symmetric Wiener-Hopf factorization (SWHF), is closely related to band topology.
Motivated by applications to non-Hermitian topology, we focus on classes AIII and DIII, although analogous statements hold for other symmetry classes.

For a gapped Hermitian Bloch Hamiltonian, $h(\beta)$ admits a WHF
\begin{align}
    h(\beta) = h_+(\beta) \Lambda(\beta) h_-(\beta),
\end{align}
which can be chosen to respect Hermiticity,
\begin{align}\label{eq: Hermitian WHF}
    h_+^\dagger = h_- \qc \Lambda^\dagger = \Lambda.
\end{align}
As defined in Eq.~(\ref{eq: HC_Top_CC}), the Hermitian conjugation reverses the signs of the partial indices, ${\kappa_j} \mapsto {-\kappa_j}$.
Together with the uniqueness of the partial indices, this implies that the set of partial indices of $\sigma$, denoted $\mathcal{K}[\sigma]$, is symmetric, so that every $\kappa_m>0$ has a corresponding partner $-\kappa_m$.
Thus, we can arrange $\Lambda$ in a symmetric form
\begin{align}
\begin{cases}
    \Lambda_{M+1-m, m} = \beta^{\kappa_m} & m \in M_+ \\
    \Lambda_{m, M+1-m} = \beta^{-\kappa_m} & m \in M_+ \\
    \Lambda_{m, m} = s_m & m \in M_0
\end{cases}
\end{align}
with $s_m \in\pm 1$ arranged in ascending order.
Here, the block assignment realizes a symmetric placement of positive/negative indices.
This type of factorization is called the symmetric Wiener-Hopf factorization (SWHF)\cite{AAlase_AnnPhys_2023_WienerHopfFactorization}.

In the presence of additional symmetries, the SWHF assumes a symmetry-constrained (modified) form.
A Hamiltonian in class AIII satisfies the chiral symmetry, which allows $h$ to be written as
\begin{align} \label{eq: Ham in off-diagonal}
    h &= \mqty(0 & A^\dagger \\ A & 0),
\end{align}
where $A$ is a general (non-Hermitian) matrix.
Then, the WHF of $A$, $A=A_+ D A_-$, yields a SWHF of $h$ in the form of
\begin{align} \label{eq: SWHF for class AIII}
    h &= h_+ \Lambda h_- \nonumber \\
    &=
    \mqty(A_-^\dagger & \\ & A_+) 
    \mqty( & D^\dagger \\ D & )
    \mqty(A_- & \\ & A_+^\dagger),
\end{align}
consistent with Eq.~(\ref{eq: Hermitian WHF}).

Similarly, Hamiltonians in class DIII are block off-diagonal and admit an SWHF structurally equivalent to the AIII form introduced in Eq.~(\ref{eq: SWHF for class AIII}).
Time-reversal and particle-hole symmetries, however, constrain $A$ to satisfy $A=-A^\top$ in an appropriate basis.
Given the skew-symmetry of $A$, the argument analogous to the Hermitian case [Eq.~\eqref{eq: Hermitian WHF}] implies that the partial indices must appear in pairs of opposite signs $(\kappa_m, -\kappa_m)$.

Choosing a suitable $U$, as introduced in Eq.~(\ref{eq:gauge_freedom_WHF}), we can determine the factors as,
\begin{align}\label{eq: A factor for class DIII}
    A_+^\top = A_- \qc D^\top = -D,
\end{align}
where $D$ takes the skew-symmetric form
\begin{align}\label{eq: D factor for class DIII}
\begin{cases}
    D_{M+1-m, m} = -\beta^{\kappa_m} & m \in M_+ \\
    D_{m, M+1-m} = \beta^{-\kappa_m} & m \in M_+ \\
    D_{M+1-m, m} = 1 & m \in M_0, m \le M/2 \\
    D_{m, M+1-m} = -1 & m \in M_0, m \le M/2
\end{cases}.
\end{align}

Concrete algorithms for constructing both WHF and SWHF are provided in the Appendix of Ref.~\cite{AAlase_AnnPhys_2023_WienerHopfFactorization}.

\subsubsection{SWHF and Hermitian band topology}
As demonstrated in Ref.~\cite{AAlase_AnnPhys_2023_WienerHopfFactorization}, the SWHF of a Hermitian Bloch Hamiltonian is directly related to its band topology via the partial indices.
For both classes AIII and DIII, the relevant topological invariants can be expressed in terms of the off-diagonal block $A$ introduced in Eq.~(\ref{eq: Ham in off-diagonal}).

For class~AIII, the topological invariant is the winding number of $A$. 
Considering the WHF $A = A_+ D A_-$, the factors $A_+$ and $A_-$ are analytic and invertible for $|\beta| \le 1$ and $|\beta| \ge 1$, respectively. 
Since $A_\pm$ have neither zeros nor poles in their corresponding regions, their winding numbers vanish, $W[A_\pm]=0$. 
Consequently, the winding number of $A$ equals that of $D$,
\begin{align}
    W[A] = W[D] = \sum_{\kappa \in \mathcal K[A]} \kappa.
\end{align}
Hence, the winding number is given by the sum of the partial indices.
Furthermore, the number of right [left] localized modes is equivalent to the sum of positive [negative] partial indices.

Similarly, the $\mathbb{Z}_2$ topological invariant for class~DIII is reduced by the WHF in Eqs.~(\ref{eq: A factor for class DIII}) and~(\ref{eq: D factor for class DIII}) to
\begin{align}
    (-1)^{\nu[A]} = \frac{\Pf D(-1)}{\Pf D(1)} 
    = (-1)^{\sum_{\kappa \in {\mathcal K_+[A]}} \;\kappa}.
\end{align}
Here, we used $\det D = 1$ and $\Pf(A_+ D A_+^\top) = (\det A_+) \Pf D$. 
Therefore, the $\mathbb{Z}_2$ invariant is determined by the parity of the sum of the positive partial indices.

In both classes, the topological invariant is thus encoded in the partial indices of the off-diagonal block $A$.

\subsubsection{WHF and non-Hermitian band topology}
The WHF framework can also be applied to non-Hermitian Hamiltonians.
Their topological classification can be carried out through Hermitization, 
i.e., by classifying the doubled Hermitian Hamiltonian at a reference energy $E$ \cite{ZGong_PRX_2018_TopologicalPhasesOf, KKawabata_PRX_2019_SymmetryAndTopology}.

A class A non-Hermitian Hamiltonian maps to a class AIII Hermitian Hamiltonian via Hermitization, 
\begin{align} \label{eq: A to AIII}
    \tilde \sigma = \mqty(\admat{\sigma^\dagger, \sigma}),
\end{align}
where $\sigma = E-h$.
Therefore, a WHF of $\sigma$ induces an SWHF of $\tilde \sigma$, and the topological information of $\sigma$ is encoded in its partial indices.

A class AII$^\dagger$ non-Hermitian Hamiltonian is mapped to a class DIII Hermitian Hamiltonian, with
\begin{align} \label{eq: AIId to DIII}
    \tilde \tau = \mqty(\admat{\tau^\dagger, \tau}),
\end{align}
where $\tau= (E-h)U_T$.
Therefore, the topological information can be obtained from the partial indices of $\tau$.

Note that the topological information obtained from Hermitian doubling corresponds to the point-gap topology derived via unitary flattening \cite{ZGong_PRX_2018_TopologicalPhasesOf, KKawabata_PRX_2019_SymmetryAndTopology}.

\section{Strong Szeg\"o limit theorem and WHF} \label{sec: Szego in multiband}

In this section, we consolidate the theoretical groundwork connecting the Szeg\"o limit theorem and the Wiener-Hopf factorization (WHF) as a preparation for our later analysis of multiband non-Hermitian systems.
We first revisit the classical and modified Szeg\"o limit theorems, interpreting them in terms of the partial indices of the WHF. This review clarifies the conditions under which the OBC and PBC spectral potentials coincide and identifies how nonvanishing partial indices lead to deviations in the OBC potential. These insights then form the basis for our generalization of the Szeg\"o limit theorem to multiband systems in the following section.

\subsection{Classical strong Szeg\"o limit theorem and partial indices}
Let $\sigma$ be a matrix-valued Laurent polynomial (Hermitian or non-Hermitian).
The classical strong Szeg\"o limit theorem \cite{GSzego_MathAnn_1915_EinGrenzwertsatzUber, HWidom_AdvMath_1974_AsymptoticBehaviorOf, HWidom_AdvMath_1976_AsymptoticBehaviorOf} 
concerns the asymptotic behavior of the determinant ratio $E_N$,
\begin{align} \label{eq: classical Szego}
    \lim_{N \to \infty} E_N[\sigma] = \lim_{N \to \infty} \frac{\det \mathcal T_N [\sigma]}{G[\sigma]^{N+1}}\;,
\end{align}
where $G[\sigma]$ is the exponential of the Ronkin function for $\mu = 0$, corresponding to the PBC spectral potential. It is written as
\begin{align}
    G[\sigma] = \exp[\int_0^{2\pi} \frac{dk}{2\pi} \ln \det \sigma(e^{ik})] = e^{R_\sigma(0)}.
\end{align}
This limit converges when $\sigma$ is gapped on the unit circle and its winding number vanishes.
Thus, the OBC spectral potential can be expressed as
\begin{align} \label{eq: OBC = PBC + E_N}
    \frac{1}{N}\ln\det\mathcal{T}_N[\sigma]
    = \frac{N+1}{N}R_\sigma(0) + \frac{1}{N}\ln E_N[\sigma],
\end{align}
indicating that its deviation from the PBC spectral potential is controlled by $E_N[\sigma]$.
The applicability of the Szeg\"o limit theorem can be expressed in terms of the partial indices of the WHF of $\sigma$ \cite{HWidom_AdvMath_1974_AsymptoticBehaviorOf}.
When all partial indices of $\sigma$ vanish, the ratio $E_N[\sigma]\to C\neq0$ converges to a nonzero constant, and the OBC and PBC spectral potentials coincide up to $\mathcal{O}(1/N)$ corrections.
Deviations from this condition arise when one or more partial indices are nonzero, in which case $E_N[\sigma]$ in Eq.~(\ref{eq: classical Szego}) decays to zero, and the OBC and PBC spectral potentials no longer coincide.
In this case, the convergence rate of $E_N[\sigma]$, 
\begin{align}
    E_N[\sigma] = e^{-\alpha N + \order{N}},
\end{align}
determines the deviation of the OBC potential by $-\alpha$ \cite{EBasor_JStatPhys_2019_ModifiedSzegoWidomAsymptotics}. 
The convergence rate $\alpha$ is related to the localization length $\mu$, on which we will focus later.

For single-band Hamiltonians, the winding number equals the partial index.
Thus, optimizing the Ronkin function is equivalent to finding a $\mu$ for which the partial index of $\sigma_\mu$ is zero, and thus 
\begin{align} \label{eq: Generalized Szego for class A}
    \lim_{N \to \infty} \frac{1}{N} \ln \det \mathcal T_N[\sigma] = \min_\mu R_\sigma(\mu).
\end{align}
This is the generalized Szeg\"o limit theorem proposed in Ref.~\cite{HWang_PRX_2024_AmoebaFormulationOf}.

In multiband systems, however, the total winding number is the sum of all partial indices, so a vanishing winding number no longer guarantees that each index individually vanishes. 
Consequently, the naive Amoeba formulation can fail unless there exists a $\mu$ for which all partial indices become zero. 
If such a $\mu$ exists, the classical Szeg\"o limit theorem still holds for $\sigma_\mu$; otherwise, the correspondence between OBC and PBC potentials breaks down.

\subsection{Modified Szeg\"o limit theorem for topologically nontrivial Hermitian systems}
As a next step, we review the modified Szeg\"o limit theorem established for topologically nontrivial Hermitian Hamiltonians \cite{EBasor_JStatPhys_2019_ModifiedSzegoWidomAsymptotics, AAlase_AnnPhys_2023_WienerHopfFactorization}, which governs the convergence behavior of $E_N[\sigma]$. 
We further summarize how these deviations are connected to boundary modes in such systems, thereby establishing the conceptual bridge used in the following section.

Let $\tilde \sigma$ be a Hermitian two-band class AIII Hamiltonian, written in the off-diagonal form of Eq.~(\ref{eq: Ham in off-diagonal}). Because of chiral symmetry, the winding number of the full Hamiltonian satisfies $W[\tilde{\sigma}]=0$.
However, the topology of $\tilde{\sigma}$ is characterized by the winding number of its off-diagonal block, $W[A]=\kappa$. When $\kappa = 0$, the system is topologically trivial and the determinant ratio $E_N[\tilde{\sigma}]$ converges to a finite, nonzero constant, as predicted by the classical Szeg\"o limit theorem.
For $\kappa\neq 0$, the Hamiltonian is topologically nontrivial, and $\mathcal K[\tilde{\sigma}]$ is $(|\kappa|, -|\kappa|)$.
The resulting decay of $E_N[\tilde{\sigma}]$ to zero reflects the contribution of boundary-localized modes in $\mathcal{T}_N[\tilde{\sigma}]$.

It has been shown in Ref.~\cite{EBasor_JStatPhys_2019_ModifiedSzegoWidomAsymptotics} that the spectrum of $\mathcal{T}_N[\tilde{\sigma}]$ is well approximated by the PBC spectrum together with a pair of boundary-localized modes.
However, due to finite-size hybridization, these two edge modes acquire small but finite energies $\pm \varepsilon$ rather than lying exactly at zero. 
The splitting $\varepsilon$ scales with the overlap between the edge modes, $\varepsilon \propto e^{-\mu N}$, where $\mu>0$ is the inverse localization length.
The modified Szeg\"o limit theorem~\cite{EBasor_JStatPhys_2019_ModifiedSzegoWidomAsymptotics} then implies
\begin{align} \label{eq: convergence ratio of E_N}
    E_N[\tilde{\sigma}] \propto \varepsilon^2 \propto e^{-2\mu N}.
\end{align}
The generalization to multiple pairs of edge modes is straightforward by
\begin{align}
    E_N[\tilde{\sigma}] \propto \prod_j e^{-2\mu_j N},
\end{align}
where $\mu_j>0$ denotes the inverse localization lengths of each pair.

As discussed in Ref.~\cite{AAlase_AnnPhys_2023_WienerHopfFactorization}, the number of zero modes is determined by $\mathcal K[\sigma]$.
A positive partial index of $\mathcal K[A]$ corresponds to a mode localized at the right boundary, while a negative partial index of  $\mathcal K[A]$ (or equivalently, a positive partial index of $\mathcal K[A^\dagger]$) corresponds to a mode localized at the left boundary.

For example, if $\kappa = 1$ (i.e., $W[A]=1$), the off-diagonal block $A$ contributes one right-localized mode, while $A^\dagger$ contributes one left-localized mode. 
Thus, the full Hamiltonian $\mathcal{T}_N[\tilde{\sigma}]$ hosts one mode on each boundary. 
The localization length of these modes is determined by the root of $\det A$ with the largest magnitude inside the unit circle, yielding $\mu = \ln|\beta|$. 
This shows that the Szeg\" o limit theorem for a topologically nontrivial Hermitian Hamiltonian must be modified by this decay exponent.

This argument extends to general nonzero $\kappa$. 
For positive (negative) $\kappa$, the block $A$ contributes $\kappa$ right- (left-) localized modes, governed by the $\kappa$ roots of $\det A$ inside (outside) the unit circle with the largest (smallest) magnitudes. 
In either case, the total decay exponent is given by the sum of their inverse localization lengths.

The above discussion illustrates how nonzero winding numbers modify the asymptotic behavior of $E_N[\tilde{\sigma}]$. 
To further clarify the role of partial indices in multiband settings, we next consider a four-band class AIII Hamiltonian where the off-diagonal block $A$ is a two-band matrix.
In this case, even when $W[A]=0$, the block $A$ may still have nontrivial partial indices, e.g., $\mathcal K[A]=(+1,-1)$.
Since $\tilde{\sigma}$ is constructed from $A$ and $A^\dagger$, the partial indices $(+1,-1)$ of $A$ lead to $\mathcal K[\tilde \sigma]=(+1,+1,-1,-1)$ for the full Hamiltonian $\tilde{\sigma}$.
Thus, $A$ supports two edge modes localized at opposite boundaries, and $\mathcal{T}_N[\tilde{\sigma}]$ hosts two pairs of edge modes in total.
These modes are controlled by the dominant roots of $\det A$: one root $\beta_2$ with the smallest absolute value outside the unit circle and one root $\beta_1$ with the largest absolute value inside.
We denote their inverse localization lengths as $\mu_2 = \ln|\beta_2| > 0$ and $\mu_1 = \ln|\beta_1| < 0$, respectively.
The contribution to $E_N[\tilde{\sigma}]$ then becomes
\begin{align} \label{eq: EN (+1, -1)}
    E_N[\tilde{\sigma}] 
    \propto e^{-2\mu_2 N}\,e^{2\mu_1 N}
    = e^{-2(\mu_2 - \mu_1)N}.
\end{align}
We note that $\mu_2-\mu_1$ is equivalent to the length of the interval where $W[\sigma_\mu]$ vanishes.

\section{Amoeba formulation for multiband systems via WHF} \label{sec: Amoeba multiband}
Building on the theoretical groundwork developed in the previous sections, we extend the Amoeba framework by incorporating the WHF to treat non-Hermitian systems with nonzero partial indices through Hermitian doubling. 
This formulation unifies the analytic structure of the Amoeba with the topological information encoded in the WHF, leading to a generalized Szeg\"o limit theorem for multiband non-Hermitian systems. 
Focusing on class~AII$^\dagger$, we further show that the WHF naturally gives rise to the symmetry-decomposed Ronkin function proposed in our previous work \cite{SKaneshiro_PRB_2025_SymplecticAmoebaFormulation}, thereby providing a rigorous foundation for the generalized Szeg\"o relation in time-reversal-symmetric non-Hermitian systems.

\subsection{Correction for non-vanishing partial indices for class A}
We now extend the preceding discussion to topologically nontrivial non-Hermitian Hamiltonians and derive the Generalized Szeg\"o limit theorem with non-vanishing partial indices.

In non-Hermitian systems, topological phases often manifest not through isolated edge states but through the extensive localization of bulk modes near the boundaries, as seen in the NHSE.
Through Hermitian doubling, each localized non-Hermitian bulk mode corresponds to a quasi-zero mode of a doubled Hermitian Hamiltonian with the same localization length \cite{NOkuma_PRL_2020_TopologicalOriginOf, NOkuma_PRB_2020_HermitianZeroModes}.
This correspondence enables us to apply the modified Szeg\"o limit theorem established for Hermitian systems to derive a generalized version valid for non-Hermitian Hamiltonians.

We first validate this approach in a single-band system and confirm that it reproduces the known results of the conventional Amoeba formulation.
As an illustrative example, let $\sigma$ be a non-Hermitian symbol whose determinant factorizes as
\begin{align}
    \det \sigma (\beta) = C \beta^{-p} \prod_{j=1}^{p+q} [\beta- \beta_j],
\end{align}
where $\beta_1, \cdots, \beta_{p+1}$ are inside the unit circle ($|\beta_j|<1$), and $\beta_{p+2}, \cdots, \beta_{p+q}$ are outside the circle ($|\beta_j|>1$).
This root distribution corresponds to a winding number $W[\sigma]=1$.
The corresponding Hermitianized symbol $\tilde \sigma$ is defined in Eq.~(\ref{eq: A to AIII}) and belongs to a two-band class AIII system.
The edge mode of this Hamiltonian is governed by $\mu_{p+1} (<0)$, which implies $E_N[\tilde \sigma] = \exp[2 \mu_{p+1}N]$ [Eq.~(\ref{eq: convergence ratio of E_N})].
Using the relation $E_N[\tilde \sigma]=|E_N[\sigma]|^2$, the modified Szeg\"o limit theorem for $\tilde \sigma$ yields the deviation for the potential of $\sigma$:
\begin{align}
    \lim_{N \to \infty} \frac 1 N \ln \abs{\det \mathcal T_N[\sigma]} &= R_\sigma(0) + \mu_{p+1} \nonumber \\
    &= \ln \abs{C} + \sum_{j=p+1}^{p+q} \mu_j,
\end{align}
which reproduces the generalized Szeg\"o limit theorem for $W[\sigma]=1$.
This argument straightforwardly generalizes to arbitrary winding numbers, corresponding to the optimization form in Eq.~(\ref{eq: Generalized Szego for class A}).

This framework for evaluating potential deviations, based on the asymptotic behavior of $E_N[\tilde\sigma]$, naturally extends to multiband systems with non-vanishing partial indices.
As another illustrative example, consider a two-band symbol $\sigma$ with vanishing total winding number $W[\sigma]=0$ but nonzero partial indices $\mathcal K[\sigma]=(1,-1)$.
Let $\det \sigma$ factorize as
\begin{align}
    \det \sigma(\beta) = C \beta^{-2p} \prod_{j=1}^{2(p+q)} [\beta-\beta_j],
\end{align}
where $\beta_1,\dots,\beta_{2p}$ lie inside the unit circle ($|\beta_j|<1$) and $\beta_{2p+1},\dots,\beta_{2(p+q)}$ lie outside it ($|\beta_j|>1$). 
This choice ensures $W[\sigma]=0$.
Using the asymptotic form of $E_N[\tilde\sigma]$ for the $(+1,-1)$ index structure, Eq.~(\ref{eq: EN (+1, -1)}), we obtain
\begin{align} \label{eq: modification for class A with (+1, -1)}
    \lim_{N \to \infty} \frac 1 N \ln \abs{\det \mathcal T_N[\sigma]} 
    &= R_\sigma(0) - (\mu_{2p+1}-\mu_{2p}),
\end{align}
where $\mu_{2p} = \ln|\beta_{2p}|$ is the logarithmic radius of the largest roots inside the unit circle and $\mu_{2p+1} = \ln|\beta_{2p+1}|$ that of the smallest root outside.

It is instructive to interpret the correction term in Eq.~(\ref{eq: modification for class A with (+1, -1)}) from a "local" perspective using the roots of $\det \sigma$ as Eq.~(\ref{eq: Ronkin function in beta}).
For $W[\sigma]=0$, the PBC spectral potential $R_\sigma(0)$ is given by the sum over roots outside the unit circle, $\ln|C| + \sum_{j=2p+1}^{2(p+q)} \mu_j$.
The OBC spectral potential $\phi(E)$ in Eq.~(\ref{eq: modification for class A with (+1, -1)}) thus becomes:
\begin{align}
\phi(E) = \ln|C| + \mu_{2p} + \sum_{j=2p+2}^{2(p+q)} \mu_j
\end{align}
Thus, the contribution from the smallest outside root ($\mu_{2p+1}$) is removed and replaced by that of the largest inside root ($\mu_{2p}$).
This result offers a clear interpretation: the nonvanishing partial indices quantify how many root contributions are exchanged across the unit circle.
In this example, $\mathcal{K}[\sigma]=(1,-1)$ means that, even though $W[\sigma]=0$, the correct OBC potential involves one such exchange, where the positive index brings in the innermost interior root and the negative index removes the outermost exterior root.

Finally, we generalize this result to the case with a nonzero winding number, $W[\sigma] \neq 0$.
Since the OBC determinant is invariant under the similarity transformation in Eq.~(\ref{eq: similarity transformation}), we can eliminate the winding number $W[\sigma_\mu]$ by introducing an appropriate shift $\mu$.
This allows us to apply Eq.~(\ref{eq: modification for class A with (+1, -1)}) using the roots of $\det\sigma_\mu$.
Therefore, we obtain a modified form of Eq.~(\ref{eq: Generalized Szego for class A}) 
that accounts for the nonvanishing partial indices $(1,-1)$:
\begin{align} \label{eq: modification for class A with winding + (+1, -1)}
    \lim_{N \to \infty} \frac 1 N \ln \abs{\det \mathcal T_N[\sigma]}
    = \min_\mu R_{\sigma}(\mu) - (\mu_{k+1} - \mu_{k}),
\end{align}
where $\mu_k$ and $\mu_{k+1}$ denote, respectively, the dominant roots located just inside and just outside the circle of radius $e^{\mu}$.

Building on the above observation, the validity of the Amoeba formulation is determined by the behavior of the partial indices after optimization.
We therefore introduce the residual partial indices,
\begin{align}
    \mathcal K^*[\sigma] \equiv \mathcal K[\sigma_{\mu^*}],
\end{align}
defined as the set of partial indices evaluated at the optimal shift $\mu^*$, chosen such that the global winding number vanishes, $W[\sigma_{\mu^*}] = 0$.

If, at this optimum, all residual partial indices vanish individually, $\mathcal K^* = (0,0,\dots)$, the conventional Amoeba optimization remains valid and the generalized Szeg\"o limit theorem applies.
However, when one or more residual indices are nonzero, the simple Amoeba formulation breaks down, and the generalized Szeg\"o limit theorem must include additional correction terms determined by $\mathcal K^*$.
Consequently, the generalized Szeg\"o limit theorem takes the compact form 
\begin{align} \label{eq: generalized Szego for class A with (+1, -1)}
    \lim_{N \to \infty} \frac{1}{N} \ln \abs{\det \mathcal T_N[\sigma]}
    = \mathcal M_{\mathcal K^*}[\sigma],
\end{align}
where $\mathcal M_{\mathcal K^*}[\sigma]$ depends on the residual partial indices as
\begin{align}
    \mathcal M_{(0,0)}[\sigma] &= \min_\mu R_{\sigma}(\mu), \\[4pt]
    \mathcal M_{(+1,-1)}[\sigma] &= \min_\mu R_{\sigma}(\mu) - (\mu_{k+1}-\mu_k).
\end{align}

\subsection{Generalized Szeg\"o limit theorem for class AII$^\dagger$}
We now demonstrate that the WHF of a non-Bloch Hamiltonian in the two-band class~AII$^\dagger$ naturally leads to a decomposition of the Ronkin function, yielding the symmetry-decomposed forms defined in Eqs.~(\ref{eq: sr Ronkin plus}) and (\ref{eq: sr Ronkin minus}).

\subsubsection{WHF and symmetry-decomposed Ronkin functions}
We focus on two-band ($M=2$) class AII$^\dagger$ systems.
Here, we define the total Ronkin function and its symmetry-decomposed counterparts using $\tau=(E-h)U_T$ instead of $\sigma=E-h$.
Since $U_T$ is a unitary matrix, $R_\tau$ is equivalent to $R_\sigma$.
This redefinition aligns more naturally with the WHF framework for class AII$^\dagger$ systems, enabling a symmetric decomposition of the Ronkin function.

As established in Sec.~\ref{sec: SWHF of Bloch}, the topological information is encoded in the partial indices of $\tau$.
The SWHF of $\tau$ can be arranged as shown in Eq.~(\ref{eq: A factor for class DIII}),
\begin{align} \label{eq: SWHF of tau}
    \tau &= \tau_+ D_{\kappa} \tau_-,
\end{align}
where $\tau_+^\top=\tau_-$ and $D_\kappa$ takes a form given by Eq.~(\ref{eq: D factor for class DIII}), 
\begin{align}
    D_{\kappa=0} = \mqty(\admat{-1, 1}) \qor
    D_{\kappa>0} = \mqty(\admat{\beta^{-\kappa}, -\beta^{\kappa}}).
\end{align}
We distinguish between the cases where the partial indices are $(0, 0)$ and $(+\kappa, -\kappa)$ with $\kappa>0$.
The $\mathbb Z_2$ invariant can be evaluated as
\begin{align}
    (-1)^{\nu[\tau]} = (-1)^{\kappa}.
\end{align}
Thus, the parity of $\kappa$ determines the $\mathbb Z_2$ invariant.

We can rewrite the SWHF of $\tau$, Eq.~(\ref{eq: SWHF of tau}), in a symplectic form
\begin{align} \label{eq: WHF in symplectic form}
    \tau = \tau_+^\mathrm{symp} J \tau_-^\mathrm{symp} \qc
    \tau_+^\mathrm{symp} = \tau_+ S_\kappa,
    \qand
    J = \mqty(\admat{1, -1}).
\end{align}
Here, $[\tau_+^\mathrm{symp}]^\top=\tau_-^\mathrm{symp}$.
The factor $S_{\kappa}$ appearing in $\tau_+^\mathrm{symp}$, takes the form
\begin{align}
S_{\kappa=0} = \mqty(\admat{1, 1}) \qor
S_{\kappa>0} = \mqty(\dmat{\beta^{-\kappa}, 1}).
\end{align}

Using this factorization, we now define the decomposed Ronkin functions:
\begin{align} \label{eq: WHF Ronkins}
    R_\tau^{(+)} (\mu) &= \int_0^{2\pi} \frac{dk}{2\pi} \ln \abs{\det [\tau_+^\mathrm{symp}]_\mu} \\
    R_\tau^{(-)} (\mu) &= \int_0^{2\pi} \frac{dk}{2\pi} \ln \abs{\det [\tau_-^\mathrm{symp}]_\mu}.
\end{align}
These quantities correspond to the contributions from $\tau_+^\mathrm{symp}$ and $\tau_-^\mathrm{symp}$ sectors, respectively.
This WHF-based definition is applicable regardless of the topological phase, providing a unified framework.

Furthermore, this formulation is consistent with the phenomenological construction in Eqs.~(\ref{eq: sr Ronkin plus}) and (\ref{eq: sr Ronkin minus}).
Since $\det\, J=1$, it follows that
\begin{equation}
\det \tau=\det (\tau_+^\mathrm{symp})\det (\tau_-^\mathrm{symp})
\end{equation}
Considering $\tau_+^\mathrm{symp} [\tau_-^\mathrm{symp}]$ has neither roots nor poles inside [outside] the unit circle, we find that
\begin{align}
    \det (\tau_+^\mathrm{symp}) &= \sqrt{C_E \prod_{j=1}^{2p} \beta_j^{-1}} \times \beta^{-\kappa} \prod_{j=1}^{2p} [\beta - \beta_j] \\
    \det (\tau_-^\mathrm{symp}) &= \sqrt{C_E \prod_{j=1}^{2p} \beta_j} \times \beta^{\kappa-2p} \prod_{j=1}^{2p} [\beta - \beta_j^{-1}].
\end{align}
By integrating these expressions, we obtain the symmetry-decomposed Ronkin functions $R_\tau^{(\pm)}$, recovering the forms given in the Eqs.~(\ref{eq: sr Ronkin plus}) and (\ref{eq: sr Ronkin minus}) for the cases where $\kappa=0$, and $1$.

\subsubsection{WHF and generalized Szeg\"o limit theorem for class AII$^\dagger$}

We now provide a proof of the generalized Szeg\"o limit theorem for class AII$^\dagger$ using the WHF-based Ronkin functions.

The TRS$^\dagger$ operator $U_T$ is independent of $\beta$, which implies that $\mathcal T_N[U_T]$ is diagonal.
Accordingly, the OBC spectral potential $\phi(E)$ can be rewritten as
\begin{align}
    \phi(E) = \lim_{N \to \infty} \frac 1N \ln \abs{\det \mathcal T_N[\sigma]}
    = \lim_{N \to \infty} \frac 1N \ln \abs{\det \mathcal T_N[\tau]}.
\end{align}
Using the symplectic form of the WHF, Eq.~(\ref{eq: WHF in symplectic form}), the spectral potential can be decomposed as
\begin{align}
    \phi(E) = \frac{2}{N} \ln \abs{\det \mathcal{T}_N[\tau_+^\mathrm{symp}]} 
    + \frac{1}{N} \ln \abs{\det \mathcal{T}_N[J]} 
    + \order{N^{-1}}.
\end{align}
The symplectic factor $J$ does not contribute to the leading term since $\det \mathcal{T}_N[J] = (\det J)^N = 1$.
The factor $\tau_+^\mathrm{symp}$ itself does not preserve TRS$^\dagger$ symmetry, since it corresponds to only one sector of the full WHF and therefore lacks the full symmetric structure.
The Kramers-paired roots have been separated into $\tau_+^\mathrm{symp}$ and $\tau_-^\mathrm{symp}$.

For $\kappa = 0$, $\tau_+$ is invertible inside the unit circle, and $\tau_+^\mathrm{symp} = \tau_+ S_0$ shares the same analytic domain. 
Thus, $\tau_+^\mathrm{symp}$ admits a trivial WHF, $\tau_+^\mathrm{symp} = \tau_+ S_0 \cdot 1 \cdot 1$,  where both the central factor $D$ and the right factor $A_-$ are the identity, so all partial indices of $\tau_+^\mathrm{symp}$ vanish.
Consequently, the classical Szeg\"o limit theorem applies directly to the symbol $\tau_+^\mathrm{symp}$:
\begin{align}
    \phi(E) = 2 \int_0^{2\pi} \frac{dk}{2\pi} \ln \abs{\det \tau_+^\mathrm{symp}} = \int_0^{2\pi} \frac{dk}{2\pi} \ln \abs{\det \tau}\;.
\end{align}

For $\kappa>0$, the factor $S_\kappa$ contains negative powers of $\beta$.
The WHF of $\tau_+^\mathrm{symp} = \tau_+ S_\kappa$ takes the form $\tau_+^\mathrm{symp} = \tau_+ \cdot S_\kappa \cdot 1$, where the central factor is $S_\kappa$, and hence $\mathcal{K}[\tau_+^\mathrm{symp}] = (0, -\kappa)$.
Consequently, we must apply the generalized Szeg\"o limit theorem, Eq.~(\ref{eq: generalized Szego for class A with (+1, -1)}), which accounts for these nonzero partial indices 
\begin{align}
    \phi(E) = 2 \mathcal M_{\mathcal K^*}[\tau_+^\mathrm{symp}],
\end{align}
where $\mathcal K^*$ denotes the residual partial indices of $\tau_+^\mathrm{symp}$,
i.e., the partial indices of the similarity-transformed symbol $[\tau_+^\mathrm{symp}]_\mu(e^{ik})=\tau_+^\mathrm{symp}(e^{\mu+ik})$ evaluated at the optimally chosen $\mu$ for which the total winding vanishes.
If the partial indices of $[\tau_+^\mathrm{symp}]_\mu$  vanish simultaneously, i.e., the residual partial indices are trivial, $\mathcal K^*[\tau_+^\mathrm{symp}]=(0,0)$, the simple optimization formulation is recovered:
\begin{align} \label{eq: generalized Szego for class AIId with trivial residual partial indices}
    \phi(E) = 2 \mathcal M_{(0, 0)}[\tau_+^\mathrm{symp}] = 2 \min_\mu R_\tau^{(+)}.
\end{align}
If the partial indices of the optimized $[\tau_+^\mathrm{symp}]_\mu$ do not vanish simultaneously, i.e., $\mathcal K^*[\tau_+^\mathrm{symp}]=(-1,1)$, the optimal value must be corrected according to Eq.~(\ref{eq: generalized Szego for class A with (+1, -1)}):
\begin{align} \label{eq: generalized Szego for class AIId with nontrivial residual partial indices}
    \phi(E) =2\mathcal M_{(+1,-1)}[\tau_+^\mathrm{symp}] &= 2(\min_\mu R_\tau^{(+)}(\mu) - (\mu_{k+1}-\mu_k)),
\end{align}
where $\mu_k$ and $\mu_{k+1}$ denote, respectively, the dominant roots of $\tau_+^\mathrm{symp}$ located just inside and just outside the circle of radius $e^{\mu}$.

Combining the results for $\kappa=0$ and $\kappa>0$, we arrive at a unified expression for the OBC spectral potential, valid for all $\kappa \ge 0$:
\begin{align} \label{eq: generalized Szego for class AIId}
    \phi(E) = 2 \mathcal{M}_{\mathcal K^*}[\tau_+^\mathrm{symp}]
\end{align}

This expression reproduces Eq.~(\ref{eq: OBC spectral potential}) for $\kappa=0$ and $1$, and extends it naturally to higher topological sectors ($\kappa\ge2$).

\section{Numerical verification} \label{sec: Numerical verification}
In this section, we validate the WHF-based formulation developed in the preceding sections.
First, we analyze class-A two-band systems with nonvanishing partial indices and examine the correction to the generalized Szeg\"o limit theorem.
We show that nonzero partial indices invalidate the simple optimization formalism, and that the necessary correction is given by Eq.~(\ref{eq: generalized Szego for class A with (+1, -1)}).
Next, we verify the WHF-based decomposition of the Ronkin function for systems with TRS$^\dagger$, considering two representative parameter sets.
We demonstrate that the WHF-based decomposed Ronkin functions reproduce the phenomenological symmetry-decomposed Ronkin functions in the regions where $\kappa = 0$ and $1$.
Finally, we investigate the parameter regime where a $\kappa = 2$ domain emerges.

\subsection{Correction for non-vanishing partial indices in two-band class A systems}
In this section, we numerically verify the correction to the generalized Szeg\"o limit theorem that arises from nonvanishing partial indices, Eq.~(\ref{eq: generalized Szego for class A with (+1, -1)}).
Our aim is to demonstrate a situation in which the conventional Amoeba formulation fails, thereby necessitating the correction term derived in Sec.~\ref{sec: Amoeba multiband}.

To this end, we consider a model in which there is a parameter regime, where the partial indices $\mathcal{K}[\sigma_\mu]$ do not simultaneously vanish for any shift $\mu$.
The simplest example with this property is a block-diagonal two-band Hamiltonian composed of two uncoupled Hatano-Nelson chains:
\begin{align} \label{eq: two-HN}
\begin{cases}
    h(\beta) &= \mqty(\dmat{h_1(\beta), h_2(\beta)}) \\
    h_1(\beta) &= 1.3\,\beta^{-1} - 0.2 + 0.7\,\beta \\
    h_2(\beta) &= 0.5\,\beta^{-1} + 0.3 + 0.1i + 1.5\,\beta
\end{cases}
\end{align}
We define the symbols of the blocks as $\sigma_{1,2} = E - h_{1,2}$ and the full symbol as $\sigma = \mathrm{diag}(\sigma_1, \sigma_2)$.
The partial indices of the full system, $\mathcal{K}[\sigma] = (\kappa_1, \kappa_2)$, are then given simply by the winding numbers of the uncoupled blocks, $\kappa_{1,2} = W[\sigma_{1,2}]$.

\begin{figure}[t]
    \centering
    \includegraphics[width=0.85\linewidth]{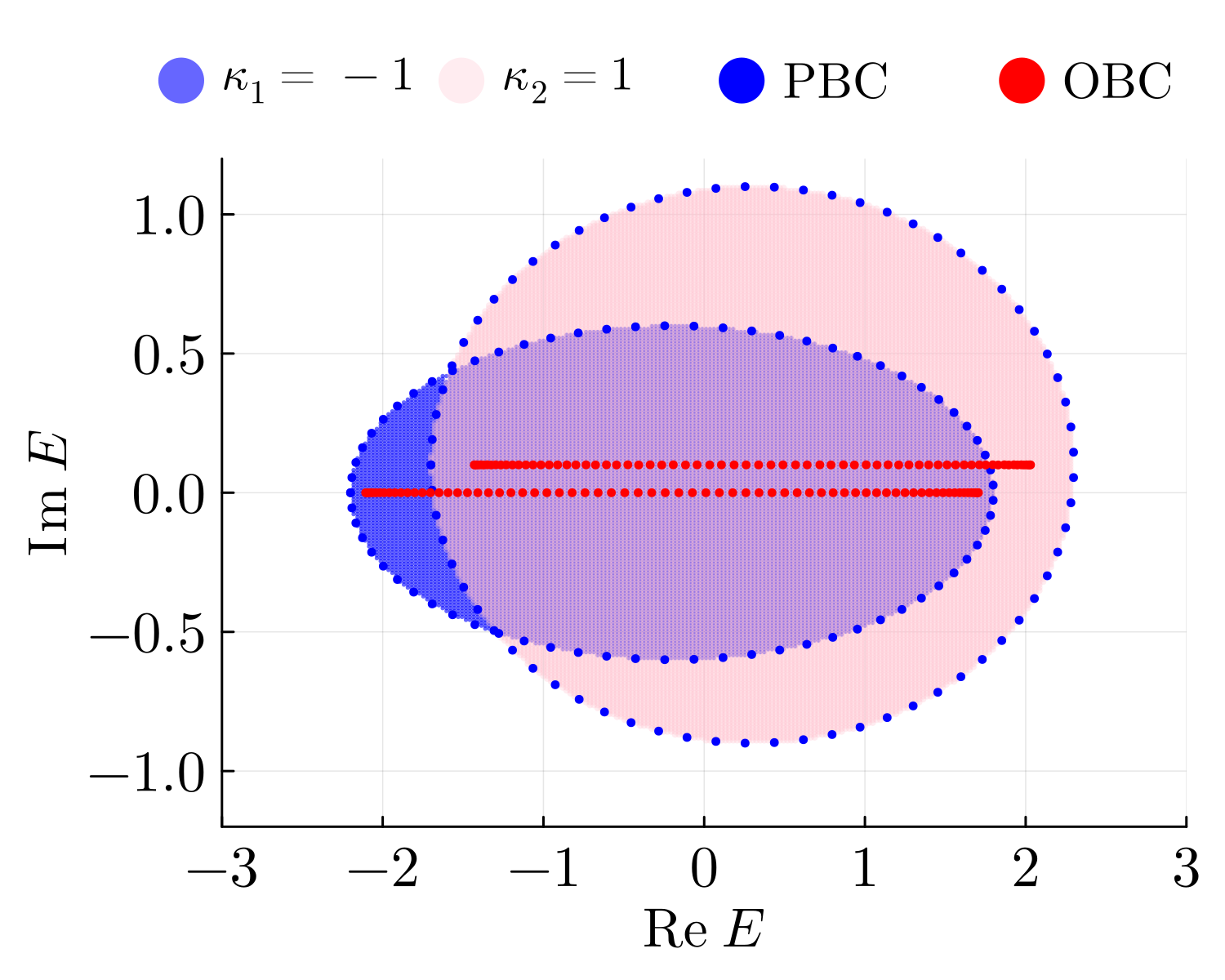}
  \caption{
        Comparison of OBC and PBC spectra for the two uncoupled Hatano-Nelson models defined in Eq.~(\ref{eq: two-HN}).
        Blue dotted lines show the PBC spectra of the individual blocks $h_1$ and $h_2$, while red lines denote the full OBC spectrum for $N=70$.
        Shaded regions in the complex energy plane indicate where the partial indices are nonzero.
        The blue-shaded region corresponds to $\kappa_1 = W[\sigma_1] = -1$ with $\kappa_2 = 0$, and the red-shaded region corresponds to $\kappa_2 = W[\sigma_2] = 1$ with $\kappa_1 = 0$.
        Their overlap (purple) represents the regime $\mathcal{K}[\sigma] = (-1, 1)$, where the conventional Amoeba optimization breaks down and the generalized formulation of Sec.~\ref{sec: Amoeba multiband} must be applied.
    }
    \label{fig: Two HN spectrum and pi}
\end{figure}

Figure~\ref{fig: Two HN spectrum and pi} depicts the topological phases in the complex energy plane, showing the PBC (blue dotted lines) and OBC (red lines) spectra.
The shaded regions indicate where the partial indices (winding numbers) take nonzero values.
In the blue-shaded area, $\mathcal{K}[\sigma]=(-1,0)$; in the red-shaded area, $\mathcal{K}[\sigma]=(0,1)$.
The overlap between them, shown in purple, marks the regime where $\mathcal K[\sigma]=(-1,1)$, in which the conventional Amoeba formulation fails and the generalized correction becomes essential.
In the white region where $\mathcal K[\sigma] = (0,0)$, the PBC and OBC potentials coincide, and the conventional Amoeba formulation holds.

\begin{figure}
    \centering
    \begin{subfigure}{0.95\textwidth}
        \centering
        \subcaption{}
        \vspace{-15pt}
        \includegraphics[width=0.85\linewidth]{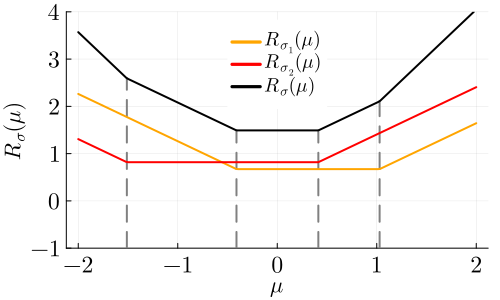}
    \end{subfigure}
    \begin{subfigure}{0.95\textwidth}
        \centering
        \subcaption{}
        \vspace{-15pt}
        \includegraphics[width=0.85\linewidth]{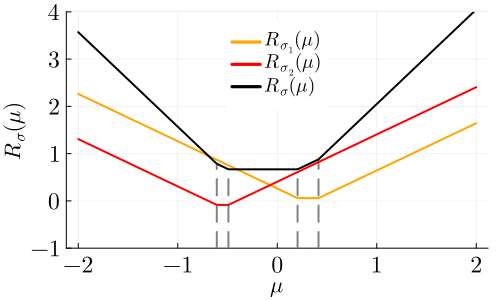}
    \end{subfigure}
    \caption{
    Ronkin functions for two representative reference energies $E$, selected from distinct topological domains in Fig.~\ref{fig: Two HN spectrum and pi}.
    The total Ronkin function $R_\sigma$ is shown in black, while the partial Ronkin functions $R_{\sigma_1}$ and $R_{\sigma_2}$ are plotted in red and orange, respectively.
    (a) Trivial case ($E = -2.0 + 1.0i$): The reference point lies in the white region, where $\mathcal{K}[\sigma] = (0, 0)$. 
    Both $R_{\sigma_1}$ and $R_{\sigma_2}$ can be minimized simultaneously, and the conventional optimization scheme applies.
    (b) Nontrivial case ($E = 0.0 + 0.2i$): The reference point lies in the purple region, where $\mathcal{K}[\sigma] = (-1, 1)$.
    Here, $R_{\sigma_1}$ and $R_{\sigma_2}$ cannot be minimized simultaneously, and the correction term in Eq.~(\ref{eq: generalized Szego for class A with (+1, -1)}) is required to obtain the correct OBC potential.
    }
    \label{fig: Two HN partial indices for white and purple}
\end{figure}

Figure~\ref{fig: Two HN partial indices for white and purple} verifies our theoretical prediction by analyzing the optimization landscape at two representative reference energies $E$ for the Hamiltonian defined in Eq.~(\ref{eq: two-HN}).
The panels show the Ronkin functions $R_{\sigma}$ (black), $R_{\sigma_1}$ (red), and $R_{\sigma_2}$ (orange), with vertical gray dashed lines marking the root positions $\mu_j$.
Fig.~\ref{fig: Two HN partial indices for white and purple}(a) shows the trivial case at $E=-2.0+1.0i$ where the partial indices vanish simultaneously, i.e., $\mathcal K[\sigma] = (0, 0)$.  Fig.~\ref{fig: Two HN partial indices for white and purple}(b) shows the nontrivial case at $E=0.0+0.2i$ where the partial indices do not vanish simultaneously, i.e., $\mathcal K[\sigma] = (-1, 1)$.
We note that since the winding number $W[\sigma]$ vanishes at both points, the conventional optimization formulation predicts the OBC potential as $R_\sigma(0)$.

For the case $E=-2.0+1.0i$ [Fig.~\ref{fig: Two HN partial indices for white and purple}(a)], the reference point lies in the white region of Fig.~\ref{fig: Two HN spectrum and pi}, and $\mathcal K[\sigma] = (0, 0)$.
The Ronkin functions $R_{\sigma_1}$ and $R_{\sigma_2}$ admit a common minimization interval, so the simple optimization scheme applies.
In contrast, for $E=0.0+0.2i$ [Fig.~\ref{fig: Two HN partial indices for white and purple}(b)], the reference point lies in the purple region with nonzero partial indices, $\mathcal K[\sigma] = (-1, 1)$.
As a result, there is no region where the Ronkin functions $R_{\sigma_1}$ and $R_{\sigma_2}$ can be jointly minimized, signaling the breakdown of the simple optimization scheme and requiring the correction term in Eq.~(\ref{eq: generalized Szego for class A with (+1, -1)}).
For this specific model and parameter set, the correction term evaluates to $-(\mu_{3}-\mu_{2})$, corresponding to the modification $\min_\mu R_{\sigma}(\mu) - (\mu_{k+1}-\mu_{k})$ in Eq.~(\ref{eq: generalized Szego for class A with (+1, -1)}). 
This correction accounts for the exchange of contributions between the largest inner root ($\mu_2$) and the smallest outer root ($\mu_3$) at the minimum of $R_\sigma$.

\begin{figure}
    \centering
    \begin{subfigure}{0.95\textwidth}
        \centering
        \subcaption{}
        \vspace{-15pt}
        \includegraphics[width=0.85\linewidth]{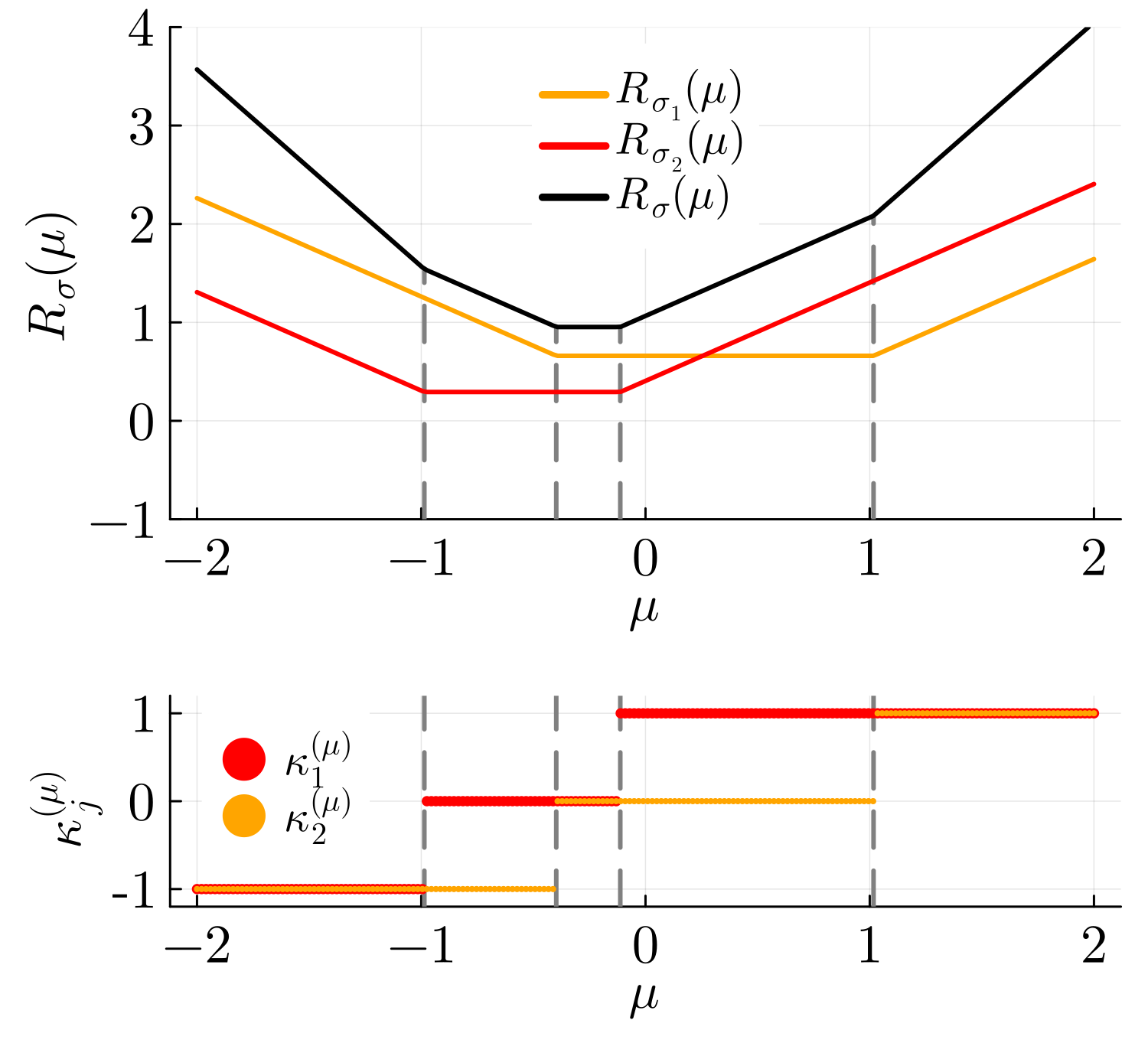}
    \end{subfigure}
    \begin{subfigure}{0.95\textwidth}
        \centering
        \subcaption{}
        \vspace{-15pt}
        \includegraphics[width=0.85\linewidth]{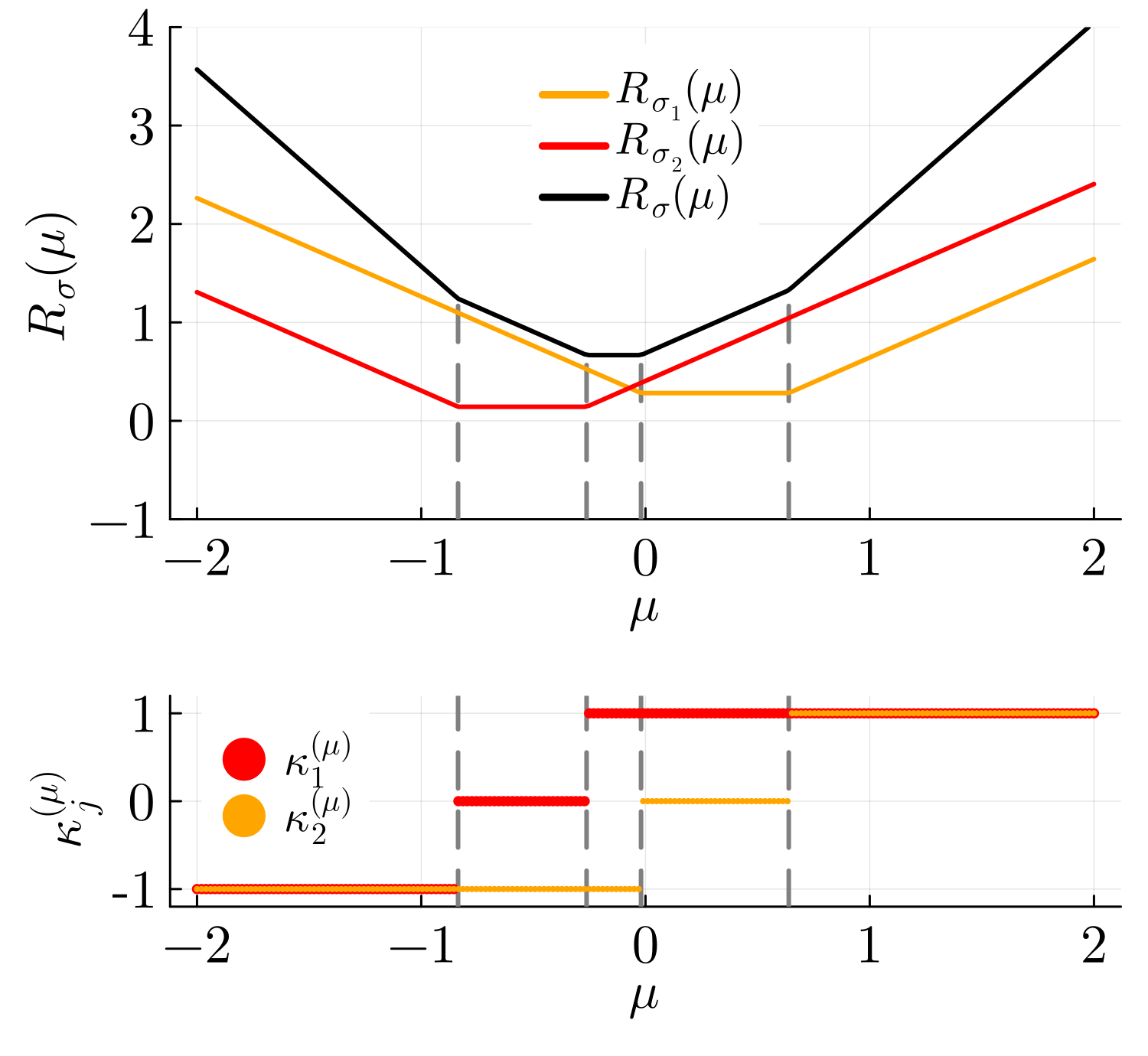}
    \end{subfigure}
    \caption{
    Ronkin functions and $\mu$-dependent partial indices for two reference energies taken from the red region in Fig.~\ref{fig: Two HN spectrum and pi}.
    Each top panel shows the Ronkin functions, similar to Fig.~\ref{fig: Two HN partial indices for white and purple}. Each bottom panel displays the $\mu$-dependent partial indices $\mathcal{K}[\sigma_\mu] = (\kappa_1^{(\mu)}, \kappa_2^{(\mu)})$.
    (a) $E = 2.2 + 0.1i$: The partial indices vanish after optimization, i.e., the residual partial indices are $\mathcal{K}^*[\sigma] = (0, 0)$, yielding the correct OBC potential through the standard optimization scheme.
    (b) $E = 0.5 + 0.6i$: The residual partial indices are nonzero even after optimization, $\mathcal{K}^*[\sigma] = (+1, -1)$, indicating that the correction in Eq.~(\ref{eq: generalized Szego for class A with (+1, -1)}) is required to obtain the correct potential.
    }
    \label{fig: Two HN partial indices for red}
\end{figure}

The above discussion extends to points with non-vanishing winding number $W[\sigma] \neq 0$ by employing the similarity transformation and using $\mu$-dependent partial indices $\mathcal K[\sigma_\mu]=(\kappa_1^{(\mu)}, \kappa_2^{(\mu)})$.
Figure~\ref{fig: Two HN partial indices for red} illustrates this for two instructive reference energies from the red region in Fig.~\ref{fig: Two HN spectrum and pi}.
In addition to the Ronkin functions, we show the partial indices $\kappa_1^{(\mu)}$ (red) and $\kappa_2^{(\mu)}$ (orange) separately.
For $E=2.2+0.1i$ [Fig.~\ref{fig: Two HN partial indices for red}(a)], the partial indices share a common vanishing interval; equivalently, the residual partial indices vanish $\mathcal K^*[\sigma]=(0, 0)$.
In this case, the simple optimization scheme applies, and the minimum of $R_\sigma(\mu)$ correctly yields the OBC potential.
For the case $E=0.5+0.6i$ [Fig.~\ref{fig: Two HN partial indices for red}(b)], the partial indices never vanish simultaneously for any $\mu$, so $\mathcal K^*[\sigma]=(+1, -1)$, and the correction in Eq.~(\ref{eq: generalized Szego for class A with (+1, -1)}) is necessitated.

\begin{figure}
    \centering
    \begin{subfigure}{0.95\textwidth}
        \centering
        \subcaption{}
        \vspace{-15pt}
        \includegraphics[width=0.85\linewidth]{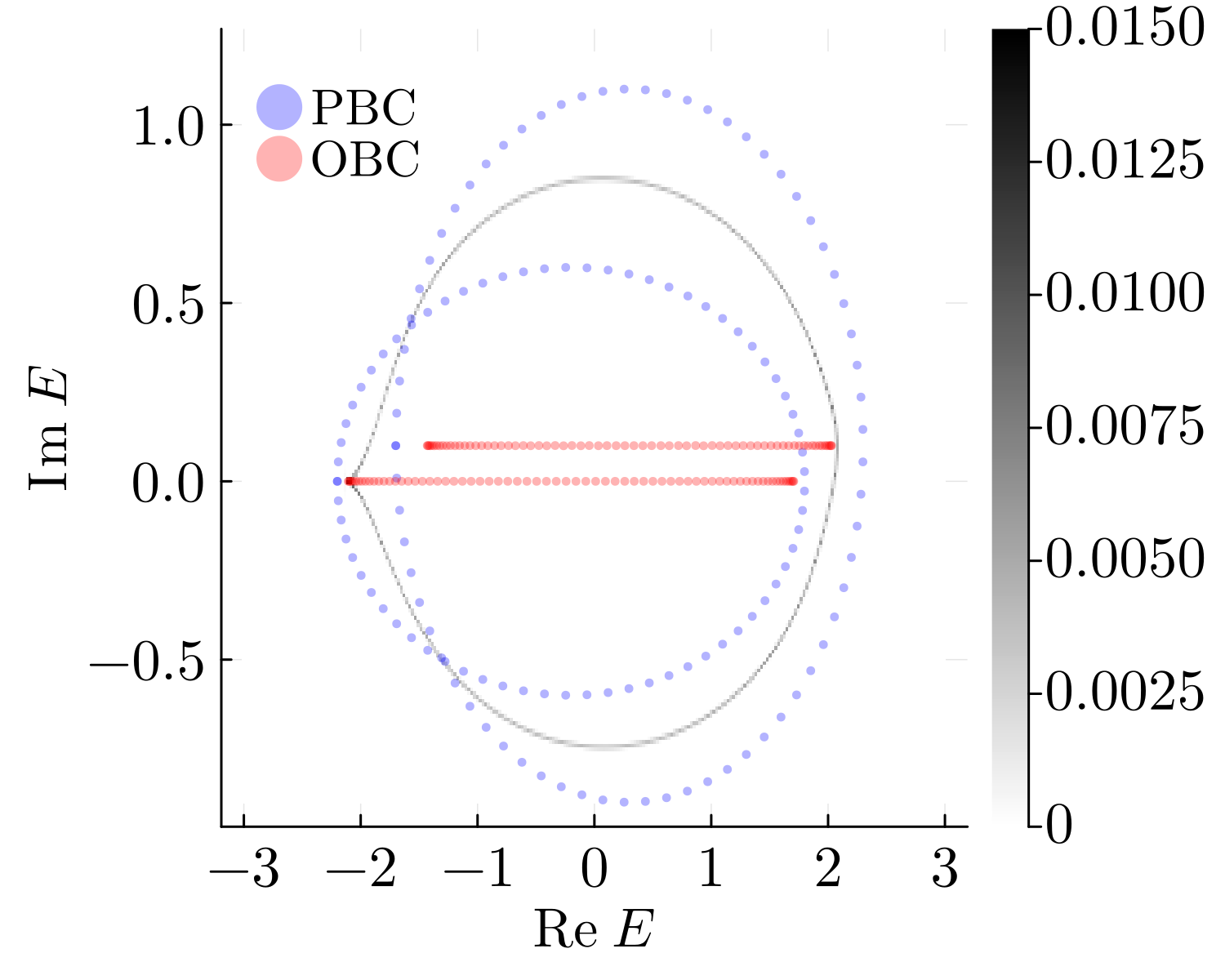}
    \end{subfigure}
    \begin{subfigure}{0.95\textwidth}
        \centering
        \subcaption{}
        \vspace{-15pt}
        \includegraphics[width=0.85\linewidth]{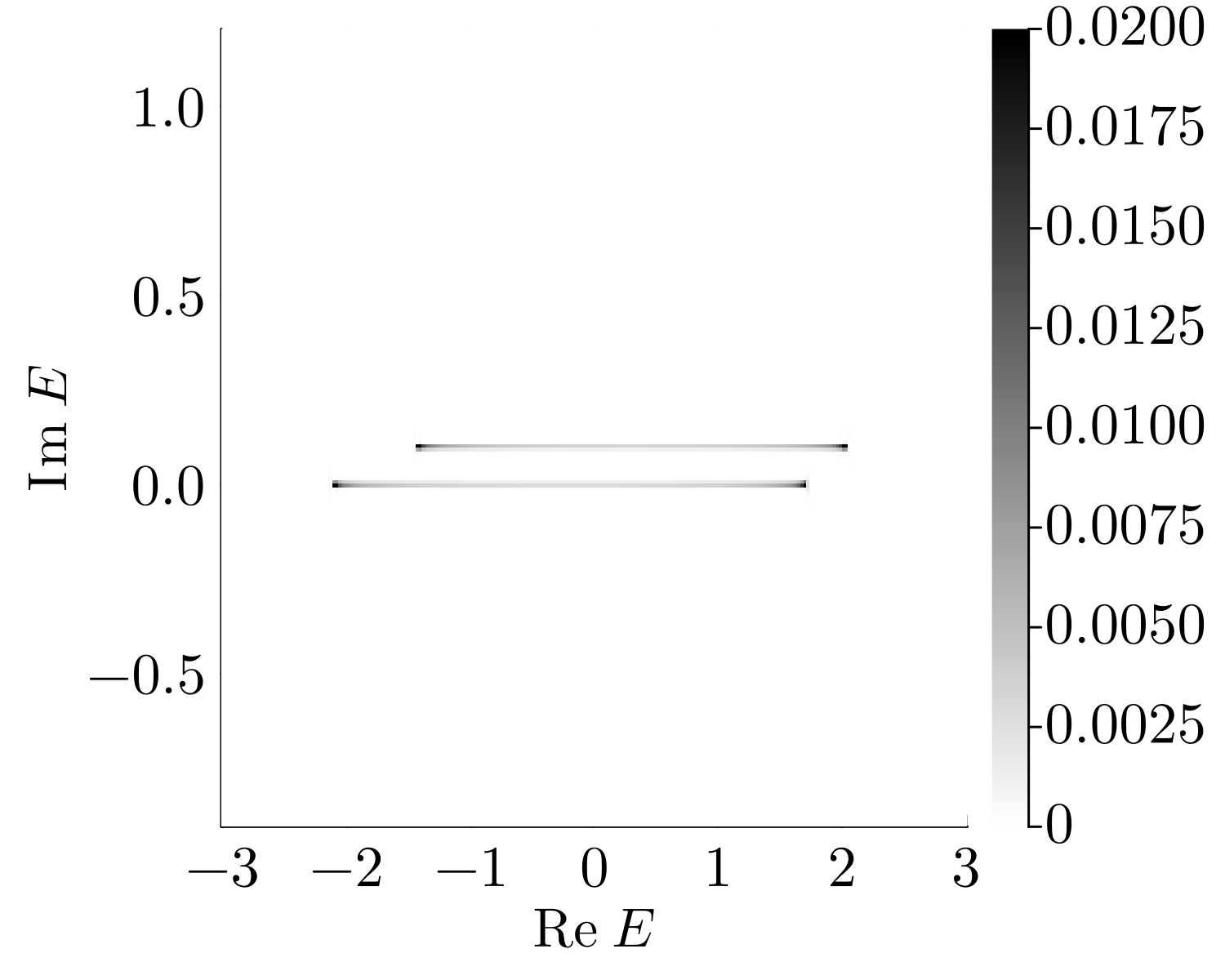}
    \end{subfigure}
    \caption{
    Comparison of the DOS evaluated by (a) the conventional formulation and (b) our WHF-based formulation.
    Values are scaled by the area of the energy cell $\Delta \mathrm{Re}~E \times \Delta \mathrm{Im}~E \simeq 2.5 \times 10^{-4}$ due to the discretizations of the Hessian in Eq.~(\ref{eq: DOS and Potential}).
    (a) The conventional simple optimization fails to reproduce the OBC spectrum (red line), as evident from the mismatch with the calculated DOS (grayscale).
    (b) The WHF-based formulation, which includes the correction for nonvanishing partial indices, accurately reproduces the OBC spectrum.
    }
    \label{fig: Two HN DOS}
\end{figure}

To test our prediction quantitatively, we compute the DOS, using the residual partial indices as a guide to determine where the correction term must be applied.
Figure~\ref{fig: Two HN DOS} compares the DOS obtained from (a) the conventional simple optimization and (b) our WHF-based formulation [Eq.~(\ref{eq: generalized Szego for class A with (+1, -1)})].
Because of the discretization of the Hessian in Eq.~(\ref{eq: DOS and Potential}), the plotted values are scaled by the energy cell area
$\Delta \mathrm{Re}~E \times \Delta \mathrm{Im}~E \simeq 2.5 \times 10^{-4}$.
As shown in Fig.~\ref{fig: Two HN DOS}(a), the simple optimization fails to reproduce the OBC DOS.
By contrast, the corrected WHF-based formulation [Fig.~\ref{fig: Two HN DOS}(b)] accurately reproduces the OBC DOS, in excellent agreement with the spectrum shown in Fig.~\ref{fig: Two HN spectrum and pi}.

\subsection{WHF framework in systems with TRS$^\dagger$}
Next, we verify the WHF-based decomposition for class AII$^\dagger$ in one dimension by analyzing the two-band symplectic Hatano-Nelson model with next-nearest-neighbor hopping for two different parameter sets. 
The first set corresponds to the model previously studied in Ref.~\cite{SKaneshiro_PRB_2025_SymplecticAmoebaFormulation}. 
We demonstrate that the WHF-based decomposition of the Ronkin function reproduces the symmetry-decomposed form introduced in that work, confirming the consistency of the two approaches.

The non-Bloch Hamiltonian is given by
\begin{align} \label{eq: Numerics_nonBloch}
    h(\beta)
    &= 
      (t_2 -i \Delta_2 X + g_2 Z) \beta^{-2}
    + (t_1 -i \Delta_1 X + g_1 Z) \beta^{-1} \nonumber \\
   &+ (t_1 +i \Delta_1 X - g_1 Z) \beta
    + (t_2 +i \Delta_2 X - g_2 Z) \beta^2
\end{align}
where $X, Y, Z$ are Pauli matrices.
The TRS$^\dagger$ operator is $Y$, which is independent of $\beta$, consistent with Eq.~(\ref{eq: HC_Top_CC}).

\subsubsection{Regime with $\kappa = 0, 1$}
First, we confirm that the WHF-based formalism reproduces the symmetry-decomposed Ronkin function for parameters yielding the positive partial index as $\kappa = 0$ or $1$.
The parameters are set to $t_1 = 0.3$, $t_2 = 0.8$, $g_1 = 0.5$, $\Delta_1 = 0.3$, and $\Delta_2 = 0.2$.

\begin{figure}
    \centering
    \includegraphics[width=0.9\linewidth]{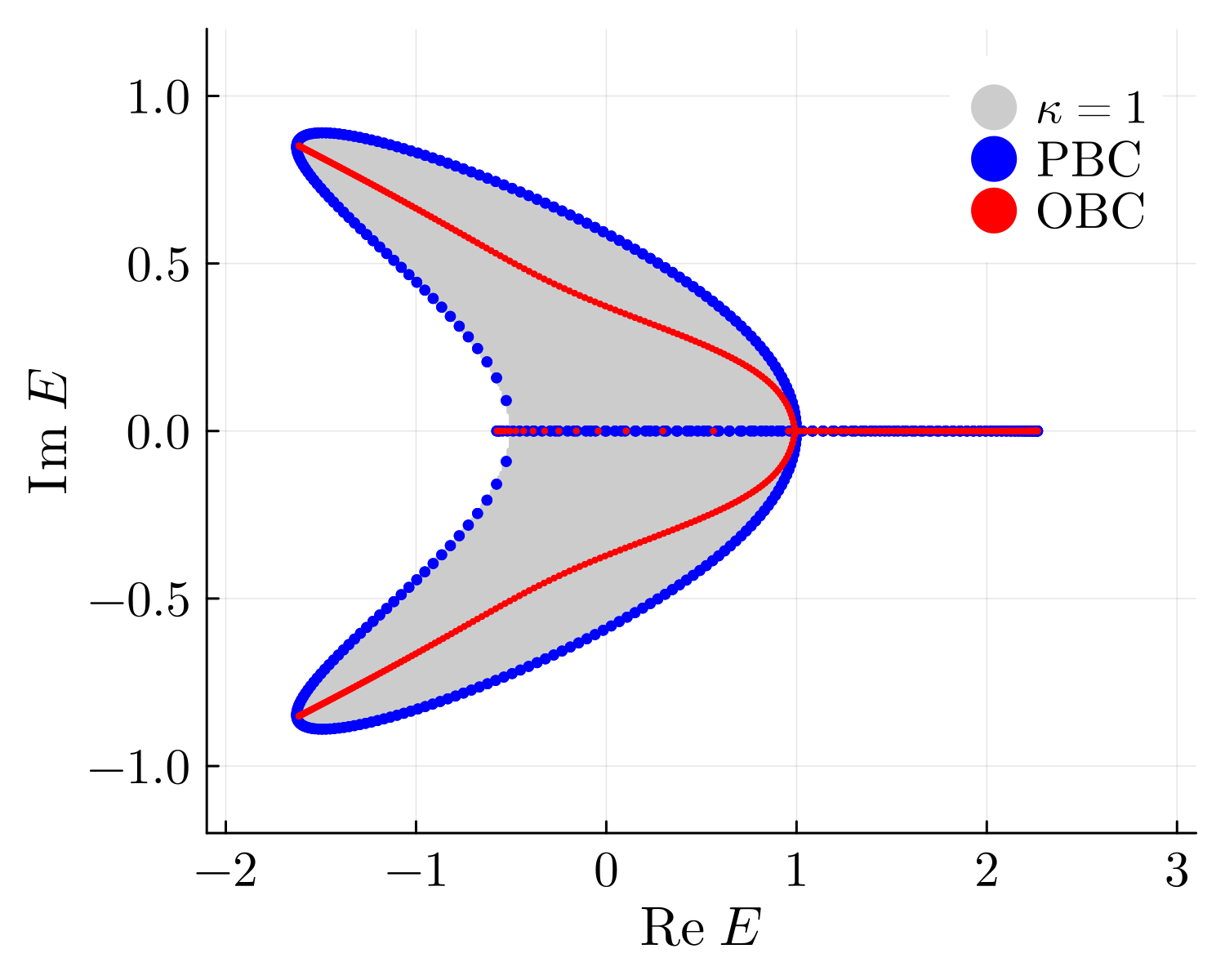}
    \caption{
        Validation of the partial index $\kappa$ as the topological invariant.
        The Hamiltonian is defined in Eq.~(\ref{eq: Numerics_nonBloch}), with parameters $t_1 = 0.3$, $t_2 = 0.8$, $g_1 = 0.5$, $\Delta_1 = 0.3$, and $\Delta_2 = 0.2$.
        Blue and red points indicate the PBC and OBC spectra, respectively ($N = 400$).
        The shaded region denotes the domain where the WHF-derived partial index is $\kappa = 1$.
        This region coincides with the $\mathbb{Z}_2$-nontrivial phase ($\nu[\tau] = 1$) obtained independently, confirming that the parity of $\kappa$ correctly captures the $\mathbb{Z}_2$ topology.
    }
    \label{fig: l_config_z2}
\end{figure}

In this parameter regime, the system is characterized by the $\mathbb Z_2$ topological invariant $\nu[\tau]$ defined in Eq.~(\ref{eq: Z2 topological invariant})~\cite{KKawabata_PRX_2019_SymmetryAndTopology}, where $\tau = (E-h)U_T$.
As established in Sec.~\ref{sec: SWHF of Bloch}, $\nu[\tau]$ is equivalent to the parity of the sum of positive partial indices.
Since the partial indices in this system take the paired form $\mathcal{K}[\tau]=(\kappa, -\kappa)$, $\nu[\tau]$ corresponds simply to the parity of $\kappa$.

Figure~\ref{fig: l_config_z2} shows the positive partial index $\kappa$ of $\tau$ across the complex energy plane.
Blue and red points indicate the PBC and OBC spectra, respectively, obtained by diagonalizing the Hamiltonian with $400$ sites. 
This model exhibits two distinct phases: $\kappa = 0$ (white) and $\kappa = 1$ (shaded). 
As evaluated in Ref.~\cite{SKaneshiro_PRB_2025_SymplecticAmoebaFormulation}, the shaded $\kappa = 1$ region coincides with the topologically nontrivial region identified by the $\mathbb{Z}_2$ topological invariant $\nu[\tau]$.
This confirms that the parity of $\kappa$ provides a consistent and physically equivalent characterization of the $\mathbb{Z}_2$ topology in class~AII$^\dagger$ systems.

\begin{figure*}
    \centering
    \begin{subfigure}{0.325\textwidth}
        \centering
        \subcaption{}
        \vspace{-15pt}
        \includegraphics[width=0.85\linewidth]{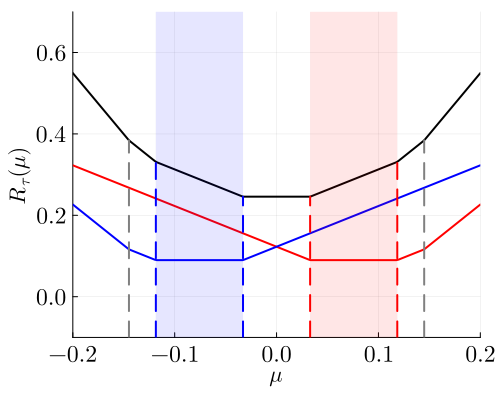}
    \end{subfigure}
    \hfill
    \begin{subfigure}{0.325\textwidth}
        \centering
        \subcaption{}
        \vspace{-15pt}
        \includegraphics[width=0.85\linewidth]{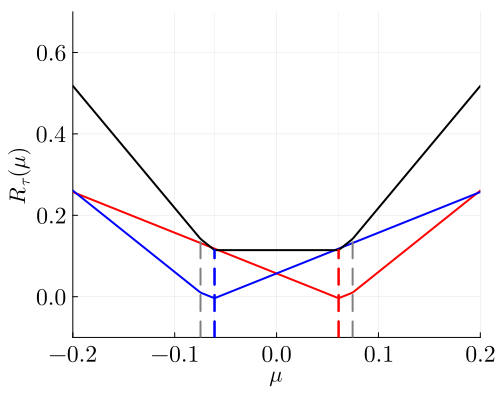}
    \end{subfigure}
    \hfill
    \begin{subfigure}{0.325\textwidth}
        \centering
        \subcaption{}
        \vspace{-15pt}
        \includegraphics[width=0.85\linewidth]{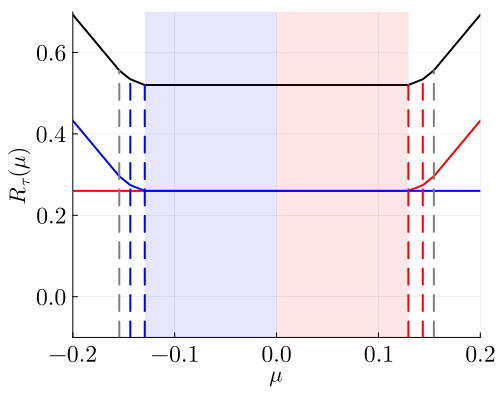}
    \end{subfigure}
    
    \vspace{1em}
    
        \begin{subfigure}{0.325\textwidth}
            \centering
            \subcaption{}
            \vspace{-15pt}
            \includegraphics[width=0.85\linewidth]{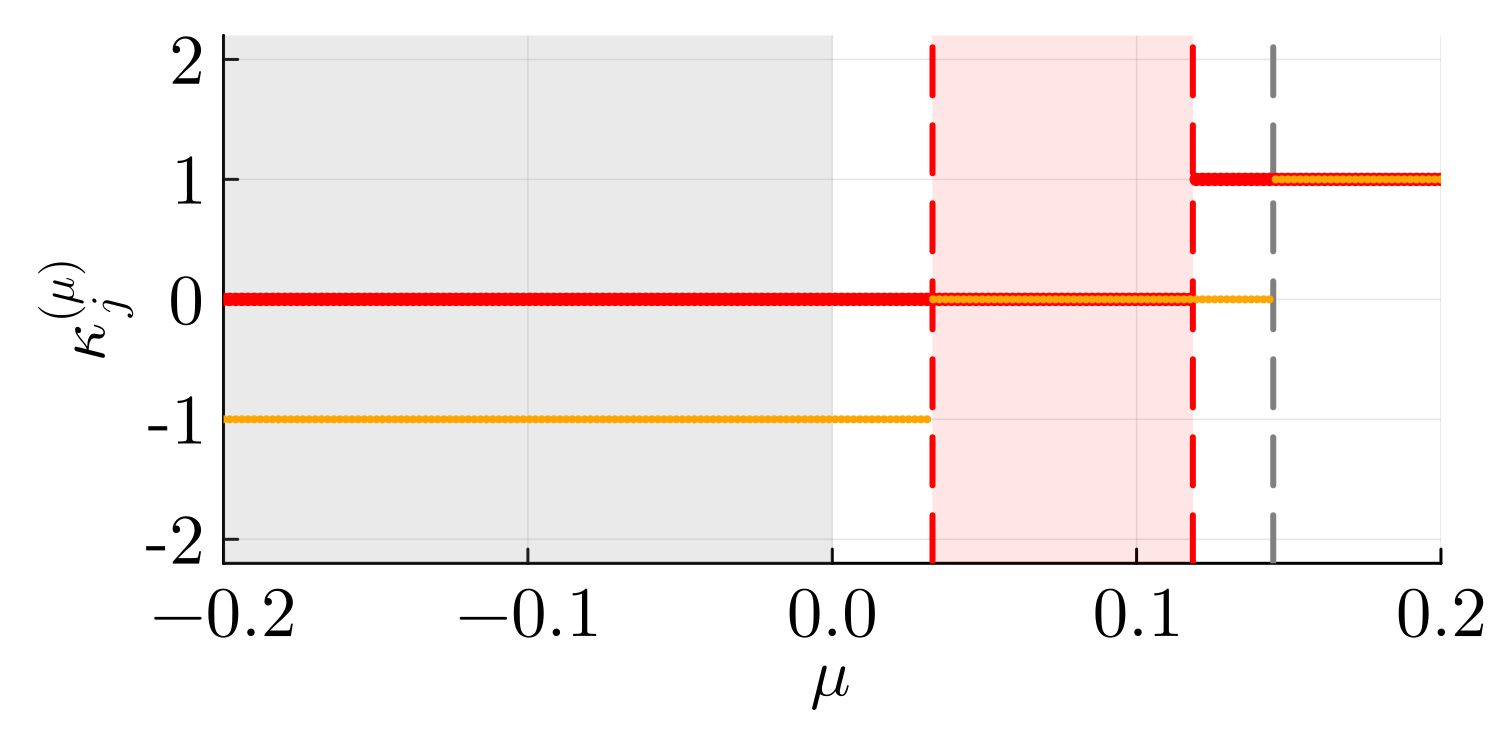}
        \end{subfigure}
    \hfill
        \begin{subfigure}{0.325\textwidth}
            \centering
            \subcaption{}
            \vspace{-15pt}
            \includegraphics[width=0.85\linewidth]{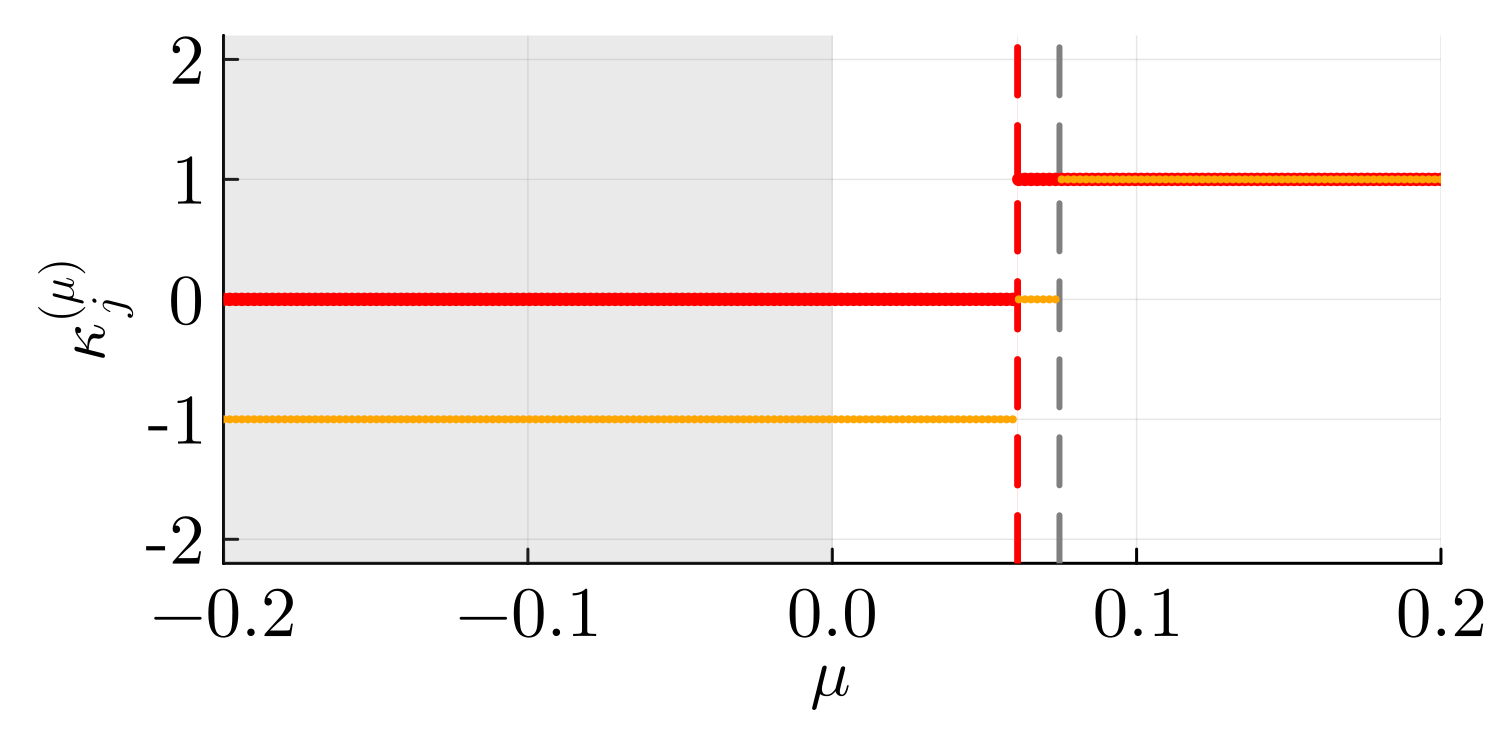}
        \end{subfigure}
    \hfill
        \begin{subfigure}{0.325\textwidth}
            \centering
            \subcaption{}
            \vspace{-15pt}
            \includegraphics[width=0.85\linewidth]{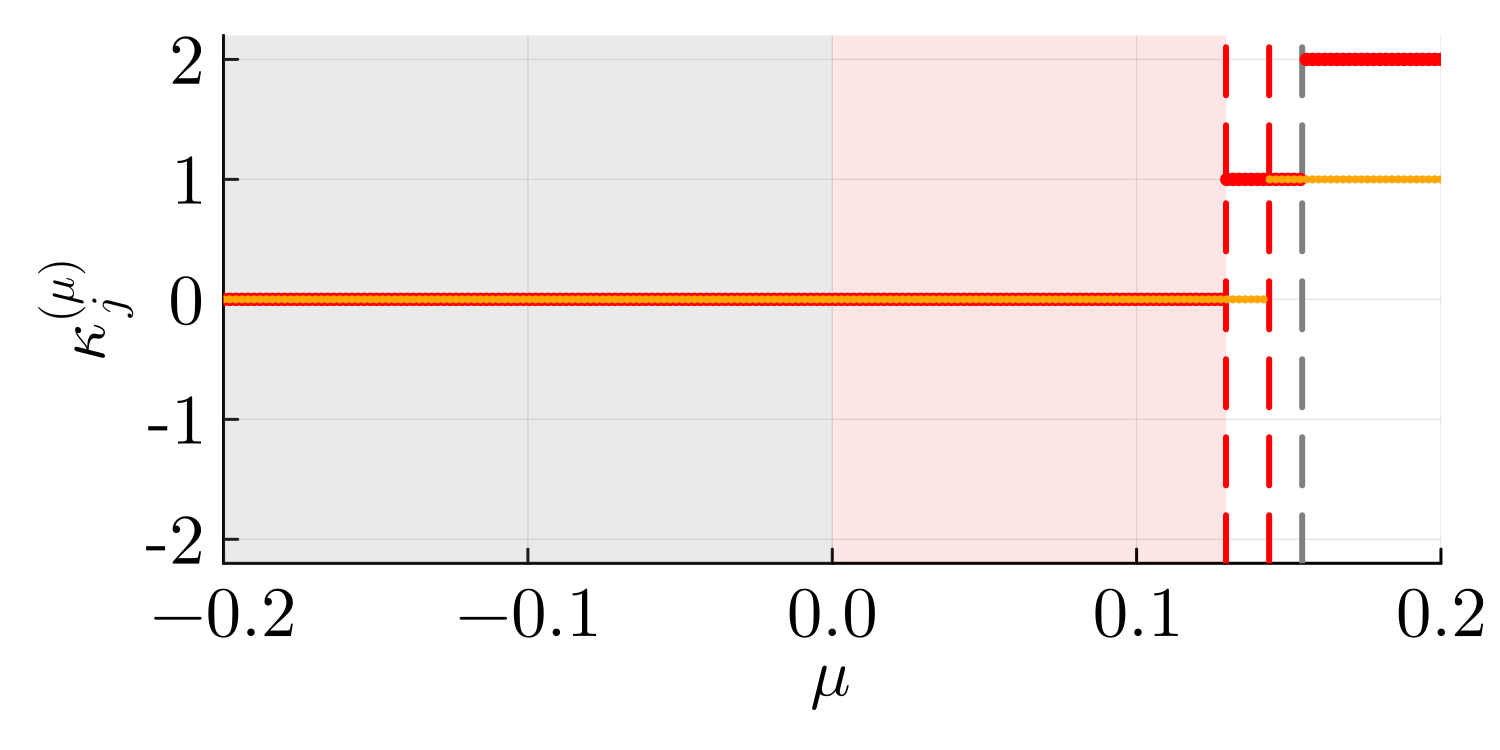}
        \end{subfigure}
    \caption{
    WHF-based decomposed Ronkin functions (top row, a-c) and partial indices of $[\tau_+^\mathrm{symp}]_\mu$, 
    $\mathcal{K}[[\tau_+^\mathrm{symp}]_\mu] = (\kappa_1^{(\mu)}, \kappa_2^{(\mu)})$ (bottom row, d-f).
    The panels correspond to three reference energies:
    (a, d) $E = -0.5 + 0.3i$ ($\kappa = 1$), 
    (b, e) $E = 0.5 + 0.265i$ ($\kappa = 1$), and 
    (c, f) $E = 1.0 + 0.5i$ ($\kappa = 0$).
    (a-c): decomposed Ronkin functions $R_\tau^{(+)}$ (red), $R_\tau^{(-)}$ (blue), and total $R_\tau$ (black).
    Vertical dashed lines mark the root positions $\mu = \pm \mu_j$, where the colored lines correspond to the GBZ condition $\mu_1=\mu_2$.
    Due to TRS$^\dagger$, $R_\tau$ is even in $\mu$, and $R_\tau^{(\pm)}$ obey the symmetry relation in Eq.~(\ref{eq: partial Ronkin symmetry}).
    (d-f): partial indices $\kappa_1^{(\mu)}$ (red) and $\kappa_2^{(\mu)}$ (orange) of $[\tau_+^{\mathrm{symp}}]_\mu$.
    Due to the symmetry, we shaded the negative region $\mu \le 0$ in gray.
    The colored shaded regions for $\mu > 0$ indicate the optimization intervals: $[0, \mu_1]$ for the $\kappa=0$ phase (c, f) and $[\mu_1, \mu_2]$ for the $\kappa=1$ phase (a, b, d, e).
    The partial indices change whenever $\mu$ crosses a root $\mu_j$ where $\sigma_\mu$ closes its gap.
    Their sum, $\kappa_1^{(\mu)} + \kappa_2^{(\mu)}$, equals the derivative of $R_\tau^{(+)}$.
    The WHF-based decomposition reproduces the phenomenological symmetry-decomposed Ronkin functions for $\kappa = 1$ and symmetrically decomposition $R_\tau$ for $\kappa = 0$.}
    \label{fig: WHF partial Ronkin}
\end{figure*}

\begin{figure}
    \centering
    \includegraphics[width=0.9\linewidth]{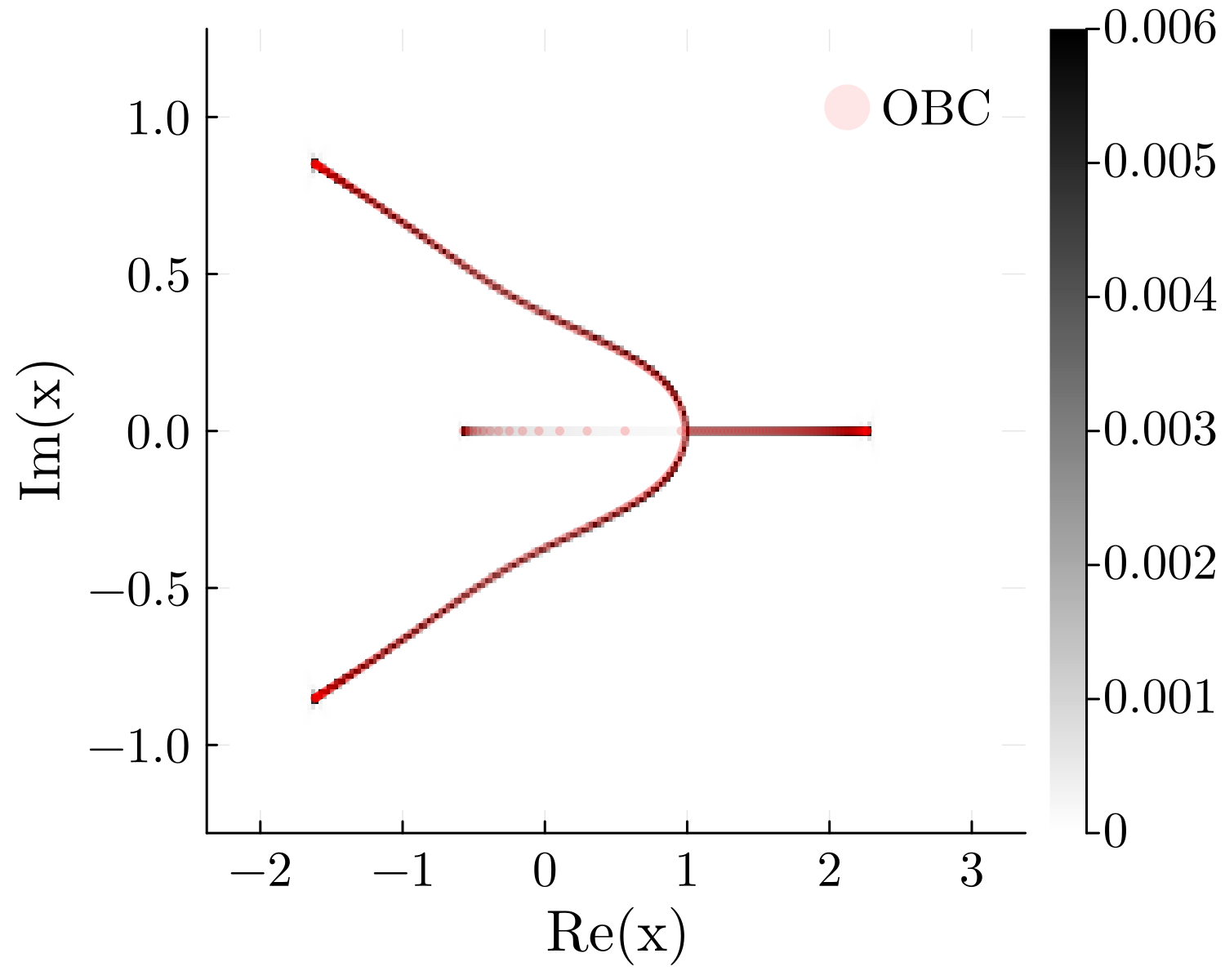}
    \caption{
    DOS multiplied by $\Delta\mathrm{Re}~E \times \Delta\mathrm{Im}~E \simeq 4 \times 10^{-4}$,
    calculated by optimizing the WHF-based decomposed Ronkin functions defined in Eq~(\ref{eq: WHF Ronkins}).
    The Hamiltonian and parameters are equivalent to those in Fig.~\ref{fig: l_config_z2}.
    The WHF-based formulation reproduces the OBC spectrum (red line, $N=400$).
    }
    \label{fig: DOS kappa=1}
\end{figure}

Having confirmed that $\kappa$ serves as the correct topological index, we now verify the central mechanism of our WHF formulation [Eq.~(\ref{eq: generalized Szego for class AIId})], which defines the OBC potential through the symplectic form of the WHF, $\tau = \tau_+ S_\kappa J [\tau_+ S_\kappa]^\top,$
as given in Eq.~(\ref{eq: WHF in symplectic form}). 
Figure~\ref{fig: WHF partial Ronkin} illustrates this test for three representative energies $E$.
The top row [Fig.~\ref{fig: WHF partial Ronkin}(a-c)] presents the WHF-based decomposed Ronkin functions $R_\tau^{(\pm)}$, 
while the bottom row [Fig.~\ref{fig: WHF partial Ronkin}(d-f)] shows the corresponding partial indices of $[\tau_+^\mathrm{symp}]_\mu = [\tau_+ S_\kappa]_\mu$.
The reference points are chosen as $E = -0.5 + 0.3i$ ($\kappa = 1$) [(a,d)], $E = 0.5 + 0.265i$ ($\kappa = 1$) [(b,e)], and $E = 1.0 + 0.5i$ ($\kappa = 0$) [(c,f)]. 
In all panels, the vertical dashed lines indicate the roots $\pm\mu_j$.
Specifically, the colored dashed lines mark the GBZ roots $\mu_1, \mu_2$ and their negatives.
The red and blue shaded regions indicate the intervals over which the minimization is performed.

The top panels [Fig.~\ref{fig: WHF partial Ronkin}(a-c)] demonstrate that the WHF-derived $R_\tau^{(\pm)}$ perfectly reproduce the phenomenological decomposed Ronkin functions [Eqs.~(\ref{eq: sr Ronkin plus}) and (\ref{eq: sr Ronkin minus})]: they are convex, piecewise linear, and satisfy the symmetry relation $R_{\tau}^{(+)}(\mu)=R_{\tau}^{(-)}(-\mu)$ [Eq.~(\ref{eq: partial Ronkin symmetry})]. 
The bottom panels [Fig.~\ref{fig: WHF partial Ronkin}(d-f)] display the partial indices $\mathcal{K}[[\tau_+^\mathrm{symp}]_\mu] = (\kappa_1^{(\mu)}, \kappa_2^{(\mu)})$ of the factor $[\tau_+^\mathrm{symp}]_\mu$.
Given that their transposed counterparts follow $\mathcal{K}[[\tau_-^\mathrm{symp}]_\mu] = (-\kappa_2^{(-\mu)}, -\kappa_1^{(-\mu)})$,
the negative region $\mu \le 0$ is shaded in gray to focus on the positive sector.
Each index changes when $\mu$ crosses a root $\mu_j$ where $\tau$ closes its gap.
The sum $\kappa_1^{(\mu)} + \kappa_2^{(\mu)}$ corresponds to the winding number of $[\tau_+^\mathrm{symp}]_\mu$, which in turn equals the derivative of $R_\tau^{(+)}$.

As shown in Fig.~\ref{fig: WHF partial Ronkin}(f) for the $\kappa=0$ phase, the partial indices vanish within the interval $0 \le \mu \le \mu_1$. 
This includes the vicinity of the origin, thereby validating the conventional optimization of the total Ronkin function.
However, for the $\kappa=1$ phase [Fig.~\ref{fig: WHF partial Ronkin}(d, e)], the behavior is distinct.
Strikingly, we find that the partial indices $\kappa_{1,2}^{(\mu)}$ vanish simultaneously within the finite interval $\mu_{1} \le \mu \le \mu_{2}$; thus, $\mathcal K^*[\tau_+^\mathrm{symp}]=(0,0)$.
The emergence of this interval provides direct numerical evidence that the WHF framework is fully consistent with the non-Bloch band theory for class~AII$^\dagger$ systems: it coincides precisely with the optimization window $[\mu_{1},\mu_{2}]$ dictated by the GBZ condition \cite{KKawabata_PRB_2020_NonBlochBand}.
From the perspective of the formulation based on the WHF, the condition $\mathcal{K}^*[\tau_+^\mathrm{symp}]=(0,0)$ precisely marks the regime in which the generalized Szeg\"o limit theorem [Eq.~(\ref{eq: generalized Szego for class AIId with trivial residual partial indices})] holds exactly, without requiring any additional correction terms.
 
Finally, as a comprehensive validation of our framework, we compute the OBC spectral DOS by evaluating the OBC potential through the WHF-based optimization [Eq.~(\ref{eq: WHF Ronkins})] across the entire complex energy plane. 
The resulting distribution is shown in Fig.~\ref{fig: DOS kappa=1}, where each point represents an energy cell of size 
$\Delta\mathrm{Re}\,E \times \Delta\mathrm{Im}\,E \simeq 4 \times 10^{-4}$, used for discretizing the Hessian in Eq.~(\ref{eq: DOS and Potential}). 
The WHF-based decomposition accurately reproduces the OBC spectrum, in full agreement with Fig.~\ref{fig: l_config_z2} and consistent with the results obtained from the symmetry-decomposed Ronkin functions. 
This excellent correspondence confirms both the validity and the practical applicability of our method across the $\kappa = 0$ and $\kappa = 1$ regimes.

\subsubsection{Regime including $\kappa = 2$}
To demonstrate that Eq.~(\ref{eq: generalized Szego for class AIId}) also holds in parameter regimes with partial index $\kappa = 2$, we analyze the same Hamiltonian using parameters $t_1 = 0.3$, $t_2 = 0.5$, $g_2 = 1.0$, $\Delta_1 = 0.0$, and $\Delta_2 = 0.9$.

\begin{figure}
    \centering
    \includegraphics[width=0.9\linewidth]{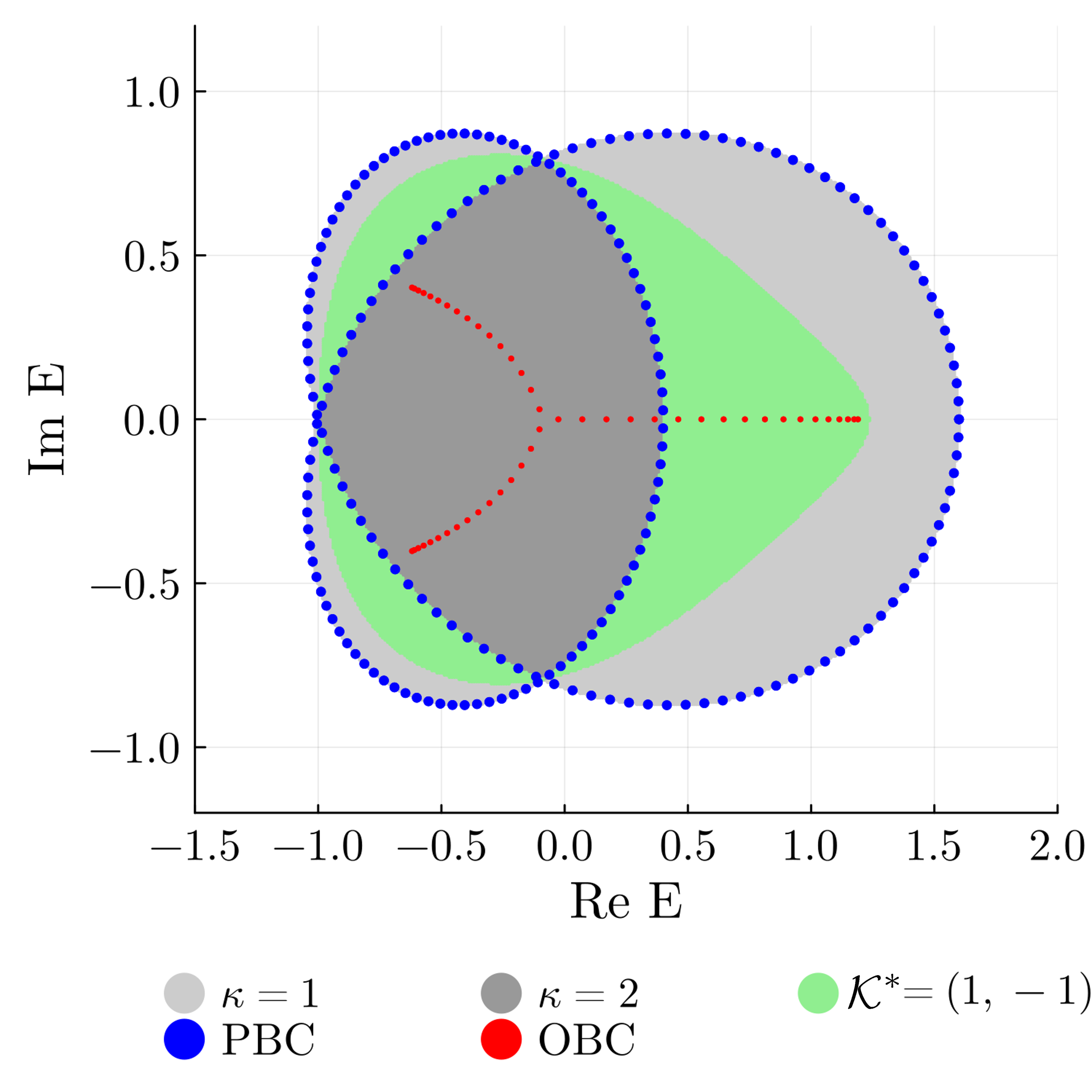}
    \caption{
    Map of the OBC and PBC spectra with WHF-derived partial indices of the regime, including a $\kappa = 2$ phase.
    Blue and red points denote the PBC ($N = 200$) and OBC ($N = 50$) spectra, respectively.
    Model parameters are $t_1 = 0.3$, $t_2 = 0.5$, $g_2 = 1.0$, $\Delta_1=0.0$, and $\Delta_2 = 0.9$.
    The shaded regions indicate the WHF-derived partial index: $\kappa = 1$ (light gray) and, notably, a $\kappa = 2$ phase (dark gray).
    The unshaded (white) region corresponds to the trivial $\kappa = 0$ phase.
    The green overlay marks a parameter domain where the partial indices do not vanish simultaneously, leading to nonvanishing residual partial indices $\mathcal{K}^* = (+1, -1)$.
    }
    \label{fig: l_config_kappa_2}
\end{figure}

Figure~\ref{fig: l_config_kappa_2} shows the spectrum and partial index in the complex energy plane.
The blue and red points denote the PBC and OBC spectra, respectively, obtained using $N=200$ (PBC) and $N=50$ (OBC) sites.
The partial index $\kappa$ of $\tau = (E - h)U_T$ takes the value $1$ in the light-gray region, $2$ in the dark-gray region, and $0$ elsewhere.
Because the numerical diagonalization of this two-band Hamiltonian under OBC becomes increasingly demanding with system size, we restrict the calculation to $50$ sites for the OBC case, which still captures the essential spectral features.

Crucially, in the $\kappa=2$ region (dark gray), the $\mathbb{Z}_2$ topological invariant [Eq.~(\ref{eq: Z2 topological invariant})] yields $\nu[\tau]=0$, implying topological triviality from the $\mathbb{Z}_2$ viewpoint.
Yet, this regime reveals a previously unexplored structure: despite its $\mathbb{Z}_2$-trivial character, the partial indices of the symbol remain nonzero, indicating the presence of exponentially localized states under OBC, as typically associated with the NHSE.
This sets the $\kappa=2$ phase apart from both the conventional trivial ($\kappa=0$) and topological nontrivial ($\kappa=1$) cases.
Further evidence for the presence of a skin effect in this regime is provided by the OBC spectrum, which differs markedly from the PBC spectrum and shows pronounced boundary sensitivity. The combination of nonvanishing partial indices (signaling localization) and the pronounced boundary sensitivity of the OBC spectrum relative to the PBC spectrum provides clear evidence that a skin effect occurs in the $\kappa=2$ regime.
This skin effect is not protected by TRS$^\dagger$ symmetry but instead originates from the nontrivial $\kappa = 2$ winding.
The resulting $\kappa = 2$ skin modes are fragile: a small TRS$^\dagger$-preserving perturbation (e.g., $\Delta_1 = 0.01$) can remove them entirely without closing the point gap, confirming that these skin modes lack symmetry protection and arise solely from the winding structure.
For further details, see Appendix~\ref{appendix: instability}.

Having established the existence of the $\kappa = 2$ phase, we now examine the optimization of the factor $[\tau_+^\mathrm{symp}]_\mu$ [from Eq.~(\ref{eq: WHF in symplectic form})], which plays a central role in decomposing the Ronkin function.
Our WHF analysis reveals a complication: a parameter region where the partial indices of $[\tau_+^\mathrm{symp}]_\mu$ do not vanish simultaneously, leaving residual partial indices $\mathcal{K}^*[\tau_+^\mathrm{symp}] \neq (0,0)$ even after finding the optimal $\mu$, as indicated by the green area in Fig.~\ref{fig: l_config_kappa_2}.
Within this green region, the optimization yields $\mathcal{K}^*[\tau_+^\mathrm{symp}] = (1,-1)$ [or $(-1,1)$], demonstrating that a straightforward optimization of the decomposed Ronkin function is insufficient and that a correction for the nonvanishing partial indices is required, analogous to the multiband class~A case.

\begin{figure*}
    \centering
    \begin{subfigure}{0.325\textwidth}
        \subcaption{}
        \vspace{-15pt}
        \includegraphics[width=0.85\linewidth]{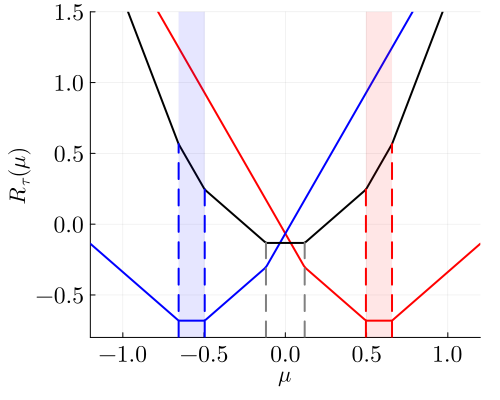}
    \end{subfigure}
    \hfill
    \begin{subfigure}{0.325\textwidth}
        \subcaption{}
        \vspace{-15pt}
        \includegraphics[width=0.85\linewidth]{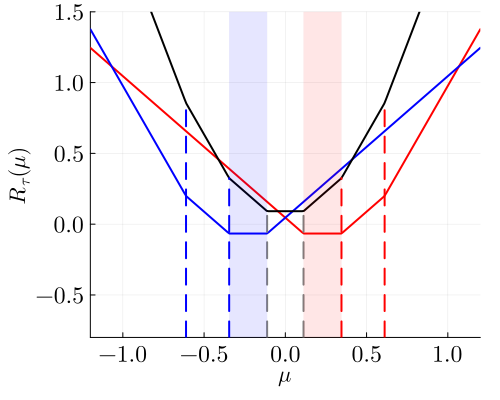}
    \end{subfigure}
    \hfill
    \begin{subfigure}{0.325\textwidth}
        \subcaption{}
        \vspace{-15pt}
        \includegraphics[width=0.85\linewidth]{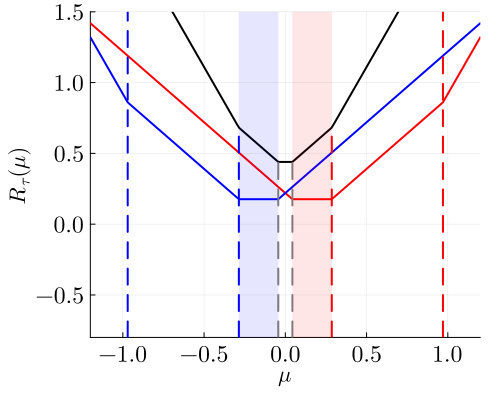}
    \end{subfigure}

    \vspace{1em}
    
    \begin{subfigure}{0.325\textwidth}
        \subcaption{}
        \vspace{-15pt}
        \includegraphics[width=0.85\linewidth]{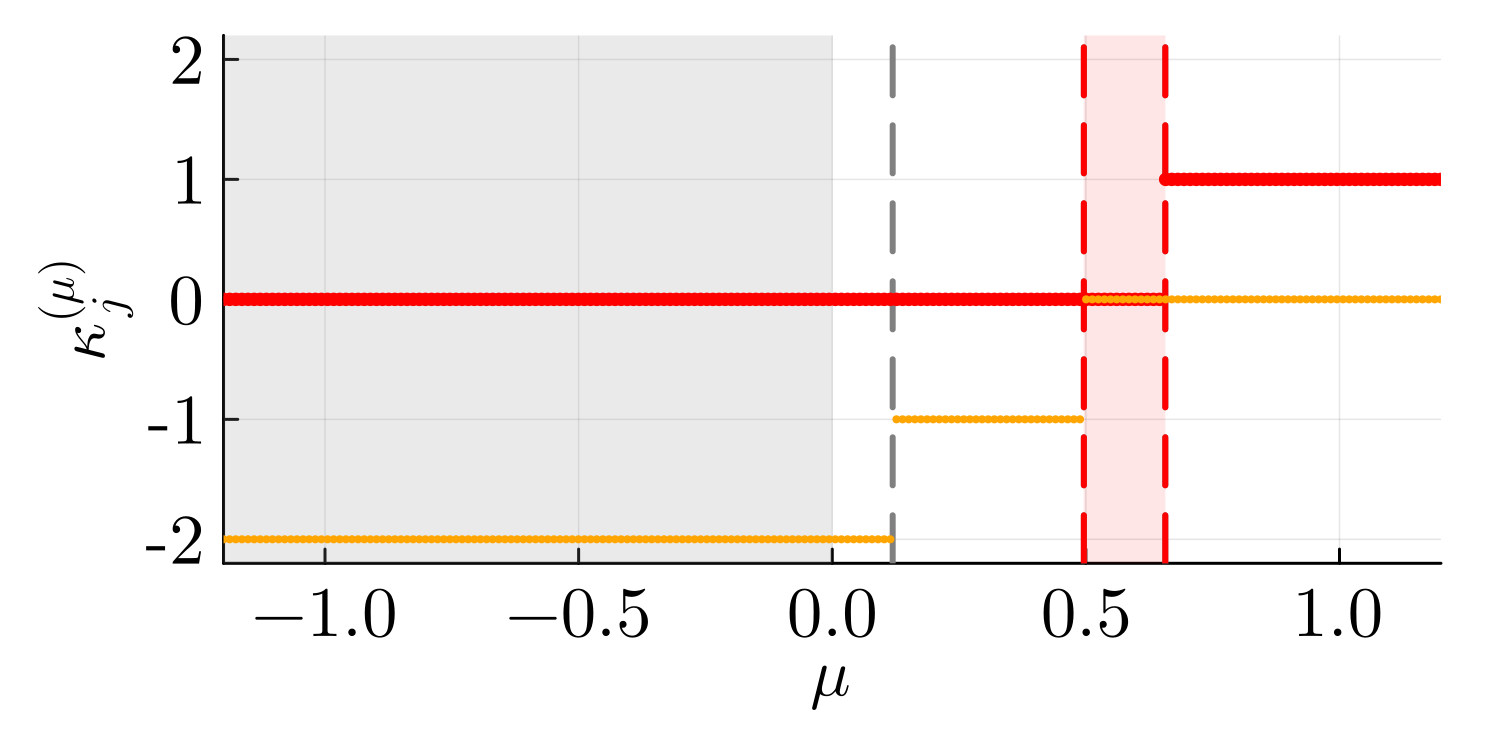}
    \end{subfigure}
    \hfill
    \begin{subfigure}{0.325\textwidth}
        \subcaption{}
        \vspace{-15pt}
        \includegraphics[width=0.85\linewidth]{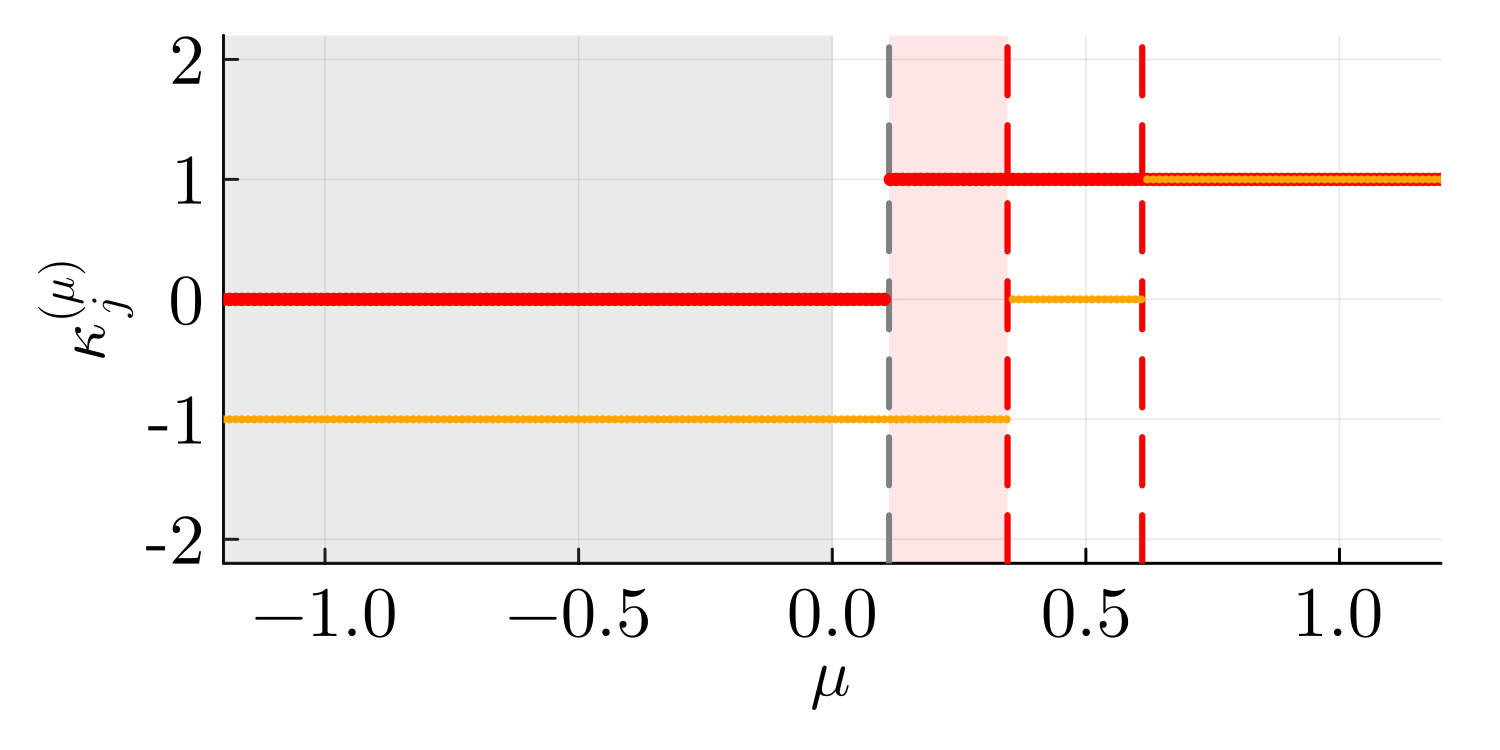}
    \end{subfigure}
    \hfill
    \begin{subfigure}{0.325\textwidth}
        \subcaption{}
        \vspace{-15pt}
        \includegraphics[width=0.85\linewidth]{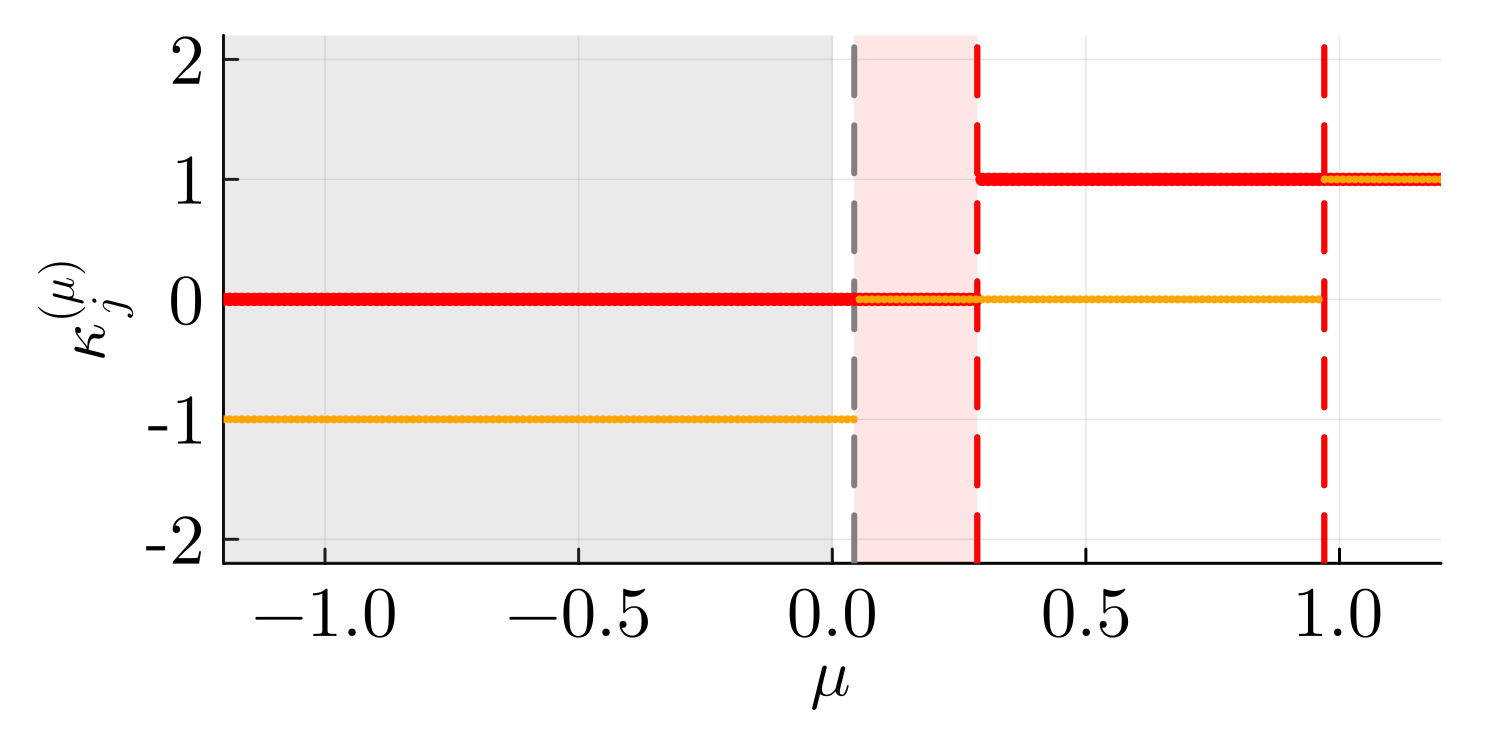}
    \end{subfigure}
    \caption{
    Partial indices of $[\tau_+^\mathrm{symp}]_\mu$ for three representative energies:
    (a) $E = -0.5 + 0.3i$ with $\kappa = 2$,
    (b) $E = 0.6 + 0.2i$ with $\kappa = 1$, and
    (c) $E = 0.9 + 0.7i$ with $\kappa = 1$.
    Red and blue vertical lines correspond to the GBZ condition $\mu_2=\mu_3$.
    The shaded regions indicate the target intervals for optimization: $[\mu_2, \mu_3]$ for the $\kappa=2$ phase [(a)] and $[\mu_1, \mu_2]$ for the $\kappa=1$ phase [(b, c)].
    Outside the green region in Fig.~\ref{fig: l_config_kappa_2} [(a), (c)], $\kappa_1^{(\mu)}$ and $\kappa_2^{(\mu)}$ vanish simultaneously within the shaded interval, resulting in vanishing residual partial indices.
    Inside the green region [(b)], however, the partial indices never vanish simultaneously, indicating the breakdown of the simple optimization formalism due to nonvanishing residual partial indices.
    }
    \label{fig: kappa=2: partial indices vs mu}
\end{figure*}

Figure~\ref{fig: kappa=2: partial indices vs mu} illustrates this behavior.
The top panels [Fig.~\ref{fig: kappa=2: partial indices vs mu}(a-c)] display the WHF-based decomposed Ronkin functions, while the bottom panels [Fig.~\ref{fig: kappa=2: partial indices vs mu}(d-f)] show the $\mu$-dependence of the partial indices, $\mathcal{K}[[\tau_+^\mathrm{symp}]_\mu] = (\kappa_1^{(\mu)}, \kappa_2^{(\mu)})$.
In each panel, the shaded regions indicate the optimization interval dictated by the topological phase: $[\mu_2, \mu_3]$ for the $\kappa=2$ phase [Fig.~\ref{fig: kappa=2: partial indices vs mu}(a, d)] and $[\mu_1, \mu_2]$ for the $\kappa=1$ phase [Fig.~\ref{fig: kappa=2: partial indices vs mu}(b, c, e, f)].

For the $\kappa=2$ case [Fig.~\ref{fig: kappa=2: partial indices vs mu}(a)] and the ordinary $\kappa=1$ case [Fig.~\ref{fig: kappa=2: partial indices vs mu}(c)], the decomposed Ronkin function $R_{\tau}^{(+)}$ (red line) is correctly minimized within the respective red shaded intervals, where the partial indices $\mathcal{K}[[\tau_+^\mathrm{symp}]_\mu]$ vanish simultaneously [Fig.~\ref{fig: kappa=2: partial indices vs mu}(d, f)]. 
This confirms that the simple optimization formalism remains valid in these regimes.
In contrast, for the intermediate $\kappa=1$ regime shown in Fig.~\ref{fig: kappa=2: partial indices vs mu}(b, e) (corresponding to the green region in Fig.~\ref{fig: l_config_kappa_2}), the partial indices do not vanish within the candidate interval $[\mu_1, \mu_2]$.
These non-vanishing residual partial indices imply that the simple optimization fails, necessitating the correction term in Eq.~(\ref{eq: generalized Szego for class A with (+1, -1)}).

\begin{figure}
    \centering
    \includegraphics[width=0.95\linewidth]{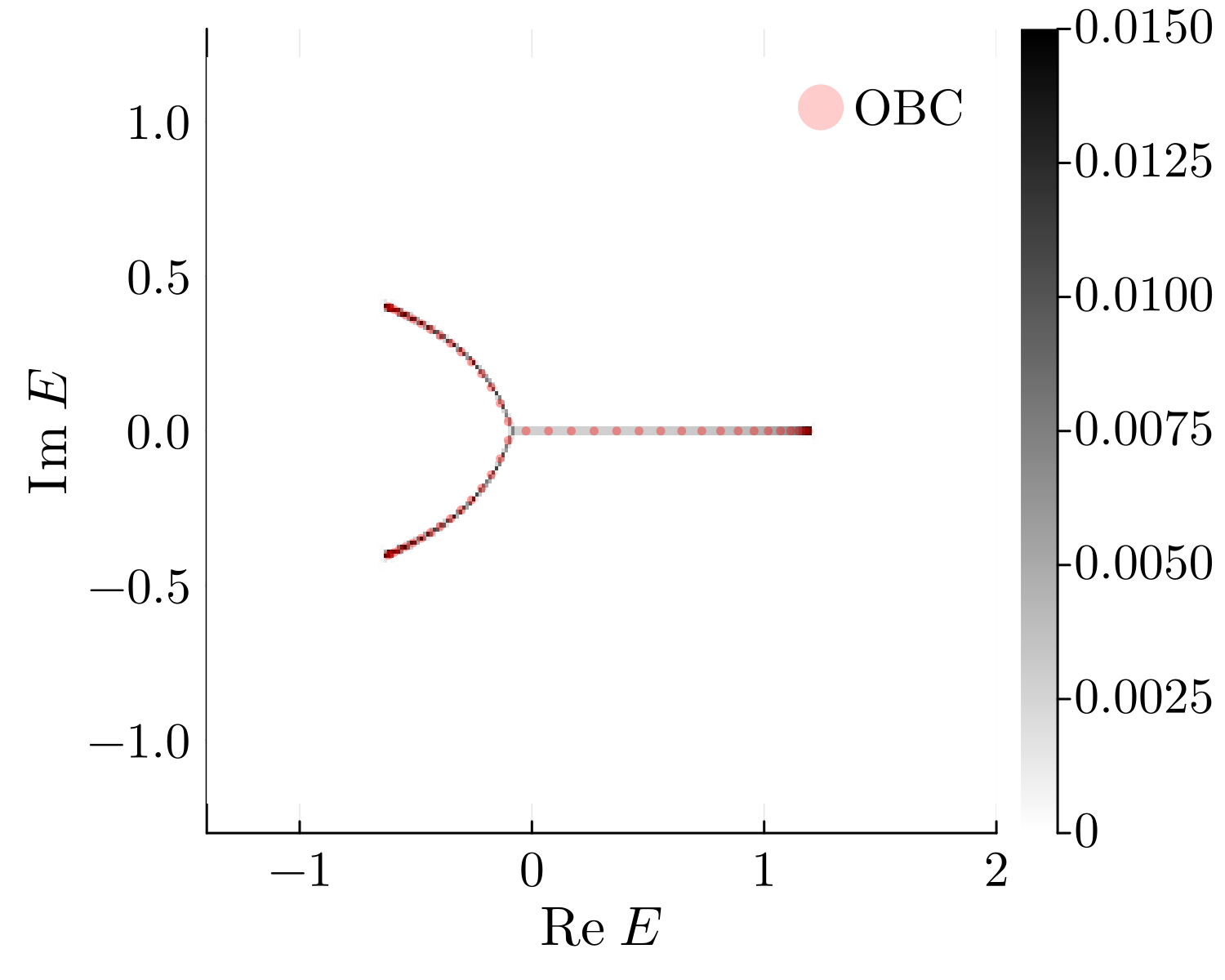}
    \caption{
    DOS multiplied by the surface element $\Delta\mathrm{Re}~E \times \Delta\mathrm{Im}~E \simeq 3 \times 10^{-4}$.
    The Hamiltonian and parameters are identical to those in Fig.~\ref{fig: l_config_kappa_2}.
    The OBC spectral potential is obtained by optimizing the WHF-based decomposed Ronkin functions [Eq.~(\ref{eq: WHF Ronkins})] together with the correction term accounting for nonvanishing partial indices [Eq.~(\ref{eq: generalized Szego for class AIId with nontrivial residual partial indices})].
    The resulting WHF-based formulation accurately reproduces the OBC spectrum (red line, $N=50$).
    }
    \label{fig: DOS kappa=2}
\end{figure}

The resulting OBC DOS, multiplied by the surface element $\Delta\mathrm{Re}~E \times \Delta\mathrm{Im}~E \simeq 3 \times 10^{-4}$, is plotted in Fig.~\ref{fig: DOS kappa=2}.
The OBC potential is obtained by optimizing the decomposed Ronkin functions defined in Eq.~(\ref{eq: WHF Ronkins}), together with the correction in Eq.~(\ref{eq: generalized Szego for class AIId with nontrivial residual partial indices}).
By combining these two relations, the WHF-based formulation successfully reproduces the OBC spectrum shown in Fig.~\ref{fig: l_config_kappa_2}.

\begin{figure}
    \centering
    \begin{subfigure}{0.95\textwidth}
        \centering
        \subcaption{}
        \vspace{-15pt}
        \includegraphics[width=0.85\linewidth]{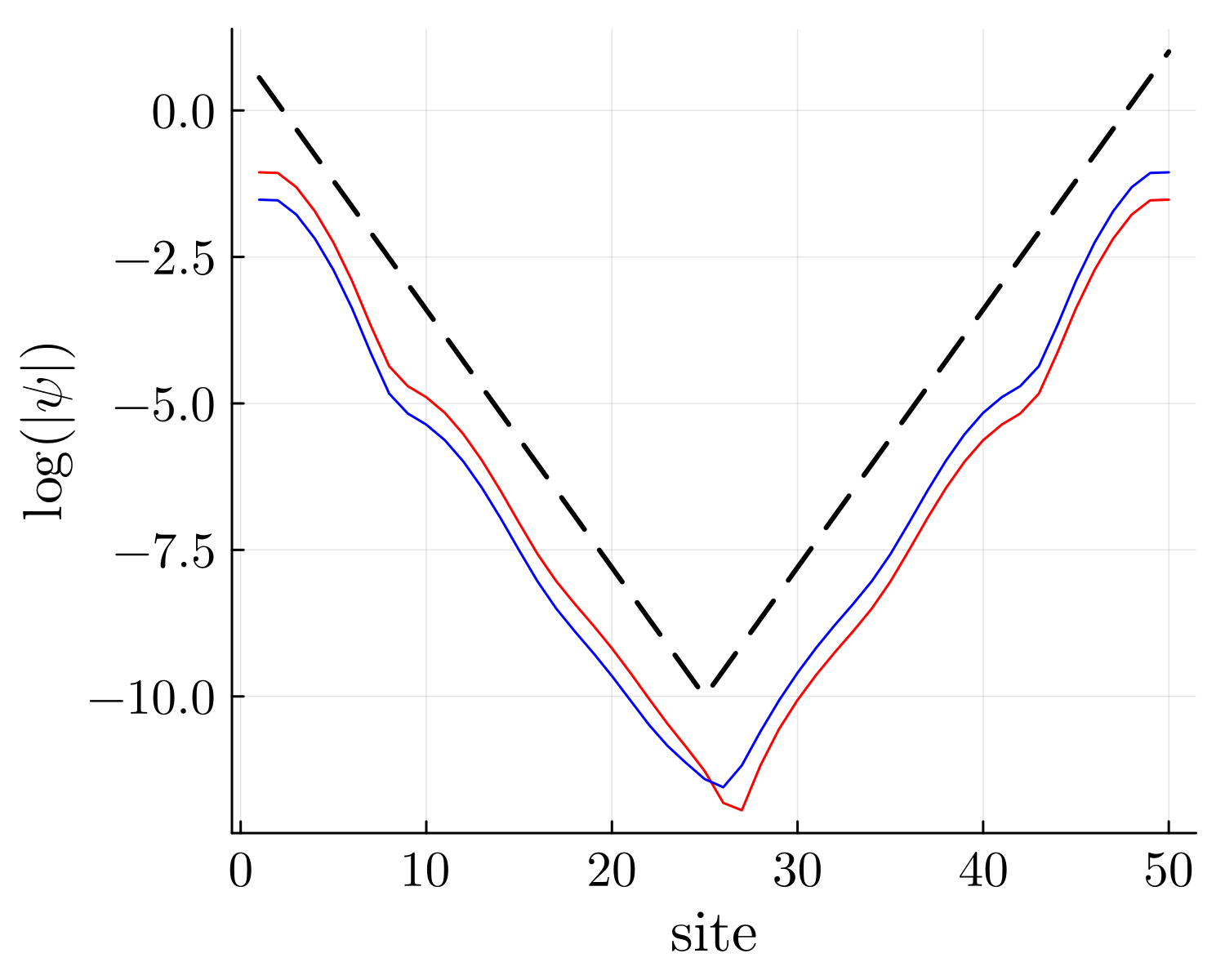}
    \end{subfigure}
    \begin{subfigure}{0.95\textwidth}
        \centering
        \subcaption{}
        \vspace{-15pt}
        \includegraphics[width=0.85\linewidth]{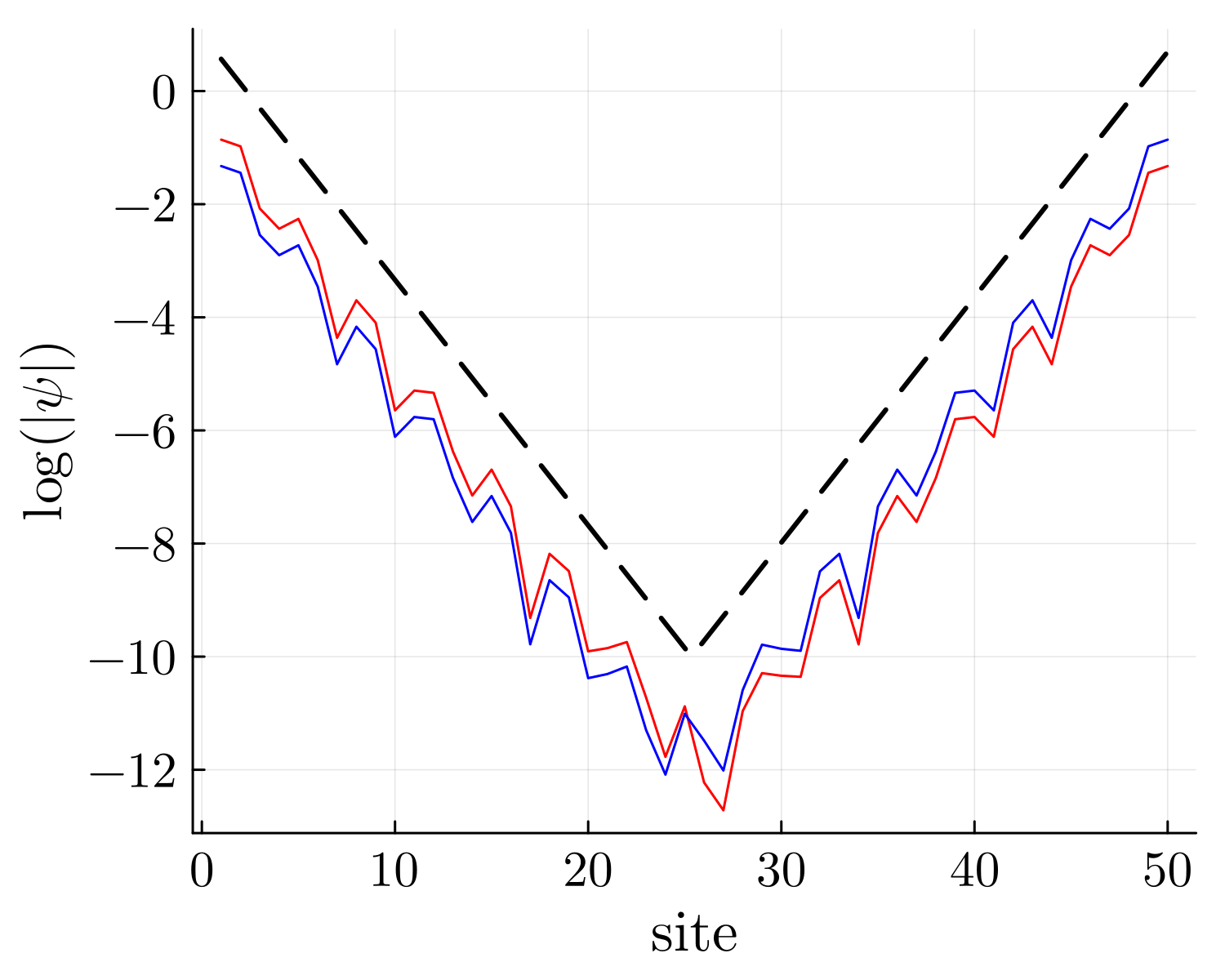}
    \end{subfigure}
    \caption{
    Fitting of wave functions for (a) $\kappa = 1$ and (b) $\kappa = 2$.
    The chain length of the OBC system is $N = 50$ sites.
    The corresponding eigenenergies are (a) $E = 1.02 + 0.06i$ and (b) $E = -0.13 + 0.08i$.
    The red and blue lines represent the wave functions of the $+1$ and $-1$ sectors of the Pauli matrix $Z$, respectively.
    The black dotted line shows the fitted profile using the localization lengths determined from the optimization, corresponding to $\mu_2 = \mu_3$.
    }
    \label{fig: WH and fitting}
\end{figure}

Our numerical analysis reveals that the conventional GBZ condition $\mu_1 = \mu_2$~\cite{KKawabata_PRB_2020_NonBlochBand} is replaced by $\mu_2 = \mu_3$.
This modified GBZ condition is consistent with the WHF decomposed Ronkin functions and directly dictates the localization length of the eigenmodes. 
We interpret this behavior as a manifestation of the $\kappa = 2$ phase, which, as discussed in Appendix~\ref{appendix: instability}, is generally fragile.

To verify the accuracy of the localization length, Fig.~\ref{fig: WH and fitting} shows fits of the OBC eigenstates ($N = 50$) using the functional form $e^{-\mu_2(x-25)} + e^{\mu_2(x-25)}$, where the values of $\mu_2 (=\mu_3)$ are obtained from the WHF-based optimization process.
The red and blue lines represent the wave functions of the $+1$ and $-1$ sectors of the Pauli matrix $Z$, respectively.
For $E = 1.02 + 0.06i$ [Fig.~\ref{fig: WH and fitting}(b)], the partial index is $\kappa = 1$, and the corresponding eigenmode exhibits a TRS$^\dagger$-protected $\mathbb{Z}_2$ skin mode.
In contrast, for $E = -0.13 + 0.08i$ [Fig.~\ref{fig: WH and fitting}(a)], the partial index is $\kappa = 2$, and the corresponding eigenmode realizes a non-protected skin mode.
The fitting form, based on the optimized $\mu_2$, accurately reproduces the spatial profiles in both regimes.

\section{Conclusion} \label{sec: conclusion}
In this work, we have established the Wiener-Hopf factorization (WHF) as a powerful framework for extending the Amoeba formulation to multiband non-Hermitian systems.
Our approach resolves a key limitation of the existing formalism: although the generalized Szeg\"o limit theorem is, in principle, valid in arbitrary dimensions, its practical application has been restricted to single-band systems due to complications arising from competing localization lengths in multiband settings, an effect that becomes unavoidable in the presence of symmetry-protected degeneracies.

By combining the WHF with Hermitian doubling, we have provided a clear physical interpretation of the generalized Szeg\"o limit theorem: edge modes in the doubled Hermitian Hamiltonian generate the shift between the OBC and PBC spectral potentials.
This perspective naturally identifies the precise criterion for the validity of the simple Amoeba optimization, i.e., that all residual partial indices must vanish.
When this condition holds, the conventional generalized Szeg\"o limit theorem applies directly; when it does not, we derived the corresponding correction terms [Eq.~(\ref{eq: generalized Szego for class A with (+1, -1)})], which account for the nonvanishing residual partial indices.

For systems with transpose-type time-reversal symmetry (class AII$^\dagger$), we showed that the WHF naturally yields the symmetry-decomposed Ronkin functions introduced phenomenologically in Ref.~\cite{SKaneshiro_PRB_2025_SymplecticAmoebaFormulation}.
Through the symplectic form of the WHF [Eq.~(\ref{eq: WHF in symplectic form})], we obtain a rigorous decomposition $R_\tau(\mu) = R_\tau^{(+)}(\mu) + R_\tau^{(-)}(\mu)$, directly linking the WHF to the non-Bloch band theory.
This decomposition is not merely a computational convenience but reflects the intrinsic structure of Kramers pairs in these systems.

Our WHF-based framework reveals a fundamental connection between the topological structure of non-Hermitian Hamiltonians and the asymptotic behavior of their OBC determinants.
The partial indices, which encode the number and localization of boundary modes, directly control the deviations between OBC and PBC spectral potentials.
This insight extends the modified Szeg\"o limit theorem, originally formulated for Hermitian topological systems~\cite{EBasor_JStatPhys_2019_ModifiedSzegoWidomAsymptotics}, into the non-Hermitian domain, as the generalized Szeg\"o limit theorem~\cite{HWang_PRX_2024_AmoebaFormulationOf}.
The breakdown of the simple optimization formalism in multiband systems is thus not a deficiency but a natural manifestation of nontrivial topological structure.
The required correction terms acquire a transparent physical interpretation: they describe the exchange of root contributions across the Brillouin zone boundary.

While this work focused on one-dimensional systems in classes A and AII$^\dagger$, the same mechanism is expected to extend to certain higher-dimensional cases~\cite{KYokomizo_PRB_2023_NonBlochBands}, though its full generalization to higher dimensions remains an open challenge.
This is primarily because the extension requires a multivariable Wiener-Hopf factorization, for which a general mathematical theory has not yet been established.
Conversely, however, the relationship between the accumulation of topological edge and skin modes and the deviation of the spectral potential may provide a physical guiding principle for constructing a generalized Szeg\"o limit theorem in higher dimensions.

Whereas we focus on the case without edge modes, our formulation remains valid even in the presence of the edge modes.
Since the number of edge modes is finite, they do not provide the relevant contributions to the OBC spectrum under the thermodynamic limit.
Notably, the residual partial indices can detect the edge modes.
For details, see Appendix~\ref{appendix: RPI and BBC}.

In conclusion, the Wiener-Hopf factorization provides a unified mathematical foundation for the Amoeba formulation in multiband systems, resolving long-standing questions about the applicability of the generalized Szeg\"o limit theorem and establishing rigorous connections to non-Hermitian topology.
This framework not only validates existing phenomenological approaches but also paves the way toward systematic extensions across symmetry classes and spatial dimensions.

\section*{Acknowledgements}
S.K. deeply appreciates the fruitful discussions with D. Nakamura, K. Shimomura, and K. Kawabata.
This work is supported by JSPS, KAKENHI Grants No. JP24KJ1353 (S.K.).

\section*{Data Availability}
The data that support the findings of this article are not publicly available upon publication because it is not technically feasible and/or the cost of preparing, depositing, and hosting the data would be prohibitive within the terms of this research project. The data are available from the authors upon reasonable request.

\appendix
\section*{appendix}
\renewcommand{\thesubsection}{\Alph{subsection}}

\subsection{Instability of the $\kappa=2$ phase and modified GBZ condition}
\label{appendix: instability}

In this Appendix, we examine the fragility of the $\kappa = 2$ phase in class AII$^\dagger$ and show that its apparent stability originates from an underlying unitary symmetry. This hidden symmetry enforces the modified GBZ condition $\mu_2 = \mu_3$, replacing the conventional relation $\mu_1 = \mu_2$.

The $\kappa = 2$ phase is inherently fragile.
Within the framework of the Hermitian-doubled Hamiltonian, this phase corresponds to the presence of two Kramers pairs.
As in the case of $\mathbb{Z}_2$ topological phases, an even number of Kramers pairs is not protected against generic TRS$^\dagger$-preserving perturbations: pairs can hybridize and gap out each other, rendering the phase topologically trivial.

\begin{figure}
    \centering
    \includegraphics[width=0.95\linewidth]{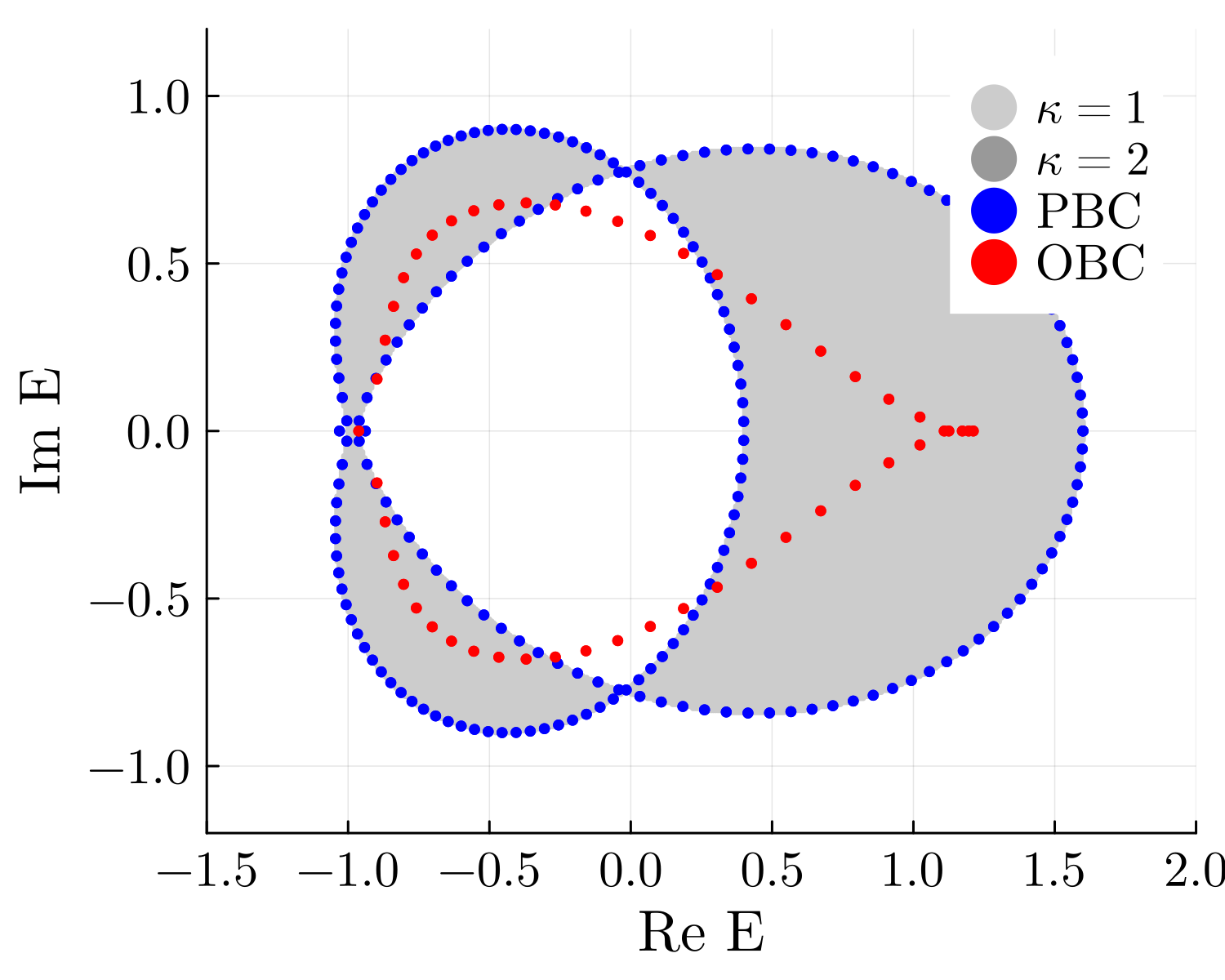}
    \caption{
    Test of the stability of the $\kappa = 2$ phase under a TRS$^\dagger$-preserving perturbation.
    The parameters are identical to those in Fig.~\ref{fig: l_config_kappa_2} except for $\Delta_1 = 0.01$.
    Introducing the perturbation $\Delta h(e^{ik}) = -2\Delta_1 X \sin k$ removes the $\kappa = 2$ phase, confirming its fragility.
    The original $\kappa=2$ phase changes to a $\kappa = 0$ phase without any qualitative change in the PBC spectrum, indicating that the bulk winding of the PBC Hamiltonian remains unchanged.
    }
    \label{fig: k2 instability}
\end{figure}

In our model, this inherent fragility can explicitly be confirmed.
The $\kappa = 2$ phase, which exists for $\Delta_1 = 0$, disappears upon introducing a TRS$^\dagger$-preserving perturbation $\Delta h(e^{ik}) = -2\Delta_1 X \sin k$, even for a small value $\Delta_1 = 0.01$, as shown in Fig.~\ref{fig: k2 instability}.
Under this perturbation, the $\kappa = 2$ in the system changes to a $\kappa = 0$ phase without appreciable change in the PBC spectrum. 
The winding in the PBC spectrum remains unchanged, indicating that the underlying bulk topology of the PBC bands is unaffected.

This behavior can be understood by analyzing the symmetries of the unperturbed Hamiltonian $h(k)$ (i.e., for $\Delta_1 = 0$):
\begin{align}
h(e^{ik}) = 2 t_1 \cos k + 2 t_2 \cos 2k - 2 (\Delta_2 X + i g_2 Z) \sin 2k.
\end{align}
The Hamiltonian $h(k)$ can be diagonalized by a $k$-independent unitary matrix $V$,
\begin{align}
V^{-1} h V = \mqty(\dmat{h_+, h_-}),
\end{align}
where each block is given by
\begin{align}
h_{\pm}(e^{ik}) = 2 t_1 \cos k + 2 t_2 \cos 2k \pm 2 i Z \sqrt{g_2^2 - \Delta_2^2} \sin 2k.
\end{align}
This diagonal form reveals an additional hidden unitary symmetry of the unperturbed Hamiltonian,
\begin{align}
[h, V Z V^{-1}] = 0.
\end{align}
Owing to this symmetry, the system decouples into two independent blocks $h_+$ and $h_-$, corresponding respectively to Kramers partners with partial indices $\pm \kappa = W[h_\pm]$.
As a result, hybridization between the Kramers pairs is prohibited, which stabilizes the $\kappa = 2$ phase for these specific parameters. 

The perturbation $\Delta h(e^{ik}) = -2\Delta_1 X \sin k$, however, breaks this hidden symmetry, since the perturbed Hamiltonian $h(e^{ik}) + \Delta h(e^{ik})$ can no longer be diagonalized by a $k$-independent matrix.
As a consequence, the two Kramers pairs hybridize and gap each other out, leading to the disappearance of the $\kappa = 2$ phase.
This demonstrates that the $\kappa = 2$ phase is not protected by TRS$^\dagger$ alone but instead relies on the additional hidden unitary symmetry $V Z V^{-1}$.

This hidden symmetry also drives the shift of the GBZ condition from the conventional $\mu_1 = \mu_2$~\cite{KKawabata_PRB_2020_NonBlochBand} to $\mu_2 = \mu_3$.
The intrinsic instability of the $\kappa = 2$ phase underlies the emergence of the green region in Fig.~\ref{fig: l_config_kappa_2}, the parameter domain where $\mathcal{K}^* = (1, -1)$ and whose boundary satisfies the conventional relation $\mu_1 = \mu_2$.
In our specific model, both the $\kappa = 2$ phase and this $\mathcal{K}^* \neq (0, 0)$ region are stabilized by a hidden unitary symmetry.
When a small TRS$^\dagger$-preserving perturbation breaks this symmetry, the $\kappa = 2$ phase disappears, and the OBC spectrum expands to the boundary of the green region, thereby restoring the standard GBZ condition.
Thus, the nonvanishing partial indices and the accompanying “abnormal” GBZ relation ($\mu_2 = \mu_3$) are fragile artifacts of this hidden symmetry rather than a robust modification of the GBZ itself.

\begin{figure}[ht]
    \centering
    \includegraphics[width=0.95\linewidth]{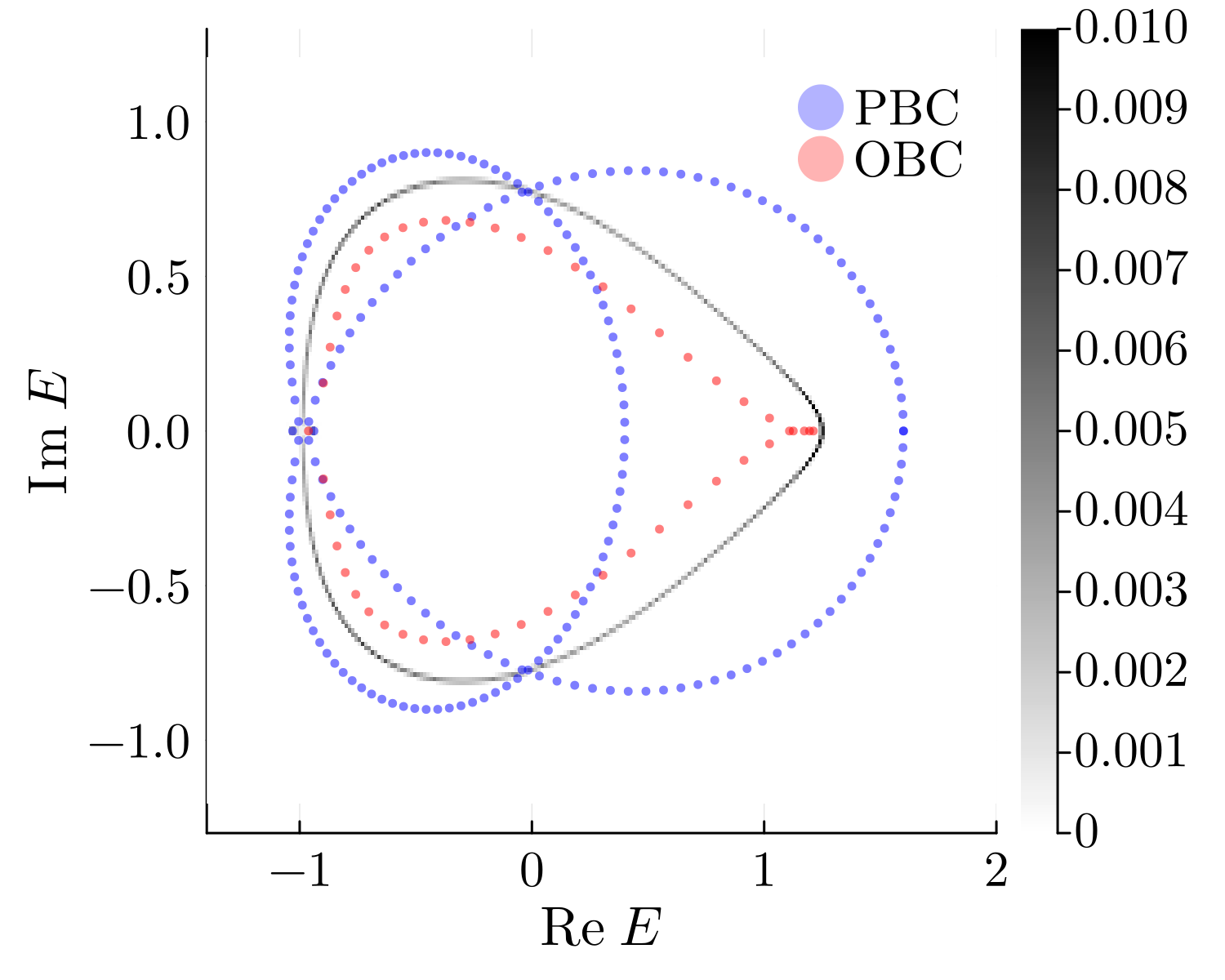}
    \caption{
    DOS multiplies by surface element $\Delta \Re~E \times \Delta \Im~E \simeq 2\times 10^{-4}$. 
    The Hamiltonian and parameters are identical to those in Fig.~\ref{fig: k2 instability}.
    Due to the finite-size effect, the OBC spectrum shifts from the Amoeba results. 
    However, these results coincide in the thermodynamic limit.
    }
    \label{fig: DoS perturbed}
\end{figure}

Figure~\ref{fig: DoS perturbed} corroborates this discussion by showing the OBC DOS of the perturbed Hamiltonian, scaled by the surface element $\Delta \mathrm{Re},E \times \Delta \mathrm{Im},E \simeq 2\times10^{-4}$.
Because the perturbation breaks the hidden symmetry, the conventional GBZ condition $\mu_1 = \mu_2$ is restored.
Due to finite-size effects, the OBC spectrum obtained by direct diagonalization ($N=50$, red dotted line) exhibits a small shift relative to the WHF-based results, which represent the $N \to \infty$ limit.
Nevertheless, both results are consistent within the thermodynamic limit.

\subsection{Residual partial indices and bulk-boundary correspondence}
\label{appendix: RPI and BBC}

\begin{figure*}
    \centering
    \begin{subfigure}{0.47\textwidth}
        \centering
        \subcaption{}
        \vspace{-15pt}
        \includegraphics[width=0.85\linewidth]{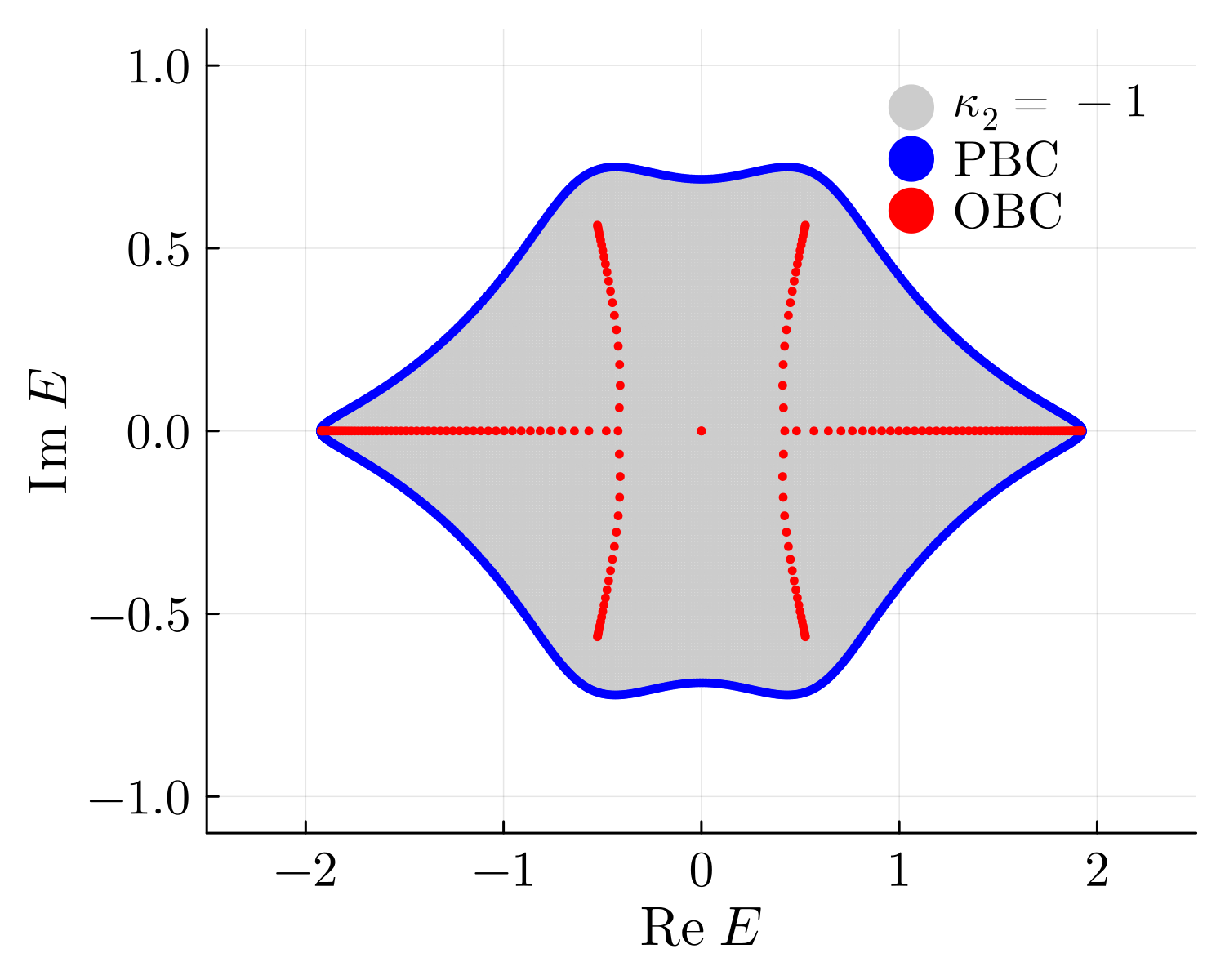}
    \end{subfigure}
    \begin{subfigure}{0.47\textwidth}
        \centering
        \subcaption{}
        \vspace{-15pt}
        \includegraphics[width=0.85\linewidth]{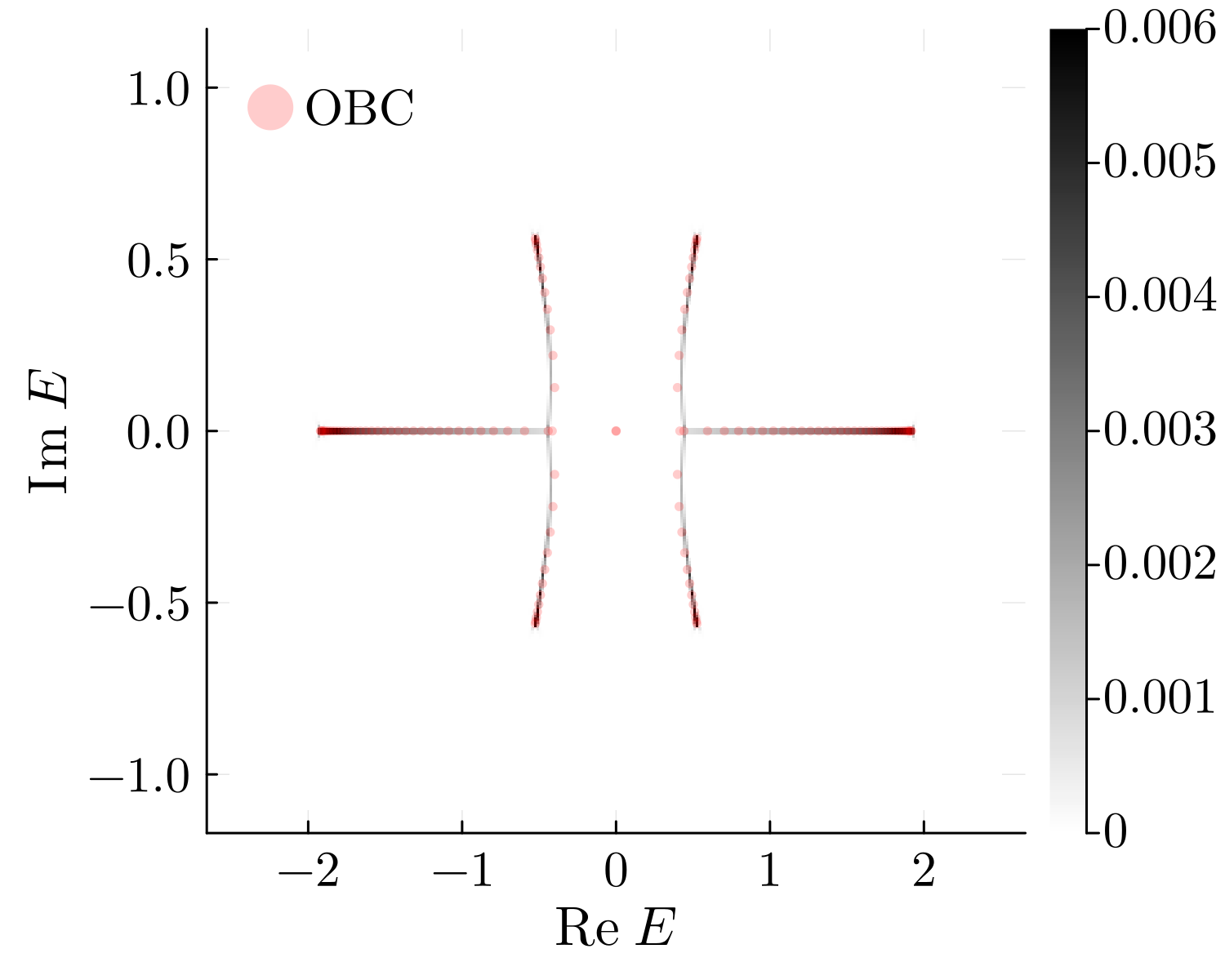}
    \end{subfigure}
    \caption{
    Numerical results for the system with SLS.
    The Hamiltonian and its parameters are defined in Eq.~\eqref{eq: Ham with SLS}.
    (a) PBC (blue) and OBC (red) spectra.
    The PBC spectrum forms a single loop, indicating the non-Hermitian skin effect, while the OBC spectrum exhibits a pair of zero-energy modes at the origin isolated from the two bulk bands.
    The partial indices $(\kappa_1, \kappa_2)$.
    $\kappa_1$ remains zero regardless of the reference point, whereas $\kappa_2$ captures the topological structure.
    (b) The DOS calculated from the OBC spectrum.
    Note that the zero-energy edge modes do not appear in the DOS, as their contribution vanishes in the thermodynamic limit.
    }
    \label{fig: results_SLS}
\end{figure*}

\begin{figure}
    \centering
    \begin{subfigure}{0.95\textwidth}
        \centering
        \subcaption{}
        \vspace{-15pt}
        \includegraphics[width=0.85\linewidth]{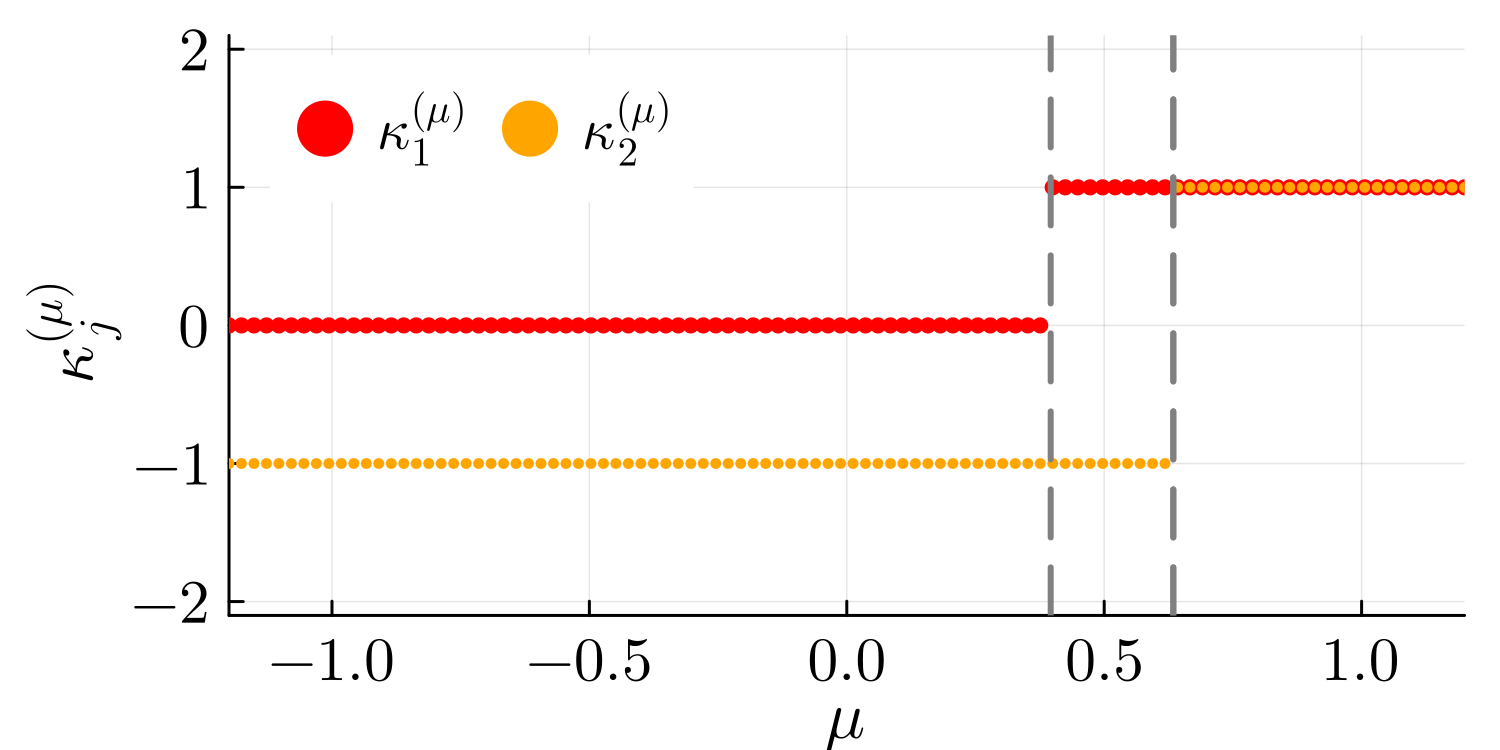}
    \end{subfigure}
    \begin{subfigure}{0.95\textwidth}
        \centering
        \subcaption{}
        \vspace{-15pt}
        \includegraphics[width=0.85\linewidth]{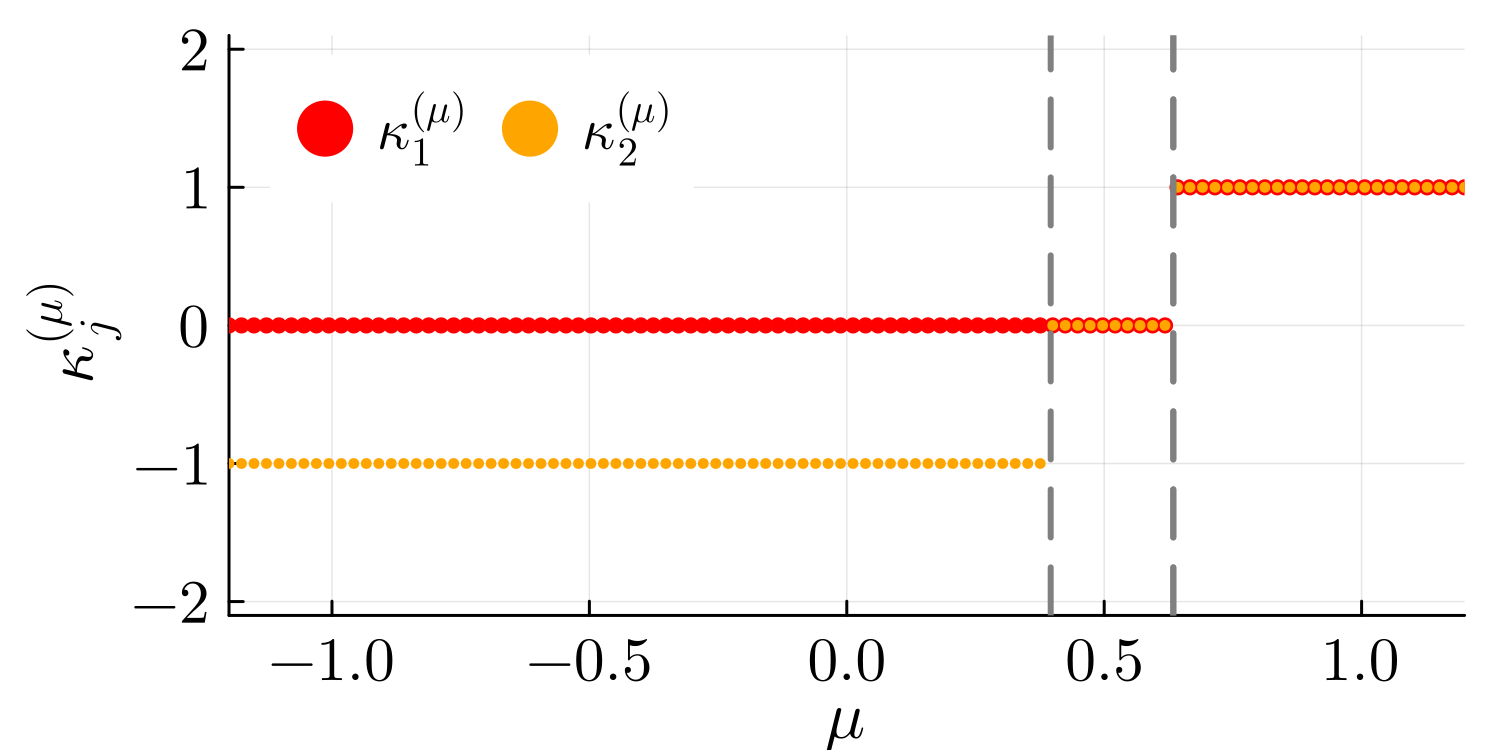}
    \end{subfigure}
    \caption{
    Residual partial indices calculated at reference energies (a) $E=0$ and (b) $E=0.01$.
    Red and orange lines represent the partial indices $\kappa_1^{(\mu)}$ and $\kappa_2^{(\mu)}$, respectively.
    Vertical dotted lines indicate the roots of $\det[h(\beta)]=0$.
    At $\mu=0$, the total winding number is non-zero ($W[h]=-1$), signaling the presence of the NHSE.
    However, nontrivial residual partial indices (quantized plateaus) are observed only at $E=0$ in (a), demonstrating that the proposed indices successfully detect the topological zero modes even in the presence of the skin effect.
    In contrast, no such topological structure is found at $E=0.01$ in (b).
    }
    \label{fig: residual_indices_SLS}
\end{figure}

In this appendix, we analyze systems with sublattice symmetry (SLS), in which topological edge modes coexist with skin modes.
We demonstrate that our formulation remains valid even in the presence of these edge modes and that the residual partial indices successfully capture their topological origin.

From the perspective of the spectral potential, a finite number of edge modes yields a vanishing contribution to the DOS in the thermodynamic limit.
Consequently, our Amoeba formulation for determining the bulk spectrum remains strictly valid.
Nevertheless, as we discuss below, the analysis based on the residual partial indices provides a powerful tool to detect and characterize these zero modes essentially.

In the following discussion, we focus on class A systems with SLS.
The SLS is defined by the relation
\begin{align}
    S h S^{-1} = -h,
\end{align}
where $S$ is a unitary matrix satisfying $S^2=1$.

In the presence of SLS, the Hamiltonian takes the canonical block-off-diagonal form in a proper basis (with $S=Z$),
\begin{align}
    h = \mqty(\admat{A, B}).
\end{align}
Accordingly, the SWHF can be generalized to respect the SLS structure as
\begin{align}
    h = \mqty(\dmat{A_+, B_-}) \mqty(\admat{D_A, D_B}) \mqty(\dmat{B_+, A_-}),
\end{align}
where $A=A_+ D_A A_-$ and $B=B_+ D_B B_-$ correspond to the WHF of the off-diagonal blocks $A$ and $B$, respectively.

Based on this block structure, the point-gap topology at $E=0$ is characterized by $\mathbb Z \oplus \mathbb Z$, with invariants corresponding to the two distinct winding numbers \cite{KKawabata_PRX_2019_SymmetryAndTopology}:
\begin{align}
(W[A], W[B]) = (W[D_A], W[D_B]).
\end{align}

Non-Hermitian systems exhibit another gap structure known as the line gap.
Here, we consider the line gap with respect to the imaginary axis, the real-line gap,
\begin{align}
    \det[h - i\eta] \neq 0
\end{align}
for any real number $\eta$.
Since the topological invariant is preserved under continuous deformations that keep the gap open, we have
\begin{align}
    W[h] = \lim_{\eta \to \infty} W[h-i\eta] = 0.
\end{align}

Topological phases characterized by a line gap are classified by continuous deformations to Hermitian Hamiltonians while preserving the gap.
In the Hermitian limit, the SLS reduces to the standard chiral symmetry, which enforces the vanishing of the total winding number, $W[h]=0$.
Therefore, the real-line gap opening ensures the Hamiltonian can be deformed to a Hermitian one.

Assuming the opening of a real-line gap, we can define the SLS-protected topological invariant as
\begin{align}
    W_\mathrm{SLS}[h] &= \int_0^{2\pi} \frac{dk}{4\pi i} \tr \qty[S h^{-1} \partial_k h] \\
    &= \frac{W[B]-W[A]}{2}.
\end{align}
which counts the number of edge modes \cite{ZGong_PRX_2018_TopologicalPhasesOf, KKawabata_PRX_2019_SymmetryAndTopology, DNakamura_PRL_2024_BulkBoundaryCorrespondence}.

However, in the presence of the NHSE, the winding number can be non-zero ($W[h] \neq 0$), which seemingly contradicts the connection to the Hermitian limit.
To resolve this consistency issue, we employ the non-Bloch band theory \cite{SYao_PRL_2018_EdgeStatesAnd, KYokomizo_PRL_2019_NonBlochBand}.
Strictly speaking, the invariant should be defined along the GBZ to enforce the vanishing winding number.
However, since the topological invariant is stable against gap-opening deformations, we can define the invariants on the circular path $|\beta|=e^\mu$, instead of the exact GBZ, such that
\begin{align}
    W[h_\mu] = W[A_\mu] + W[B_\mu] = 0.
\end{align}
This vanishing total winding ensures that the modified Hamiltonian $h_\mu$ is homotopic to a Hermitian Hamiltonian.

Based on this formulation, the SLS-protected winding number is given by
\begin{align}
    W_\mathrm{SLS}[h_\mu] = \frac{W[B_\mu]-W[A_\mu]}{2}.
\end{align}
Consequently, the topological properties are entirely determined by the residual partial indices defined in the main text.

We numerically analyze the non-Hermitian SSH model \cite{CHou_PRR_2022_DeterministicBulkBoundary}.
The Hamiltonian is defined as 
\begin{align}
\label{eq: Ham with SLS}
    h(\beta) &= \begin{pmatrix}   & R_+(\beta) \\ R_-(\beta) &   \end{pmatrix}, \\
    R_+ (\beta) &= t_{2R} \beta^{-1} + t_{1L} + t_{3L} \beta, \\
    R_- (\beta) &= t_{2L} \beta + t_{1R} + t_{3R} \beta^{-1}.
\end{align}
where $t_{1L}=1/5, t_{1R}=9/5$, $t_{2L}=6/5, t_{2R}=4/5$, $t_{3L}=9/40$ and $t_{3R}=1/40$.
The SLS operator is given by $S=Z$.

The OBC and PBC spectra are plotted in Fig.~\ref{fig: results_SLS}(a).
While the PBC spectrum forms a single loop, the OBC spectrum splits into two distinct bands and hosts a pair of zero modes at the origin.
The partial indices $(\kappa_1, \kappa_2)$, corresponding to the respective off-diagonal blocks, are shown in Fig.~\ref{fig: results_SLS}(b).
Since $\kappa_1$ vanishes regardless of the reference point, we only plot $\kappa_2$, which takes $-1$ inside the PBC spectrum.
The DOS is presented in Fig.~\ref{fig: results_SLS}(c).
As expected, since the finite number of edge modes yields a vanishing contribution to the OBC spectral potential in the thermodynamic limit, these zero modes are not visible in the DOS.

Figure~\ref{fig: residual_indices_SLS} illustrates the residual partial indices calculated at (a) $E=0$ and (b) $E=0.01$.
The red and orange lines represent the partial indices $\kappa_1^{(\mu)}$ and $\kappa_2^{(\mu)}$, respectively.
The vertical dotted lines indicate the roots of $\det h(\beta) =0$.
The total winding number $W[h] = \kappa_1 + \kappa_2$ takes a value of $-1$ at $\mu=0$, signaling the presence of the NHSE.
Crucially, however, nontrivial residual partial indices exist only at $E=0$ [Fig.~~\ref{fig: residual_indices_SLS}(a)].

Finally, we comment on cases where edge modes deviate from zero energy, such as in the non-Hermitian Rice-Mele model \cite{LLi_PRB_2024_DualBulkBoundary}.
While our formulation for determining the bulk spectrum remains valid, the detection of edge modes via residual partial indices requires additional care.
Since the partial indices characterize the kernel and cokernel of the Hamiltonian under the semi-infinite boundary condition $\mathcal{T}_\text{SIBC}[E-h]$, detecting non-zero edge modes requires performing the WHF specifically for the shifted symbol $E_\mathrm{edge}-h$.
A systematic framework for counting such dispersive edge modes is left for future work.

\bibliography{Z2_amoeba_2d}

\end{document}